\documentclass[aps,prc,twocolumn,showpacs,preprintnumbers,nofootinbib,float,superscriptaddress,longbibliography]{revtex4-2}
\usepackage{graphicx,fancybox}
\usepackage{amsmath,amssymb}
\usepackage[colorlinks=true, pdfstartview=FitV, linkcolor=red, citecolor=blue, urlcolor=blue]{hyperref}
\usepackage{color}

\begin{document}

\title{The effect of recoils on soft-drop-groomed observables in $\gamma$-tagged jets in a multistage approach}

\newcommand{\AiuAddress}{Akita International University, Yuwa, Akita-city 010-1292, Japan}
\newcommand{\McGillAddress}{Department of Physics, McGill University, Montr\'{e}al, Qu\'{e}bec H3A\,2T8, Canada}
\newcommand{\WaynePhysAddress}{Department of Physics and Astronomy, Wayne State University, Detroit, Michigan 48201, USA}
\newcommand{\ReginaAddress}{Department of Physics, University of Regina, Regina, Saskatchewan S4S 0A2, Canada}
\newcommand{\LLNLAddress}{Lawrence Livermore National Laboratory, Livermore, California 94550, USA}
\newcommand{\TexasAMCompAddress}{Research Computing Group, University Technology Solutions, The University of Texas at San Antonio, San Antonio, Texas 78249, USA}
\newcommand{\DukePhysAddress}{Department of Physics, Duke University, Durham, North Carolina 27708, USA}
\newcommand{\ShandongAddress}{Institute of Frontier and Interdisciplinary Science, Shandong University, Qingdao, Shandong 266237, China}
\newcommand{\MITNuclAddress}{Laboratory for Nuclear Science, Massachusetts Institute of Technology, Cambridge, Massachusetts 02139, USA}
\newcommand{\MITPhysAddress}{Department of Physics, Massachusetts Institute of Technology, Cambridge, Massachusetts 02139, USA}
\newcommand{\VandyAddress}{Department of Physics and Astronomy, Vanderbilt University, Nashville, Tennessee 37235, USA}
\newcommand{\UTNAddress}{Department of Physics and Astronomy, University of Tennessee, Knoxville, Tennessee 37996, USA}
\newcommand{\OrkRidgeAddress}{Physics Division, Oak Ridge National Laboratory, Oak Ridge, Tennessee 37830, USA}
\newcommand{\UCBerkeleyAddress}{Department of Physics, University of California, Berkeley, California 94270, USA}
\newcommand{\LBLAddress}{Nuclear Science Division, Lawrence Berkeley National Laboratory, Berkeley, California 94270, USA}
\newcommand{\GSIAddress}{GSI Helmholtzzentrum f\"{u}r Schwerionenforschung, 64291 Darmstadt, Germany}
\newcommand{\GoetheAddress}{Institute for Theoretical Physics, Goethe University, 60438 Frankfurt am Main, Germany}
\newcommand{\FIASAddress}{Frankfurt Institute for Advanced Studies, 60438 Frankfurt am Main, Germany}
\newcommand{\TexasAMCyclAddress}{Cyclotron Institute, Texas A\&M University, College Station, Texas 77843, USA}
\newcommand{\TexasAMPhysAddress}{Department of Physics and Astronomy, Texas A\&M University, College Station, Texas 77843, USA}
\newcommand{\SCNUKeyLabAddress}{Guangdong Provincial Key Laboratory of Nuclear Science, Institute of Quantum Matter, South China Normal University, Guangzhou 510006, China}
\newcommand{\SCNUJointLabAddress}{Guangdong-Hong Kong Joint Laboratory of Quantum Matter, Southern Nuclear Science Computing Center, South China Normal University, Guangzhou 510006, China}
\newcommand{\SCUTAddress}{
School of Physics and Optoelectronics, South China University of Technology, Guangzhou 510640, China}
\newcommand{\OhioAddress}{Department of Physics, The Ohio State University, Columbus, Ohio 43210, USA}
\newcommand{\DukeStatAddress}{Department of Statistical Science, Duke University, Durham, North Carolina 27708, USA}
\newcommand{\CCNUKeyAndIOPPAddress}{Key Laboratory of Quark and Lepton Physics (MOE) and Institute of Particle Physics, Central China Normal University, Wuhan 430079, China}
\newcommand{\BNLAddress}{Department of Physics, Brookhaven National Laboratory, Upton, New York 11973, USA}
\newcommand{\LANLAddress}{Los Alamos National Laboratory, Theoretical Division, Los Alamos, New Mexico 87545, USA}
\newcommand{\WayneCompAddress}{Department of Computer Science, Wayne State University, Detroit, Michigan 48202, USA}
\newcommand{\INTAddress}{Institute for Nuclear Theory, University of Washington, Seattle, Washington 98195, USA}
\newcommand{\SaoPauloAddress}{Instituto de F\`{i}sica, Universidade de S\~{a}o Paulo, C.P. 66318, 05315-970 S\~{a}o Paulo, S\~{a}o Paulo, Brazil}
\newcommand{\RIKENBNLAddress}{RIKEN BNL Research Center, Brookhaven National Laboratory, Upton, New York 11973, USA}
\newcommand{\KentAddress}{Department of Physics, Kent State University, Kent, Ohio 44242, USA}
\newcommand{\DaresburyAddress}{STFC Daresbury Laboratory, Daresbury, Warrington WA4 4AD, United Kingdom}
\newcommand{\OliverLodgeAddress}{
Oliver Lodge Laboratory, Department of Physics, University of Liverpool, Oxford Street, Liverpool L69 7ZE, United Kingdom}
\newcommand{\JYUAddress}{University of Jyv\"{a}skyl\"{a}, Department of Physics, P.O.\ Box 35, FI-40014 University of Jyv\"{a}skyl\"{a}, Finland}
\newcommand{\HelsinkiAddress}{Helsinki Institute of Physics, P.O.\ Box 64, FI-00014 University of Helsinki, Finland}

\author{Y.~Tachibana}
\email[Contact author: ]{ytachibana@aiu.ac.jp}
\affiliation{\AiuAddress}

\author{C.~Sirimanna}
\affiliation{\DukePhysAddress}
\affiliation{\WaynePhysAddress}

\author{A.~Majumder}
\affiliation{\WaynePhysAddress}

\author{A.~Angerami}
\affiliation{\LLNLAddress}

\author{R.~Arora}
\affiliation{\WayneCompAddress}

\author{S.~A.~Bass}
\affiliation{\DukePhysAddress}

\author{Y.~Chen}
\affiliation{\VandyAddress}
\affiliation{\MITNuclAddress}
\affiliation{\MITPhysAddress}

\author{R.~Datta}
\affiliation{\WaynePhysAddress}

\author{L.~Du}
\affiliation{\McGillAddress}
\affiliation{\UCBerkeleyAddress}
\affiliation{\LBLAddress}

\author{R.~Ehlers}
\affiliation{\UCBerkeleyAddress}
\affiliation{\LBLAddress}

\author{H.~Elfner}
\affiliation{\GSIAddress}
\affiliation{\GoetheAddress}
\affiliation{\FIASAddress}

\author{R.~J.~Fries}
\affiliation{\TexasAMCyclAddress}
\affiliation{\TexasAMPhysAddress}

\author{C.~Gale}
\affiliation{\McGillAddress}

\author{Y.~He}
\affiliation{\SCUTAddress}

\author{B.~V.~Jacak}
\affiliation{\UCBerkeleyAddress}
\affiliation{\LBLAddress}

\author{P.~M.~Jacobs}
\affiliation{\UCBerkeleyAddress}
\affiliation{\LBLAddress}

\author{S.~Jeon}
\affiliation{\McGillAddress}

\author{Y.~Ji}
\affiliation{\DukeStatAddress}

\author{F.~Jonas}
\affiliation{\UCBerkeleyAddress}
\affiliation{\LBLAddress}

\author{L.~Kasper}
\affiliation{\VandyAddress}

\author{M.~Kordell~II}
\affiliation{\TexasAMCyclAddress}
\affiliation{\TexasAMPhysAddress}

\author{A.~Kumar}
\affiliation{\ReginaAddress}
\affiliation{\McGillAddress}

\author{R.~Kunnawalkam-Elayavalli}
\affiliation{\VandyAddress}

\author{J.~Latessa}
\affiliation{\WayneCompAddress}

\author{Y.-J.~Lee}
\affiliation{\MITNuclAddress}
\affiliation{\MITPhysAddress}

\author{R.~Lemmon}
\affiliation{\DaresburyAddress}

\author{M.~Luzum}
\affiliation{\SaoPauloAddress}

\author{S.~Mak}
\affiliation{\DukeStatAddress}

\author{A.~Mankolli}
\affiliation{\VandyAddress}

\author{C.~Martin}
\affiliation{\UTNAddress}

\author{H.~Mehryar}
\affiliation{\WayneCompAddress}

\author{T.~Mengel}
\affiliation{\UTNAddress}

\author{C.~Nattrass}
\affiliation{\UTNAddress}

\author{J.~Norman}
\affiliation{\OliverLodgeAddress}

\author{C.~Parker}
\affiliation{\TexasAMCyclAddress}
\affiliation{\TexasAMPhysAddress}

\author{J.-F.~Paquet}
\affiliation{\VandyAddress} 

\author{J.~H.~Putschke}
\affiliation{\WaynePhysAddress}

\author{H.~Roch}
\affiliation{\WaynePhysAddress}

\author{G.~Roland}
\affiliation{\MITNuclAddress}
\affiliation{\MITPhysAddress}

\author{B.~Schenke}
\affiliation{\BNLAddress}

\author{L.~Schwiebert}
\affiliation{\WayneCompAddress}

\author{A.~Sengupta}
\affiliation{\TexasAMCyclAddress}
\affiliation{\TexasAMPhysAddress}

\author{C.~Shen}
\affiliation{\WaynePhysAddress}
\affiliation{\RIKENBNLAddress}

\author{M.~Singh}
\affiliation{\VandyAddress} 

\author{D.~Soeder}
\affiliation{\DukePhysAddress}

\author{R.~A.~Soltz}
\affiliation{\WaynePhysAddress}
\affiliation{\LLNLAddress}

\author{I.~Soudi}
\affiliation{\JYUAddress}
\affiliation{\HelsinkiAddress}
\affiliation{\WaynePhysAddress}

\author{J.~Velkovska}
\affiliation{\VandyAddress} 

\author{G.~Vujanovic}
\affiliation{\ReginaAddress}

\author{X.-N.~Wang}
\affiliation{\CCNUKeyAndIOPPAddress}
\affiliation{\UCBerkeleyAddress}
\affiliation{\LBLAddress}

\author{X.~Wu}
\affiliation{\McGillAddress}
\affiliation{\WaynePhysAddress}

\author{W.~Zhao}
\affiliation{\UCBerkeleyAddress}
\affiliation{\LBLAddress}
\affiliation{\WaynePhysAddress}

\collaboration{JETSCAPE Collaboration}

\begin{abstract}
We investigate medium-induced modifications to jet substructure observables that characterize hard components in central Pb-Pb collisions at $\sqrt{s_{NN}}=5.02$~TeV. 
Using a multistage Monte Carlo simulation of in-medium jet shower evolution, we explore flavor-dependent medium effects through simulations of inclusive and $\gamma$-tagged jets. 
The results show that quark jets undergo a nonmonotonic modification compared with gluon jets in observables such as the Pb-Pb to $p$-$p$ ratio of the soft drop prong angle $r_g$, the relative prong transverse momentum $k_{T,g}$, and the groomed mass $m_g$ distributions.  
Due to this nonmonotonic modification, $\gamma$-tagged jets, enriched in quark jets, provide surprisingly clear signals of medium-induced structural modifications, distinct from effects dominated by selection bias. 
Further systematic studies demonstrate that these effects are dominated by recoil medium response. 
This work highlights the potential of hard substructures in $\gamma$-tagged jets as powerful tools for probing the jet-medium interactions in high-energy heavy-ion collisions,
in particular by enabling detailed investigations of jet-medium parton scatterings via their associated medium response. 
All simulations for $\gamma$-tagged jet analyses carried out in this paper used triggered events containing at least one hard photon, which highlights the utility of these observables for future Bayesian analysis. 
\end{abstract}

\maketitle

\section{Introduction}
\label{Section:Intro}

High-energy heavy-ion collision experiments at the BNL Relativistic Heavy Ion Collider (RHIC) and the CERN Large Hadron Collider (LHC) have established jets as essential tools to probe the quark-gluon plasma (QGP), a strongly coupled, deconfined state of matter created in these collisions~\cite{Bjorken:1982tu:Manual,Appel:1985dq,Baier:1996kr,Baier:1996sk,Zakharov:1996fv,Gyulassy:1999zd,Gyulassy:2000fs,Gyulassy:2000er,Wiedemann:2000za,Wiedemann:2000tf,Guo:2000nz,Wang:2001ifa,Majumder:2009ge,Arnold:2001ba,Arnold:2002ja,Majumder:2010qh,Blaizot:2015lma,Qin:2015srf,Cao:2020wlm,Cao:2024pxc}. 
Large transverse momentum partons generated in the initial hard scatterings traverse the QGP medium, undergoing significant interactions that lead to observable modifications compared with proton-proton ($p$-$p$) collisions, where a sufficiently large medium is not formed~\cite{JETSCAPE:2024dgu}. 
Early measurements at RHIC, such as the suppression of high transverse momentum hadrons~\cite{PHENIX:2001hpc,PHENIX:2003djd,PHENIX:2003qdj,STAR:2002ggv,STAR:2003fka} and the disappearance of di-hadron correlations~\cite{STAR:2002svs,STAR:2005ryu,PHENIX:2007yjc}, provided the first experimental evidence of parton energy loss due to the strong interactions with the QGP medium. 
Recent advances at RHIC and LHC have allowed more detailed studies of reconstructed jets~\cite{ALICE:2013dpt,ATLAS:2010isq,CMS:2011iwn,STAR:2020xiv}, providing deeper insights into the fundamental mechanisms underlying jet-medium interactions, such as medium-induced radiation, scattering with medium constituents, in-medium thermalization, the hydrodynamic evolution of energy deposited by jets, etc.

The loss of jet transverse momentum ($p^{\mathrm{jet}}_{T}$), as revealed by the nuclear modification factor $R_{AA}$, has been a central focus of measurements and analyses, establishing $R_{AA}$ as a primary observable. 
However, the extent of $p^{\mathrm{jet}}_{T}$ loss in reconstructed jets depends only on the energy and momentum that has been scattered or radiated outside the jet cone. 
Partons in hard showers lose energy by scattering and radiation. 
The extra scattering in the medium changes not only the energy radiated outside the cone but also the energy-momentum distribution within the cone. 
The latter can only be accessed via jet substructure observables.  
Investigating jet substructures also necessitates triggering on jets based on their $p^{\mathrm{jet}}_{T}$, making an accurate calculation of $p^{\mathrm{jet}}_{T}$ loss essential. 
To explore jet-medium interactions effectively, it is crucial to study both the $p^{\mathrm{jet}}_{T}$ loss, reflected in the $R_{AA}$, and the internal structure modifications, captured by substructure observables.

Recent years have seen a proliferation of observables and analysis techniques referred to as jet substructure, with examples including the jet fragmentation function, jet mass, and jet shape. 
These reveal medium effects on the internal structure of jets~\cite{ATLAS:2010isq,CMS:2011iwn,ATLAS:2012tjt,ALICE:2013dpt,ATLAS:2014ipv,CMS:2016uxf,STAR:2016dfv,ATLAS:2018gwx,ALICE:2019qyj,CMS:2021vui,STAR:2020xiv,CMS:2013lhm,CMS:2016cvr,CMS:2018zze,CMS:2018jco,CMS:2018fof,ALICE:2019whv,CMS:2021nhn,CMS:2014jjt,ATLAS:2014dtd,ATLAS:2017nre,ATLAS:2018bvp,ATLAS:2019dsv,ATLAS:2019pid,Apolinario:2024equ,CMS:2024zjn:Manual,ALICE:2024fip,ALICE:2024jtb:Manual}. 
Advanced grooming techniques~\cite{Butterworth:2008iy,Krohn:2009th,Ellis:2009me,Dasgupta:2013ihk,Larkoski:2014wba,Larkoski:2015lea,Mehtar-Tani:2019rrk} extract imprints of hard partonic branchings from final-state hadrons forming reconstructed jets, even though such branchings cannot be directly observed experimentally. 
Understanding these hard branchings, which form the skeleton of a jet's internal structure, is crucial to uncovering how jet structures are formed and modified by medium effects, such as modifications of the radiation pattern or scatterings with medium constituents, ultimately leading to $p^{\mathrm{jet}}_{T}$ loss. 
In vacuum, hard branchings are relatively well-described using perturbative methods, offering robust baselines for studying medium-induced modifications in heavy-ion collisions. 
Hard branchings have also attracted significant theoretical interest, as they provide a promising avenue to explore jet-medium interactions~\cite{Chien:2016led,Mehtar-Tani:2016aco,Chang:2017gkt, Milhano:2017nzm,Casalderrey-Solana:2019ubu,Ringer:2019rfk,Caucal:2019uvr,Mehtar-Tani:2019rrk,Caucal:2021bae,Caucal:2021cfb,JETSCAPE:2023hqn,Cunqueiro:2023vxl}.

Here, it is important to note that, especially in the presence of medium effects, the notion of a hard branching extracted by grooming analyses does not necessarily correspond to an explicit parton radiation. 
The reconstructed hard branching may instead be strongly affected by recoil particles in scatterings with medium constituents~\cite{Milhano:2017nzm}, contributions from hadronization processes, or other effects that, in principle, cannot be disentangled experimentally. 
Accordingly, throughout this paper, the term ``hard branching'' denotes the splitting identified by the grooming analysis, whereas the hard substructure of the jet refers to observables characterizing the distribution of the reconstructed prongs associated with that branching.

Soft drop~\cite{Larkoski:2014wba}, one of the most widely used grooming methods in heavy-ion physics, identifies the two hard prongs with the largest angular separation within a jet while discarding low-momentum prongs, thereby defining the hard branching. 
In our previous study~\cite{JETSCAPE:2023hqn}, we analyzed the soft-drop observables $z_g$ (momentum fraction) and $r_g$ (radial separation) for inclusive jets triggered by their $p^{\mathrm{jet}}_T$, using Monte Carlo simulations of in-medium jet shower evolution with the \textsc{matter}$+$\textsc{lbt} multistage model~\cite{Cao:2017qpx,Majumder:2013re,He:2015pra,Cao:2016gvr,He:2018xjv,Luo:2018pto,JETSCAPE:2017eso,Cao:2024pxc} within the \textsc{jetscape} framework~\cite{Putschke:2019yrg,JETSCAPE:2019udz,JETSCAPE:2020shq,JETSCAPE:2020mzn,JETSCAPE:2021ehl:Manual,JETSCAPE:2022cob,JETSCAPE:2022jer,JETSCAPE:2022hcb,JETSCAPE:2023hqn,JETSCAPE:2023ikg,JETSCAPE:2024dgu,JETSCAPE:2024cqe,JETSCAPE:2024nkj}. 
The $z_g$ distribution showed negligible medium-induced modifications, consistent with experiments, while the $r_g$ distribution exhibited suppression that increases monotonically with $r_g$. 
This monotonic behavior was reproduced by incorporating modified coherence effects~\cite{Kumar:2019uvu,JETSCAPE:2022jer,Kumar:2025rsa}, where high-virtuality partons, due to finer resolution, perceive the medium as dilute and experience reduced interactions. 
As a result, jets with larger $r_g$ were strongly suppressed, with no trace of medium effects on the actual structure of hard components associated with the parton shower.

These findings suggest that the suppression of the $r_g$ distribution may arise primarily from selection bias in inclusive jet analyses, rather than from direct structural modifications of soft-drop-reconstructed hard branchings. 
Jets with large-angle hard branchings are more likely to have lost soft constituents outside the jet cone, leading to a bias toward triggering on jets with smaller angular separations and reduced overall energy. 
Thus, inclusive jet measurements are less sensitive to medium effects, causing the structural modification of the parton shower, underscoring the need for targeted observables to probe these effects.

To investigate medium effects in detail through soft-drop-reconstructed jet substructure, it is essential to mitigate the selection bias inherent in inclusive jet analyses. 
This can be achieved using $\gamma$-tagged or $Z$-tagged jets, produced in photon-jet or $Z$-boson-jet pair production events, often referred to as ``golden channels.'' 
Photons and $Z$ bosons do not interact strongly with the QGP medium. Thus, photons and $Z$ bosons that are pair-produced with a jet in the initial hard scattering are measured with transverse momenta ($p_T$) that closely approximate the initial $p_T$ of the partons generating the jets, at leading order. 
Triggering on the $p_T$ of the photon or $Z$ boson, instead of the jet, selects jets with similar initial $p_T$, independent of $p^{\mathrm{jet}}_{T}$ loss, reducing selection bias and enabling cleaner investigations of medium effects on jet structures. 
Additionally, partons paired with photons or $Z$ bosons are predominantly quarks, minimizing flavor-dependent selection bias. 
Comparing these tagged jets with inclusive jets allows systematic studies of flavor-dependent effects.

In this paper, we investigate $\gamma$-tagged jets as a means to mitigate the selection bias effect, enabling a clearer examination of medium modifications to a subset of soft-drop observables that characterize the hard substructure of jets. 
Building on our previous work~\cite{JETSCAPE:2022jer, JETSCAPE:2022hcb, JETSCAPE:2023hqn}, which established a wealth of benchmark results, we employ the \textsc{matter}$+$\textsc{lbt} multistage model within the \textsc{jetscape} framework to provide predictions under realistic high-energy heavy-ion collision configurations. 
We demonstrate that fixing the photon $p_T$ and triggering on associated jets, even those with lower $p^{\mathrm{jet}}_T$, eliminates the suppression of jets with broad hard branchings observed in inclusive jet measurements. 
This confirms that the suppression of inclusive jets arises from selection bias rather than from intrinsic structural modifications of the hard component reconstructed as part of a jet. 
Moreover, we show that $\gamma$-tagged jets exhibit pronounced broadening of soft-drop-reconstructed hard branchings compared with inclusive jets, driven by their quark-jet dominance. 
With simplified simulations, we reveal that quark jets are more susceptible to medium-induced modifications, particularly during low-virtuality evolution, leading to significant modifications in their soft-drop-reconstructed hard branching structure. 
These results establish $\gamma$-tagged jets, and similarly $Z$-tagged jets, as powerful tools for revealing clear signals of direct medium effects on the hard component reconstructed as part of a jet.

The paper is organized as follows. 
Section~\ref{Section:Model} outlines the simulation framework and methodologies employed, including the multistage \textsc{matter}$+$\textsc{lbt} model within the \textsc{jetscape} framework. 
In Sec.~\ref{Section:Analyses}, we detail the analysis procedures for $\gamma$-tagged jets as well as inclusive jets, including soft-drop grooming. 
Section~\ref{Section:Results} presents the results from the \textsc{jetscape} simulations, highlighting the flavor-dependent medium effects and selection biases on $r_g$, $k_{T,g}$, and $m_g$ distributions, as well as the role of $\gamma$-tagged jets in revealing intrinsic medium modifications.
Subsequently, in Sec.~\ref{Section:VsExp}, we compare the results from the same event sets with available experimental data from Pb-Pb collisions at the LHC. 
Finally, Sec.~\ref{Section:Summary} concludes with a summary of our findings.

\section{Model}
\label{Section:Model}
In this paper, the \textsc{jetscape} code package is utilized, which offers a flexible framework for modular integration in Monte Carlo event generation for heavy-ion collisions. 
We specifically employ our current default configuration with the \textsc{matter}$+$\textsc{lbt} setup referred to as the JETSCAPEv3.5 AA22 tune, as detailed in our previous work \cite{JETSCAPE:2022jer, JETSCAPE:2022hcb, JETSCAPE:2023hqn}, to perform simulations of jet events. 
All parameters in the tune are chosen to fit only the $R_{AA}$ s for single high-$p_{T}$ particles and reconstructed jets in Ref.~\cite{JETSCAPE:2022jer} and are not retuned for any other observables, including all the observables presented in this paper. 
In this section, we provide a brief overview of the components that constitute our simulation setup. 
For readers interested in delving deeper into the specific physical elements integrated into the multistage \textsc{matter}$+$\textsc{lbt} simulation within the \textsc{jetscape} framework, we direct them to Refs.~\cite{JETSCAPE:2022jer, JETSCAPE:2022hcb, JETSCAPE:2023hqn}. 
Additional insights into the software components of the \textsc{jetscape} framework are detailed in Ref.~\cite{Putschke:2019yrg}. 
The foundational concept concerning the multistage description of jet evolution within the medium is explained in Ref.~\cite{JETSCAPE:2017eso}.

\subsection{Overview}
\label{Section:Overview} 
For the efficient generation of jet events, the \textsc{jetscape} framework provides an option to embed a single hard scattering event from a nucleon-nucleon collision into a heavy-ion collision event using a pre-generated space-time background profile of the QGP medium. 
In the JETSCAPEv3.5 AA22 tune, we employ the background medium profile obtained from event-by-event calculations of $(2+1)$-dimensional [$(2+1)$-D] free-streaming pre-equilibrium evolution~\cite{Liu:2015nwa}, followed by viscous hydrodynamic evolution using $(2+1)$-D \textsc{vishnu}~\cite{Shen:2014vra} and hadronic scattering and decay simulated by \textsc{urqmd}~\cite{Bass:1998ca, Bleicher:1999xi}, with fluctuating initial conditions from \textsc{trento}~\cite{Moreland:2014oya}. 
Here, the best-fit parametrization, determined through maximum \textit{a posteriori} (MAP) from Bayesian methods~\cite{Bernhard:2019bmu} applied to observables measured at the LHC, is used.

In jet simulations, high-energy partons are generated according to the hard process settings described later in Sec.~\ref{Section:InitialHard}, and then these partons evolve into parton showers. 
In the multistage setup of \textsc{matter}$+$\textsc{lbt}, partons are initially subjected to splittings driven primarily by their large virtualities, with minor medium effects incorporated in the \textsc{matter} module~\cite{Majumder:2013re,Cao:2017qpx}. 
In \textsc{matter}, the splittings are simulated in a virtuality-ordered manner, while the space-time trajectory of the shower is constructed by estimating the formation time via the uncertainty principle and sequentially determining the splitting locations. 
The description of the splittings for a parton in the \textsc{matter} module is terminated when its virtuality becomes sufficiently small, approximately on par with the accumulated transverse momentum gain via scatterings in the medium. 
This termination is attributed to the applicable limit of a model that relies on virtuality as the primary mechanism for partonic splittings. 
The partons with reduced virtuality are then transferred to the \textsc{lbt} module~\cite{Wang:2013cia,He:2015pra,Cao:2016gvr}, which simulates elastic and inelastic scatterings with medium constituents based on kinetic theory with the on-shell approximation.

In the \textsc{jetscape} framework, this virtuality-based switching between the \textsc{matter} and \textsc{lbt} modules is performed bidirectionally at a parton-by-parton level using a switching parameter denoted as $Q^2_{\mathrm{sw}}$. 
If the parton's virtuality $Q^2 = p^\mu p_\mu - m^2$ drops below $Q^2_{\mathrm{sw}}$, it is transferred from \textsc{matter} to \textsc{lbt}, and it returns to \textsc{matter} when the virtuality exceeds $Q^2_{\mathrm{sw}}$ again or when it exits the QGP medium boundary at a temperature $T_{c} = 0.16$~GeV. 
Below $T_{c}$, \textsc{matter} performs vacuum-like splitting down to the cutoff scale $Q_{\mathrm{min}}^2 = 1~{\mathrm{GeV}}^2$ without any medium effects. 
Throughout this study, the switching parameter is set to $Q^2_{\mathrm{sw}}=4~{\mathrm{GeV}}^2$ ($Q_{\mathrm{sw}}=2~{\mathrm{GeV}}$).

In both the \textsc{matter} and \textsc{lbt} phases, the medium effects are calculated based on the local temperature and flow velocity from the pre-generated background medium profile. 
The medium response is described by following the subsequent evolution of recoil partons scattered out from the medium during the in-medium collisions simulated in \textsc{matter} and \textsc{lbt} in the same way as the other showering partons. 
The recoil partons are assumed to be on-shell at their generation and are then passed to \textsc{lbt} for the following elastic and inelastic processes within the medium. 
On the other hand, for each in-medium collision event that generates a recoil parton, a deficit of energy and momentum is left in the medium. 
This deficit is also tracked and treated as a freestreaming particle, referred to as a hole parton.

All jet partons undergo hadronization via the \textsc{colorless hadronization} module, in which the Lund string model of \textsc{pythia} 8 is utilized after escaping the QGP medium with $T>T_{c}$ and reaching the virtuality cutoff scale $Q_{\mathrm{min}}^2$. 
The jet shower partons, including the recoils and further accompanying daughters, are collectively hadronized, whereas the hole partons are hadronized separately. 
Using those hadrons formed by clustering hole partons, appropriate subtraction is performed depending on the observable of interest. 
In this study, the hole hadrons are tagged for identification and, together with the other hadrons, sent to a modified anti-$k_{t}$ jet reconstruction routine, in which the four-momentum of hole hadrons or reconstructed subjets dominated by hole hadrons is subtracted during the merging process in the $E$-scheme~\cite{Luo:2023nsi}.

Incidentally, in the later stages of in-medium jet evolution, as a jet parton energy becomes close to the scale of the ambient temperature, the approach based on kinetic theory becomes less practical due to the short mean free paths. 
Such soft components of jets are expected to be thermalized and then transported hydrodynamically through the bulk medium flow~\cite{Casalderrey-Solana:2004fdk,Stoecker:2004qu,Tachibana:2019hrn,Cao:2020wlm,Schlichting:2020lef,Luo:2021iay,Mehtar-Tani:2022zwf:Manual}. 
The hydrodynamic description of the evolution of the soft components of jets, as proposed in Refs.~\cite{Chaudhuri:2005vc,Renk:2005si,Satarov:2005mv,Neufeld:2008fi,Noronha:2008un,Qin:2009uh,Betz:2010qh,Neufeld:2011yh,Schulc:2014jma,Tachibana:2014lja,Tachibana:2015qxa,Tachibana:2017syd,Okai:2017ofp,Chen:2017zte,Chang:2019sae,Tachibana:2020mtb,JETSCAPE:2020uew,Casalderrey-Solana:2020rsj,Yang:2021qtl,Yang:2022nei,Pablos:2022piv,Yang:2022yfr}, can be implemented, through a source term, by coupling with the hydrodynamic equation for the bulk evolution. 
However, this approach necessitates a comprehensive $(3+1)$-D hydrodynamic simulation for each jet event and makes the computational cost of conducting a systematic and exhaustive study, as demonstrated in this paper, outstandingly expensive. 
Therefore, this paper primarily focuses on the hard part of the jets, deferring a detailed study of the effects of the soft components to future research.

In this paper, to investigate the impact of medium response effects via recoils and holes, we also present results from calculations in which both recoils and holes are excluded. 
For this setup, the parton shower evolution is carried out identically to the default simulations with recoils and holes included; however, at the hadronization stage, all recoil and hole partons are discarded and do not contribute to final-state particles. 
Since all $s$-, $t$-, and $u$-channels contribute to the jet-medium scatterings, it is in principle impossible to distinguish whether an outgoing parton originates from the jet shower or from the medium. 
In practice, the scattered parton with the smaller energy is tagged as a recoil. 
It should also be noted that calculations without recoils and holes correspond to an unphysical setup that explicitly violates energy-momentum conservation and are intended solely for theoretical assessments.

Calculations for $p$-$p$ collisions are also necessary to establish a baseline for the heavy-ion collision simulations. 
Our $p$-$p$ calculations are also performed using the \textsc{jetscape} framework, with all medium effects turned off, where the parton shower evolution is managed solely by \textsc{matter} without jet-medium coupling. 
This configuration is known as the JETSCAPE PP19 tune and is described in detail in Ref.~\cite{JETSCAPE:2019udz}.

\subsection{Initial hard process}
\label{Section:InitialHard}
In this paper, we compare our results from the \textsc{jetscape} simulations with several different settings for the initial hard process generation. 
Below, we explain each setting. 
With any setting, the geometric position in the transverse plane at mid-space-time rapidity [$\eta_{s}=(1/2)\ln [(t+z)/(t-z)]=0$], where the hard process occurs, is determined by sampling the $N_{\mathrm{coll}}$ distribution calculated using the \textsc{trento} initial condition module for the Pb-Pb collisions.

\subsubsection{PGun}
\label{Section:PGun}
To serve as a test for systematic studies with minimal extraneous contributions and simplified settings, we generate events with just a single high-energy parton and simulate its jet shower development. 
Within the \textsc{jetscape} package, such a setup is enabled by the \textsc{pgun} module. 
In \textsc{pgun}, one specifies the species of the parton and its initial energy and then shoots it onto the transverse plane at mid-space-time rapidity ($\eta_{s}=0$) with a randomly assigned azimuthal direction. 
The generated partons are then passed to a module handling high-virtuality parton development, \textsc{matter} in this study, initiating the jet shower. 
In this study, we compare scenarios with either a gluon or a massless light quark as the initial parent parton.

\subsubsection{PythiaGun}
\label{Section:PythiaGun}
For the generation of realistic jet events in hadron-hadron collisions, the \textsc{jetscape} framework provides the \textsc{pythiagun} module, which serves as a wrapper for \textsc{pythia} 8~\cite{Sjostrand:2019zhc}. 
Within the \textsc{pythiagun} module, utilizing the functionalities of \textsc{pythia} 8, hard scatterings are generated based on leading-order perturbation calculations. 
Subsequently, final-state radiation (FSR) is disabled by default, and the resulting hard partons are directly passed to a module responsible for the evolution of highly virtual partons, \textsc{matter} in this study, to facilitate the development of jet showers. 
Simultaneously, when initial-state radiation (ISR) or multiparton interaction (MPI) is turned on, the partons produced through such a process are also passed to the jet shower evolution. 
Throughout this study, both ISR and MPI are always on for initial scattering generation using \textsc{pythiagun}.

Hard $\gamma$-jet pair production is a very rare process. 
In a fully inclusive event generation scheme, most events are not triggered, making it extremely challenging to accumulate sufficient statistics for the differential observables of jet substructure, which are the focus of this study. 
To address this, we adopt a strategy that selectively generates only the initial hard processes with leading-order contributions involving prompt-photon production, by configuring \textsc{pythiagun} with \texttt{HardQCD:all=off} and \texttt{PromptPhoton:all=on}, thereby enhancing computational efficiency. 
While this method does not fully account for the contribution from jet-photon production via fragmented photons, our main predictions focus on the region where these effects are minimal ($x_{J\gamma}<1$, i.e., $p^{\gamma}_{T} > p^{\mathrm{jet}}_{T}$), as demonstrated in our previous study \cite{JETSCAPE:2024nkj}. 
In contrast, for the inclusive jet analysis, we generate events by enabling all hard QCD 2-to-2 processes through the settings \texttt{HardQCD:all=on}.

In the actual experimental analysis, the isolation cut mentioned later in Sec.~\ref{Subsection:GammaTagged} is applied to select $\gamma$-tagged jets to enhance the leading-order contributions in initial hard scatterings involving prompt photon production. 
However, in practice, perfect extraction is not achievable, and contributions beyond leading-order prompt photons unavoidably seep in. 
For example, if a photon is emitted at a large angle from a parton jet shower, it may pass the isolation cut requirement.

Further discussions concerning the evaluation of the contributions of photons radiated from parton showers to the $\gamma$-tagged jet can be found in our separate study~\cite{JETSCAPE:2024nkj}. 
Additionally, the behaviors observed in the $\gamma$-tagged jet results presented in this paper are also expected to appear in $Z$-tagged jet measurements, as both are boson-tagged observables that offer similar advantages: cleaner access to the initial parton kinematics and strong quark-jet dominance. 
Due to their large masses, $Z$-boson radiations from parton showers are drastically suppressed, making those produced at leading order in the initial hard scatterings completely dominant in measurements.

\section{Analyses}
\label{Section:Analyses}
This section provides a detailed description of the soft-drop-grooming procedure used for calculating observables that characterize the hard component structures of jets, as well as the method for constructing $\gamma$-tagged jets in our Monte Carlo simulations of high-energy heavy-ion collisions.

\subsection{Soft-drop-grooming procedure}
\label{Subsection:SoftDrop}
In this study, we focus on observables that characterize the structure formed by the parton shower and eventually captured as the hard component of reconstructed jets, as identified through the soft-drop-grooming algorithm~\cite{Larkoski:2014wba}. 
The soft-drop-grooming identifies the hard splitting, while removing the soft branchings, at as large an angle as possible.

In the soft-drop-grooming procedure, an angular-ordered clustering tree is rebuilt using the Cambridge-Aachen (C/A) algorithm~\cite{Dokshitzer:1997in,Wobisch:1998wt} for constituents of a triggered jet reconstructed by a standard jet-finding algorithm, such as the anti-$k_{t}$ algorithm~\cite{Cacciari:2008gp}, with a jet cone size $R$. 
Next, we traverse back, i.e., from the branch with the largest angle, through the C/A tree. At each branching, we check whether the two prongs of the branching satisfy the soft-drop condition given by
\begin{align}
\label{eq:soft_drop}
\frac{p_{T,2}}{p_{T,1} + p_{T,2}} > z_{\mathrm{cut}}\left(\frac{\Delta R_{12}}{R}\right)^{\beta},
\end{align}
where $p_{T,1}$ and $p_{T,2}$~($<p_{T,1}$) are the transverse momenta of the prongs, and $\Delta R_{12} = [(\eta_{1}-\eta_{2})^{2} + (\phi_{1}-\phi_{2})^{2}]^{1/2}$ is the radial distance between the prongs in the rapidity-azimuthal angle plane. 
The soft-drop parameters $z_{\mathrm{cut}}$ and $\beta$ control the grooming procedure.

If the soft-drop condition is satisfied, we stop the procedure, and the two prongs of the branching are used to calculate the groomed jet observables. 
If the condition is not met, we continue traversing the tree by following the prong with the larger $p_{T}$ and repeat the same procedure. 
It should be noted that there may be cases where a prong pair satisfying the soft-drop condition cannot be found eventually. 
Throughout this work, the jet reconstruction and soft-drop grooming are performed using the \textsc{fastjet} package~\cite{Cacciari:2005hq, Cacciari:2011ma} with \textsc{fastjet-contrib-1.045}~\cite{fjcontrib_code}.

\subsection{$\boldsymbol{\gamma}$-tagged jet}
\label{Subsection:GammaTagged}
Photons do not interact strongly with the QGP medium. 
This characteristic is particularly notable in the leading-order contribution to $\gamma$-jet pair production in the initial hard scattering in high-energy heavy-ion collisions. 
The photons produced in such events are typically measured with transverse momentum nearly identical to the initial transverse momentum of the jet's parent parton. 
This correlation enables the estimation of the jet's energy loss due to medium interactions.

However, experimentally, $\gamma$-jet pairs can also arise from processes other than the initial hard $\gamma$-jet pair creation, such as photon radiation in final-state parton showering. 
Therefore, to suppress contamination from such non-targeted processes, specific cuts are imposed in the experimental analyses: the isolation requirement for photons and the relative azimuth angle cut. 
To enhance quantitative accuracy in our comparison to experimental data, we impose these cuts on our simulated events.

The isolation requirement, specifically the selection of photons with minimal energy emissions in their vicinity, is introduced to enhance the leading-order contributions. 
Photons are isolated based on the accumulated transverse energy or momentum inside a cone of fixed radius of size $R_{\mathrm{iso}}$ centered on the photon's direction after subtracting background contributions. 
In this study, they are calculated as follows: 
\begin{align}
E^{\mathrm{iso}}_{T}
&= 
\left(\sum_{\substack{i\in \mathrm{shower}\\ \Delta r_i < R_{\mathrm{iso}}}}E_{T,i}\right) - 
\left(\sum_{\substack{i\in \mathrm{holes}\\ \Delta r_i < R_{\mathrm{iso}}}}E_{T,i}\right)
-E_{T}^{\gamma}
, \\
p^{\mathrm{iso}}_{T}
&= 
\left(\sum_{\substack{i\in \mathrm{shower}\\ \Delta r_i < R_{\mathrm{iso}}}}p_{T,i}\right) - 
\left(\sum_{\substack{i\in \mathrm{holes}\\ \Delta r_i < R_{\mathrm{iso}}}}p_{T,i}\right)
-p_{T}^{\gamma}, 
\end{align}
where $\Delta r_i = [(\eta_i - \eta_{\gamma})^2+(\phi_i - \phi_{\gamma})^2]^{1/2}$ 
is the radial distance from the isolated photon candidate. 
On the right-hand sides, the sums are taken over final-state particles from hadronization of jet shower partons, including the recoils in the first terms and over particles hadronized from hole partons in the second terms, respectively. 
The transverse energy and momentum of the isolated photon candidate in the last terms are introduced to eliminate self-contribution. 
If the candidate satisfies either $E^{\mathrm{iso}}_{T} < E^{\mathrm{iso,cut}}_{T}$ 
or $p^{\mathrm{iso}}_{T} < p^{\mathrm{iso,cut}}_{T}$, 
based on the predetermined cut parameter 
$E^{\mathrm{iso,cut}}_{T}$ 
or $p^{\mathrm{iso,cut}}_{T}$, depending on the analysis method used in the experimental results for comparison, it qualifies as an isolated photon.

The relative azimuth angle cut 
\begin{align}
\lvert\phi_{\mathrm{jet}}-\phi_{\gamma}\rvert > \Delta\phi_{\mathrm{cut}} 
\end{align}
is introduced to ensure that the photon and jet are back-to-back in the azimuthal angle plane, thereby enhancing the leading-order contribution from the initial hard $\gamma$-jet pair production. 
The cut value most commonly used in various experiments is $\Delta\phi_{\mathrm{cut}}=7\pi/8$.

\section{Results} 
\label{Section:Results}
This section systematically investigates the modification of the soft-drop substructures from simulations with simplified initial events generated by the \textsc{pgun} module. A single quark or gluon is generated at one point in the medium and allowed to propagate outward. Strict control over the flavor and energy of the parton allows for a detailed investigation of the flavor and energy dependence of specific substructure observables.

Following this, we present theoretical predictions for $\gamma$-tagged jets using realistic initial events generated by \textsc{pythiagun}. These are compared with the results from inclusive jets, and novel, experimentally measurable signals of medium effects in these observables are discussed. 
To establish the reliability of the model calculations and provide a baseline for discussion, comparisons with available experimental results are presented in the subsequent section, Sec.~\ref{Section:VsExp}.

\subsection{Jet splitting momentum fraction}
\label{Subsection:zG}
We begin with the study of the medium modification of the jet splitting momentum fraction $z_{g}$, which is defined as 
\begin{align}
z_{g} &= \frac{p_{T,2}}{p_{T,1}+p_{T,2}}, 
\label{eq:zg}
\end{align}
where $p_{T,1}$ and $p_{T,2}$~($<p_{T,1}$) are transverse momenta of the pair prongs passing the soft-drop condition.

\begin{figure*}[htb]
\centering
\includegraphics[width=0.98\textwidth]{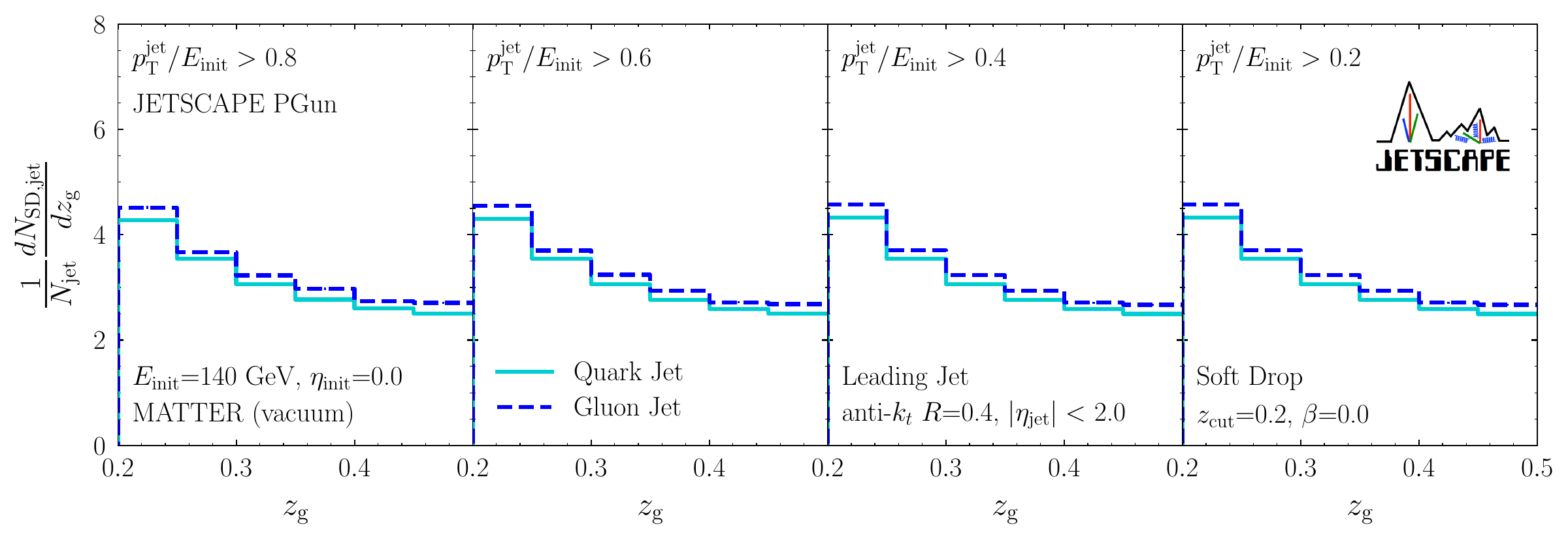}
\includegraphics[width=0.98\textwidth]{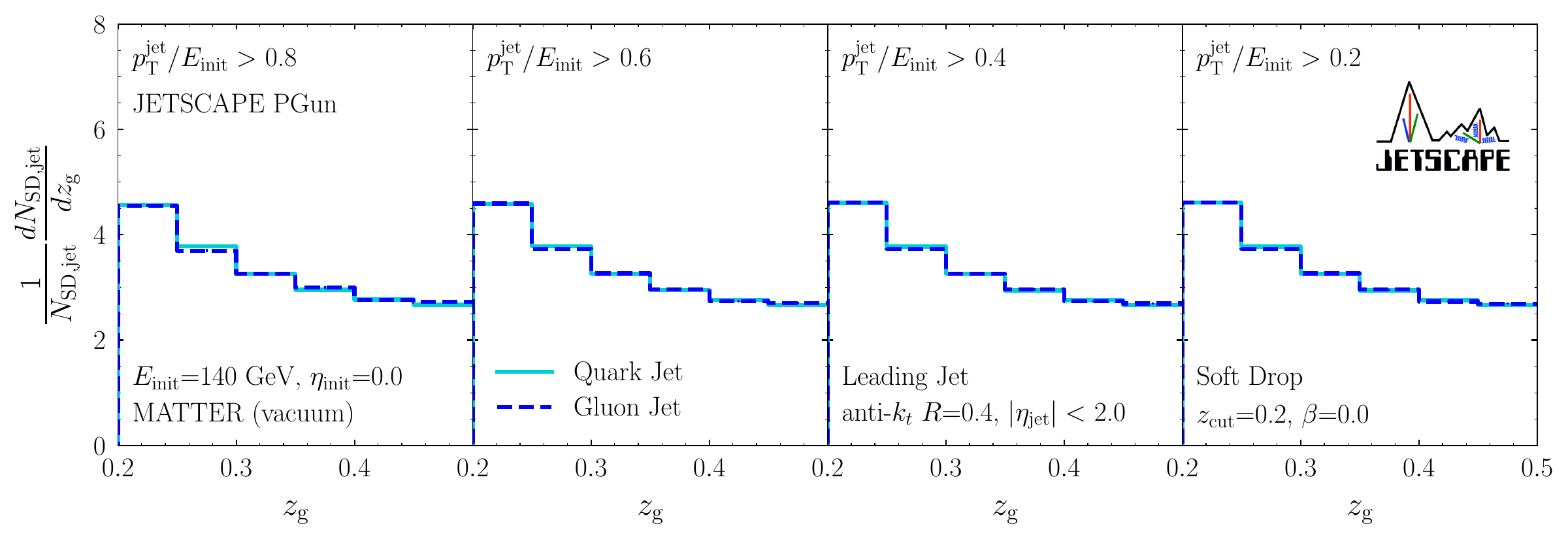} 
\caption{Distributions of jet splitting momentum fraction $z_{g}$ 
normalized by the number of all triggered jets (upper panel) and the number of jets passing the soft-drop condition (lower panel) for the leading jets in events generated with \textsc{pgun}. 
The jet shower evolution is performed by vacuum \textsc{matter} for the parent parton having $E_{\mathrm{init}} = 140$~GeV. 
Jets are reconstructed with $R=0.4$ at midrapidity $|\eta_{\mathrm{jet}}|<2.0$. 
The results are shown for quark jets (solid) and 
gluon jets (dashed) with different $p^{\mathrm{jet}}_{T}$ triggers, 
112, 84, 56, and 28~GeV. 
The soft-drop parameters are $z_{\mathrm{cut}}=0.2$ and $\beta = 0$. 
}
\label{fig:pgun_zg_vacuum}
\end{figure*}
In Fig.~\ref{fig:pgun_zg_vacuum}, the results for leading jets from \textsc{pgun} simulations for the vacuum case are shown. 
The initial energy of the parent parton is fixed at $E_{\mathrm{init}} = 140$~GeV. 
In the upper panel, the $z_{g}$ distribution normalized by the number of all triggered jets, including those that did not pass the soft-drop condition, 
\begin{align}
\frac{1}{N_{\mathrm{jet}}}\frac{d N_{\mathrm{SD,jet}}}{d z_{g}},
\label{eq:zg-dist-norm-njet}
\end{align}
is shown, whereas in the lower panel, the one normalized by the number of jets passing the soft-drop condition, 
\begin{align}
\frac{1}{N_{\mathrm{SD,jet}}}\frac{d N_{\mathrm{SD,jet}}}{d z_{g}},
\label{eq:zg-dist-norm-njetsd}
\end{align}
is shown. 
Here, $N_{\mathrm{jet}}$ and $N_{\mathrm{SD,jet}}$ are the number of triggered jets and the number of jets passing the soft-drop condition, respectively.

Gluon jets pass the soft-drop condition more often than quark jets, and as a result, for the case of the normalization by the number of all triggered jets, the gluon jet consistently exhibits larger values than the quark jet across the entire $z_{g}$ region. 
However, when normalized by the number of jets passing the soft-drop condition, the difference between quark jets and gluon jets is nearly negligible.

\begin{figure*}[htb]
\centering
\includegraphics[width=0.98\textwidth]{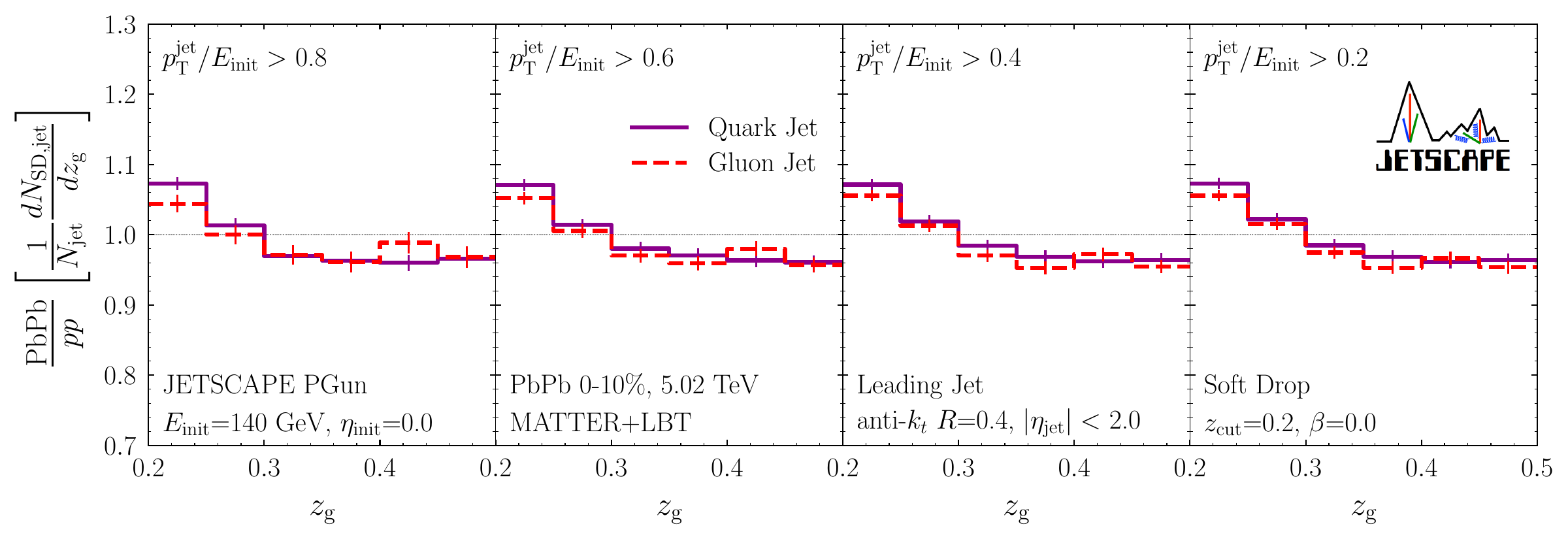}   
\caption{Ratios of $z_{g}$ distributions for the leading jets in events with the parent parton having a fixed initial energy $E_{\mathrm{init}} = 140$ GeV generated by \textsc{pgun}. 
The jet shower evolution in the QGP medium produced in 0\%--10\% Pb-Pb collisions at $\sqrt{s_{NN}}=5.02$~TeV is performed by \textsc{matter}$+$\textsc{lbt}. 
The results are shown for quark jets (solid) and gluon jets (dashed) with different $p^{\mathrm{jet}}_{T}$ triggers, 112, 84, 56, and 28~GeV. 
The soft-drop parameters are $z_{\mathrm{cut}}=0.2$ and $\beta = 0$. }
\label{fig:pgun_zg_in_medium}
\end{figure*} 
Figure~\ref{fig:pgun_zg_in_medium} shows the modification of the $z_{g}$ distribution for jets generated by \textsc{pgun} as they pass through the medium created in central (0\%--10\%) Pb-Pb collisions at $\sqrt{s_{NN}}=5.02$~TeV. 
For both quark jets and gluon jets, the modification pattern of a slight shift from small to large $z_{g}$ behavior is observed without any clear dependence on the value of the $p^{\mathrm{jet}}_{T}$ trigger. 
Thus, there is no modification attributed to the bias, solely due to energy loss, in the $z_{g}$ distribution, at least when considering quark jets and gluon jets separately.

\begin{figure*}[htb]
\centering
\includegraphics[width=0.98\textwidth]{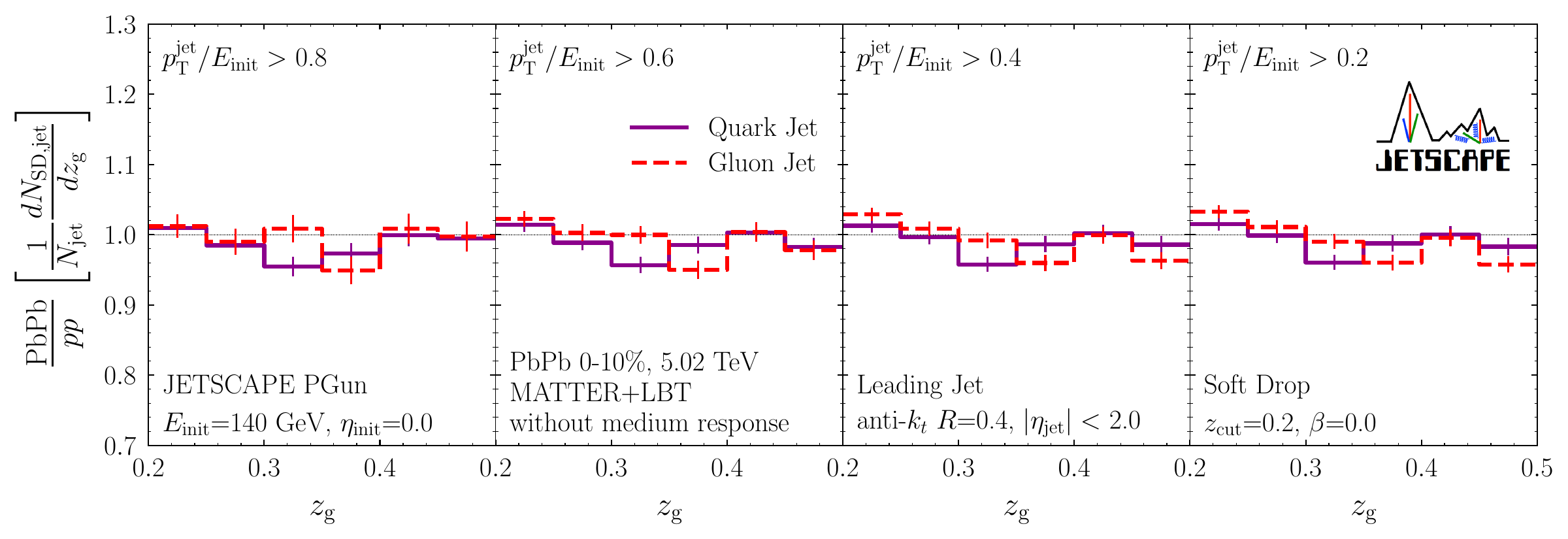}
\caption{Same as Fig.~\ref{fig:pgun_zg_in_medium} for \textsc{matter}$+$\textsc{lbt} simulations in which the medium effects from recoils and holes are artificially removed. }
\label{fig:pgun_zg_in_medium_no_recoil}
\end{figure*} 
Next, we explore the impact of the medium response effects via recoils and holes. 
Figure~\ref{fig:pgun_zg_in_medium_no_recoil} shows the results of the \textsc{pgun} simulations with \textsc{matter}$+$\textsc{lbt}, in which the effects of medium response are artificially excluded by discarding recoil and hole partons without passing them to the hadronization module. 
In this case, for both quark and gluon jets, the modification of the $z_{g}$ distribution is nearly invisible, indicating that the modifications observed in the full \textsc{matter}$+$\textsc{lbt} results are primarily governed by medium response: recoils tend to populate the smaller momentum-fraction branch of the prong pair identified as the hard branching by soft drop. 
This is consistent with findings reported in studies employing \textsc{jewel}~\cite{Milhano:2017nzm}, a Monte Carlo model for in-medium jet shower evolution that includes medium response effects from recoils.

\begin{figure*}[htb]
\centering
\includegraphics[width=0.98\textwidth]{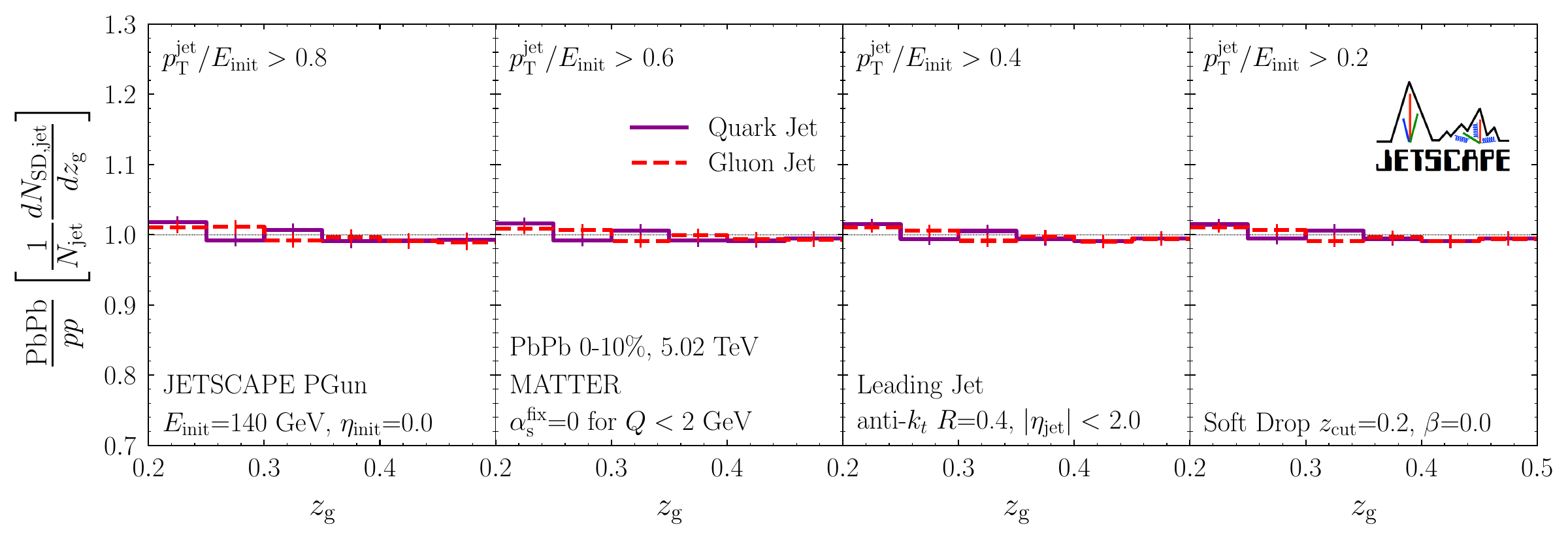} 
\caption{Same as Fig.~\ref{fig:pgun_zg_in_medium} for 
\textsc{matter} alone simulations, where the medium effect is turned off for jet partons with virtuality $Q<2$~GeV. }
\label{fig:pgun_zg_in_medium_lbt_off}
\end{figure*} 
Then, we delve into the detailed exploration of how the medium effects at high virtuality and low virtuality, respectively, bring about the hard splitting modification. 
To achieve this, we conduct simulations incorporating only the medium effects of \textsc{matter}, by turning off the medium effects at low virtuality ($Q<2$~GeV), and present the results of these simulations with the \textsc{pgun} initial hard process in Fig.~\ref{fig:pgun_zg_in_medium_lbt_off}. 
Note that in this configuration, recoils and holes from the \textsc{matter} phase are included. 
In this case, no medium effects are seen for either quark or gluon jets, and the recoils contributing to the soft-drop-reconstructed hard branching arise mainly in the low-virtuality phase, as modeled by the \textsc{lbt} module.

\begin{figure*}[htb]
\centering
\includegraphics[width=0.98\textwidth]{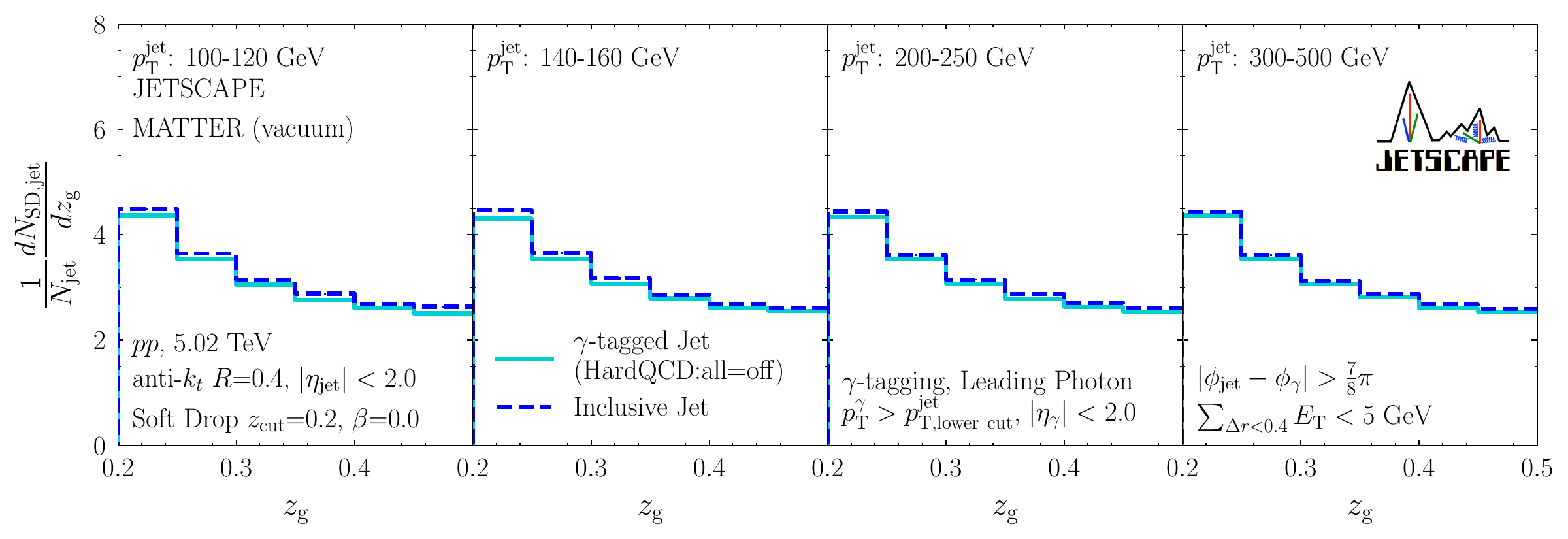}
\includegraphics[width=0.98\textwidth]{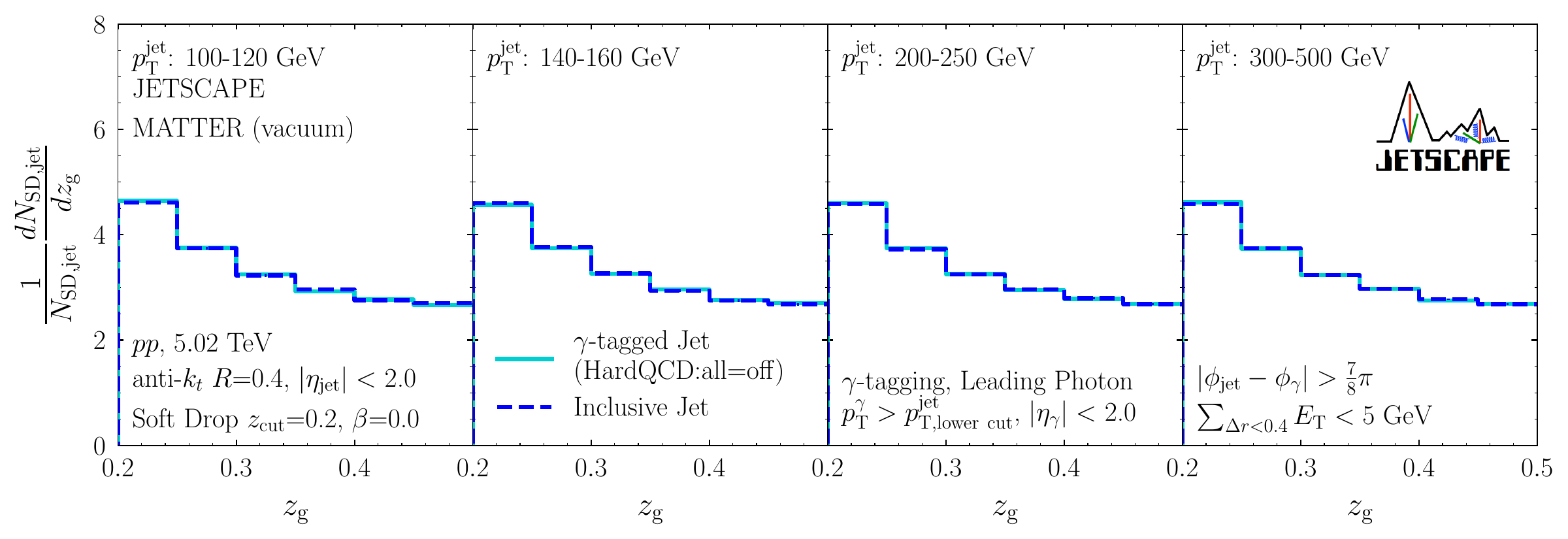}
\caption{Same as Fig.~\ref{fig:pgun_zg_vacuum} for jets from hard scatterings in $p$-$p$ collisions at $\sqrt{s}=5.02$~TeV, generated by \textsc{pythiagun} with ISR and MPI. 
The results are shown for $\gamma$-tagged jets (solid) from prompt photon-generating hard processes (\texttt{HardQCD:all=off}+\texttt{PromptPhoton:all=on}) and inclusive jets (dashed) from inclusive hard processes (\texttt{HardQCD:all=on}+\texttt{PromptPhoton:all=on}) generated at leading order by \textsc{pythia} 8 and with different $p^{\mathrm{jet}}_{\mathrm{T}}$ triggers. 
For $\gamma$-tagged jets, the isolation requirement, the relative azimuth angle cut, and the additional cut of $p^{\mathrm{jet}}_{\mathrm{T}} < p^{\gamma}_{\mathrm{T}}$ are imposed. 
}
\label{fig:inclusive_gamma_tagged_zg_vacuum}
\end{figure*}
Finally, we examine jets in more realistic events involving hard scattering in $p$-$p$ collisions at $\sqrt{s} = 5.02$ TeV, generated using the \textsc{pythiagun} module. 
The results for $\gamma$-tagged jets and inclusive jets with the same $p^{\mathrm{jet}}_{\mathrm{T}}$ triggers are compared in Fig.~\ref{fig:inclusive_gamma_tagged_zg_vacuum}. 
For the $\gamma$-tagged jet analysis, only events including prompt photon production 
at leading order in the initial hard processes are selectively generated with the \textsc{pythia} option \texttt{HardQCD:all=off} along with \texttt{PromptPhoton:all=on}.

The analysis of $\gamma$-tagged jets involves first identifying the leading photon with the highest $p_{T}$ in each event that satisfies the trigger conditions. All jets associated with this leading photon and meeting the trigger criteria are then counted. 
To mimic the realistic experimental analysis, the isolation requirement $E^{\mathrm{iso}}_{\mathrm{T}}<5$~GeV and the relative azimuth angle cut $\lvert \phi_{\mathrm{jet}} - \phi_{\mathrm{\gamma}}\rvert<7\pi/8$ 
are imposed. 
To further suppress the unwanted contribution from photon radiation from jets, we employ the additional cut of $p^{\mathrm{jet}}_{\mathrm{T}} < p^{\gamma}_{\mathrm{T}}$.

For the inclusive jet analysis, events with initial scatterings, including all hard QCD 2-to-2 events (\texttt{HardQCD:all=on}+\texttt{PromptPhoton:all=on}) are used. 
In the distributions normalized by the number of all triggered jets (upper panel), it can be observed that inclusive jets are slightly more likely to pass the soft-drop condition compared with $\gamma$-triggered jets, due to the larger fraction of gluon jets. 
When the normalization is done by the number of jets passing the soft-drop condition, the difference between $\gamma$-triggered jets and inclusive jets disappear, similar to the erasing of differences between gluon and quark jets as seen in the \textsc{pgun} results.

\begin{figure*}[htb]
\centering
\includegraphics[width=0.98\textwidth]{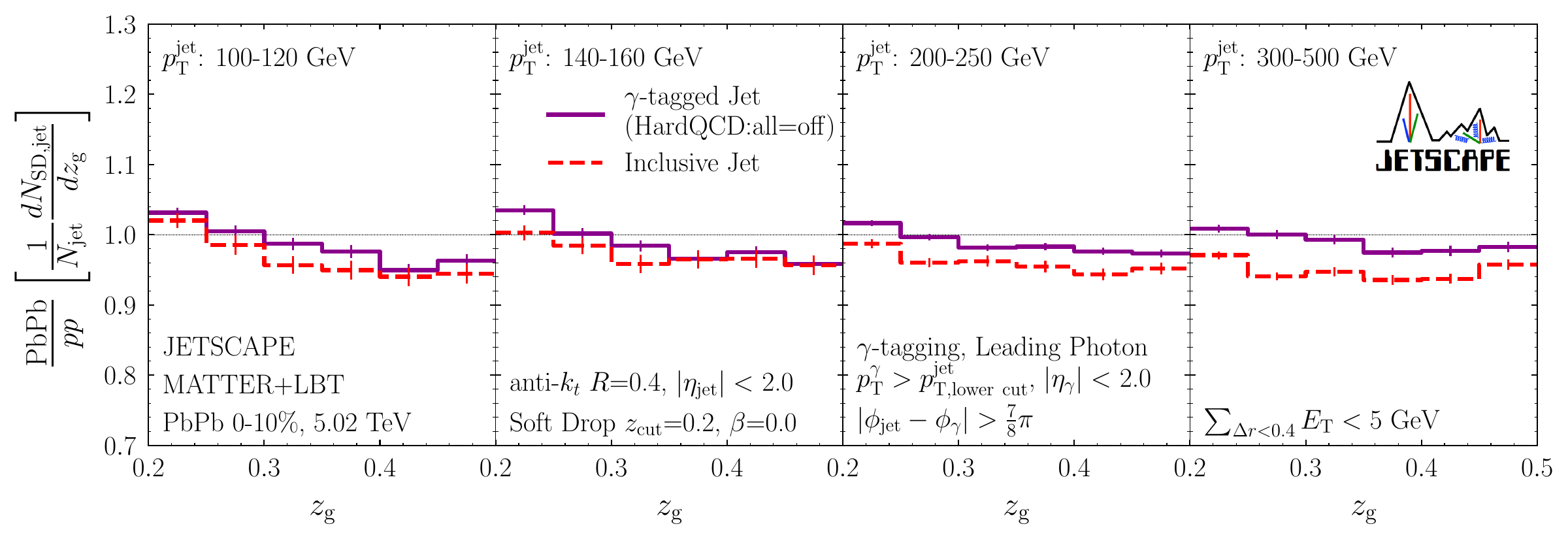}
\caption{Same as Fig.~\ref{fig:pgun_zg_in_medium} for jets from hard scatterings at $\sqrt{s_{NN}}=5.02$~TeV, generated by \textsc{pythiagun} with ISR and MPI. 
The results are shown for $\gamma$-tagged jets (solid) from prompt photon-generating hard processes (\texttt{HardQCD:all=off}+\texttt{PromptPhoton:all=on}) and inclusive jets (dashed) from inclusive hard processes (\texttt{HardQCD:all=on}+\texttt{PromptPhoton:all=on}) 
generated at leading order by \textsc{pythia} 8 with different $p^{\mathrm{jet}}_{\mathrm{T}}$ triggers. 
For $\gamma$-tagged jets, the isolation requirement, the relative azimuth angle cut, and the additional cut of $p^{\mathrm{jet}}_{\mathrm{T}} < p^{\gamma}_{\mathrm{T}}$ are imposed.}
\label{fig:inclusive_gamma_tagged_zg_in_medium}
\end{figure*}
Figure~\ref{fig:inclusive_gamma_tagged_zg_in_medium} shows the medium modification for the $\gamma$-tagged and inclusive jets in Pb-Pb collisions at $\sqrt{s_{NN}}=5.02$~TeV with the same triggers. 
For both $\gamma$-tagged jets and inclusive jets, modification becomes less significant as $p^{\mathrm{jet}}_{T}$ increases. 
Especially for jets with $p^{\mathrm{jet}}_{T}>300$~GeV, almost no changes in the shape of the $z_{g}$ distribution are observed. 
In all $z_{g}$ regions, jets passing the soft-drop condition are almost uniformly reduced due to medium effects, with the reduction being slightly more pronounced for inclusive jets.

\begin{figure*}[htb]
\centering
\includegraphics[width=0.98\textwidth]{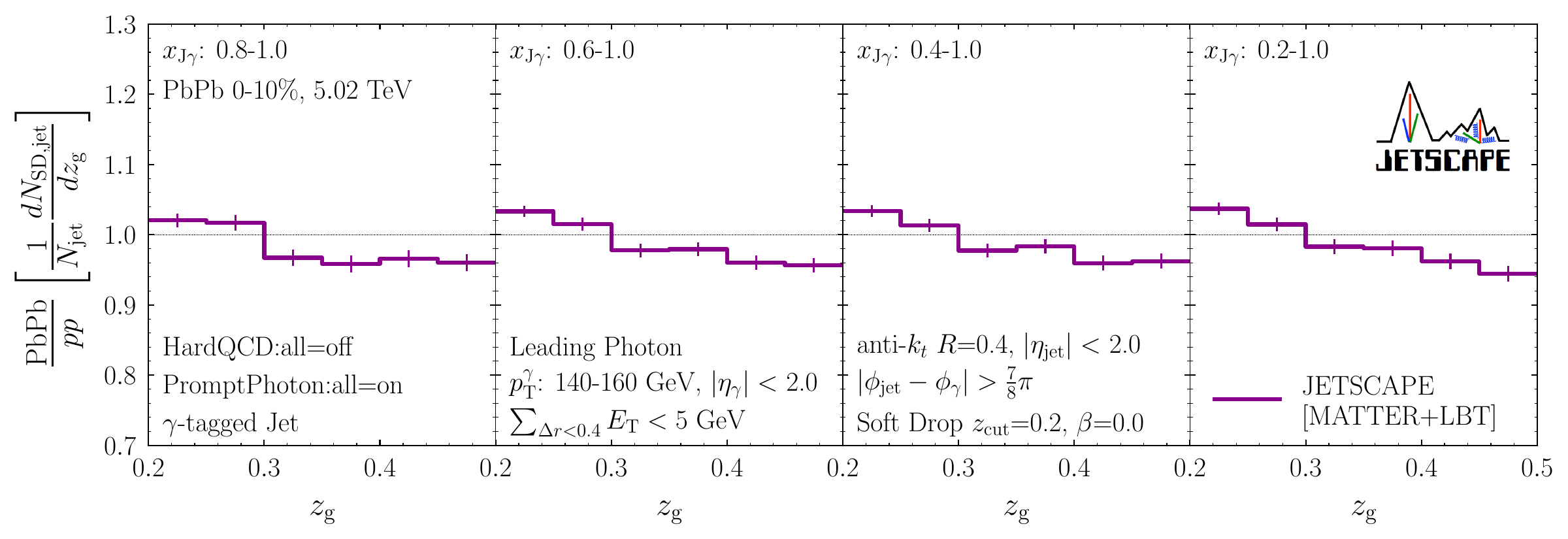}
\caption{Same as Fig.~\ref{fig:pgun_zg_in_medium} for $\gamma$-tagged jets from prompt photon-generating hard processes (\texttt{HardQCD:all=off}+\texttt{PromptPhoton:all=on}) generated at leading order by \textsc{pythia} 8 at $\sqrt{s_{NN}}=5.02$~TeV 
for different $x_{J\gamma}$ ranges. 
The photon of $140<p^{\gamma}_{\mathrm{T}} < 160$~GeV is triggered with the isolation requirement, and the relative azimuth angle cut.}
\label{fig:gamma_tagged_zg_xdep}
\end{figure*}
Figure~\ref{fig:gamma_tagged_zg_xdep} shows our prediction for the $x_{J\gamma}$ dependence on the modification of the $z_{g}$ distribution for $\gamma$-tagged jets in Pb-Pb collisions at $\sqrt{s_{NN}}=5.02$~TeV. 
As seen in the \textsc{pgun} results, there is no significant 
$p^{\mathrm{jet}}_{T}$-selection bias effect on the $z_{g}$ distribution, and thus, the modification remains almost consistent across all presented $x_{J\gamma}$ ranges.

\subsection{Jet splitting radius}
\label{Subsection:rG}
Next, we investigate the medium modification of the jet splitting radius $r_{g}$, which is defined as
\begin{align}
r_{g} = \sqrt{\left(\eta_{1}-\eta_{2}\right)^{2}+\left(\phi_{1}-\phi_{2}\right)^{2}},
\label{eq:rg}
\end{align}
where $\eta_{1}$, $\eta_{2}$, $\phi_{1}$, and $\phi_{2}$ are the rapidities and azimuthal angles of the pair prongs passing the soft-drop condition.

\begin{figure*}[htb]
\centering
\includegraphics[width=0.98\textwidth]{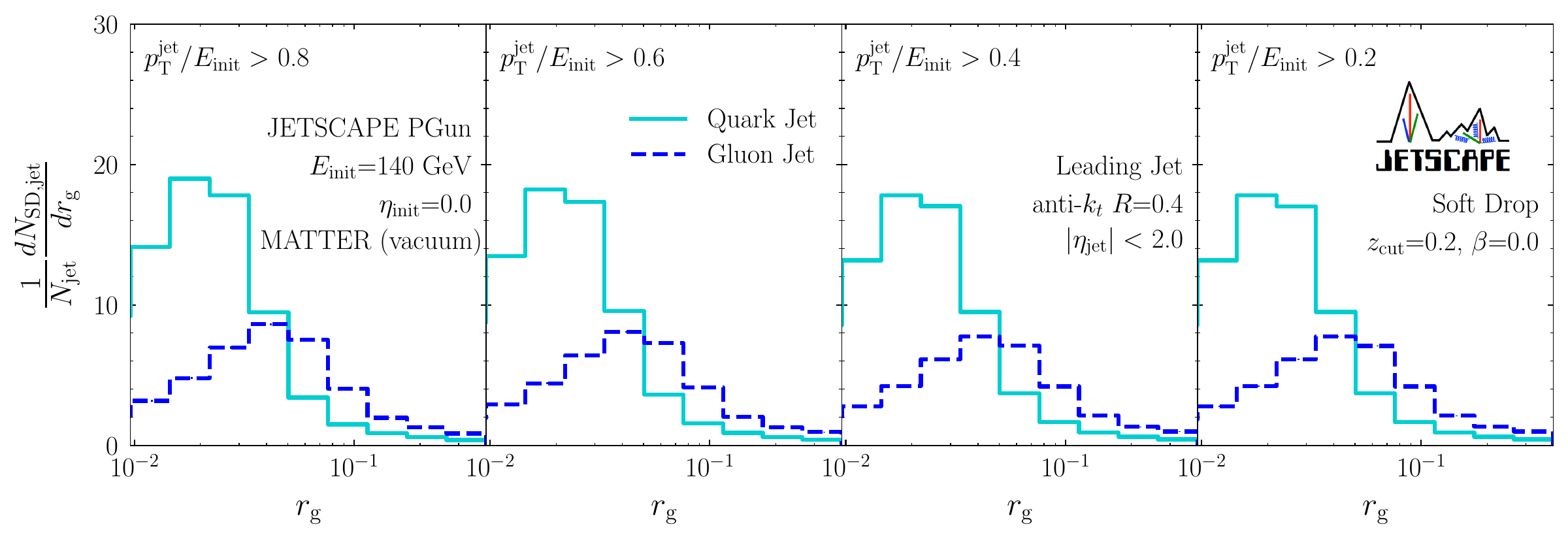}
\caption{Distributions of jet splitting momentum radius $r_{g}$ 
normalized by the number of all triggered jets for the leading jets in events generated with \textsc{pgun}. 
The jet shower evolution is performed by vacuum \textsc{matter} for the parent parton having $E_{\mathrm{init}} = 140$~GeV. 
Jets are reconstructed with $R=0.4$ at midrapidity $|\eta_{\mathrm{jet}}|<2.0$. 
The results are shown for quark jets (solid) and gluon jets (dashed) with different $p^{\mathrm{jet}}_{T}$ triggers, 
112, 84, 56, and 28~GeV. 
The soft-drop parameters are $z_{\mathrm{cut}}=0.2$ and $\beta = 0$. 
}
\label{fig:pgun_rg_vacuum}
\end{figure*} 
In Fig.~\ref{fig:pgun_rg_vacuum}, the $r_{g}$ distributions normalized by $N_{\mathrm{jet}}$, 
\begin{align}
\frac{1}{N_{\mathrm{jet}}}\frac{d N_{\mathrm{SD,jet}}}{d r_{g}},
\label{eq:rg-dist-norm-njet}
\end{align} 
for leading gluon jets and quark jets from the vacuum \textsc{pgun} simulations are compared. 
The gluon jets exhibit much wider distributions with a peak at larger $r_{g}$ values than quark jets. 
This is due to the fact that gluon jets, having a larger Casimir factor and radiating more, tend to be produced with a larger virtuality compared with quark jets. 
For the jet cone size $R=0.4$, almost no $p^{\mathrm{jet}}_{T}$ cut dependence can be observed in the vacuum case with fixed $E_{\mathrm{init}} = 140$~GeV.

\begin{figure*}[htb]
\centering
\includegraphics[width=0.98\textwidth]{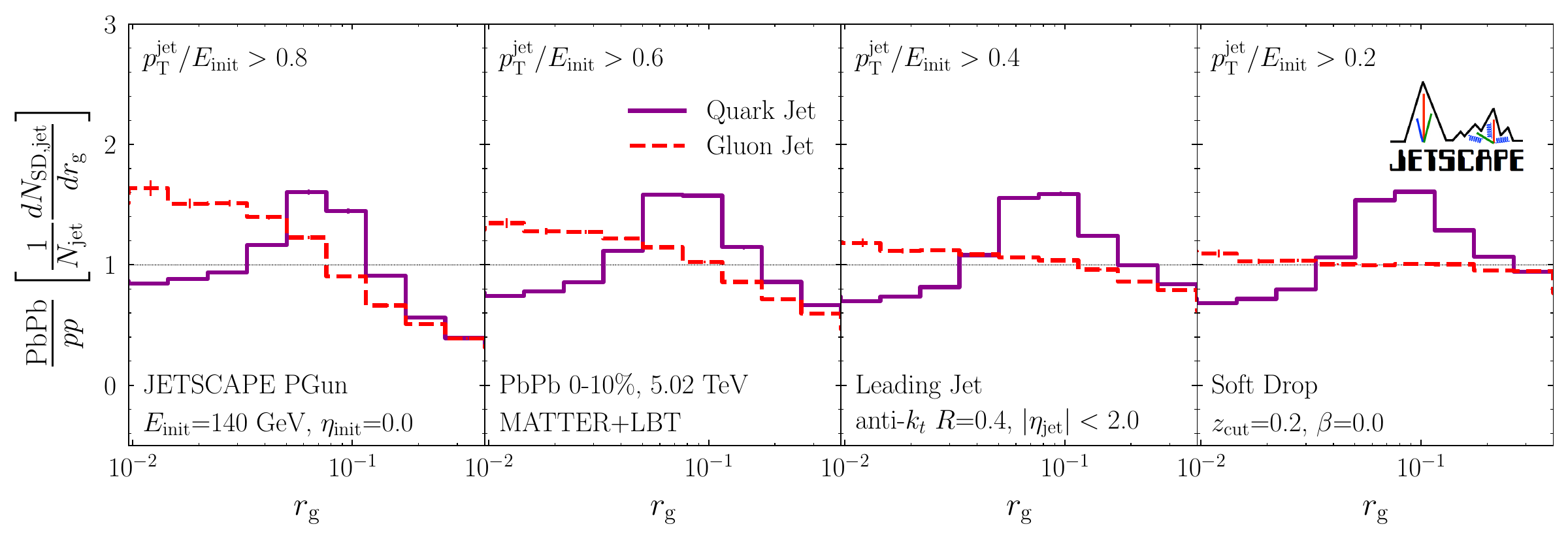} 
\caption{Ratios of $r_{g}$ distributions for the leading jets in events with the parent parton having a fixed initial energy $E_{\mathrm{init}} = 140$ GeV generated by \textsc{pgun}. 
The jet shower evolution in the QGP medium produced in 0\%--10\% Pb-Pb collisions at $\sqrt{s_{NN}}=5.02$~TeV is performed by \textsc{matter}$+$\textsc{lbt}. 
All the setups are the same as in Fig.~\ref{fig:pgun_zg_in_medium}. }
\label{fig:pgun_rg_in_medium}
\end{figure*} 
Figure~\ref{fig:pgun_rg_in_medium} shows the modification of the $r_{g}$ distribution for the \textsc{pgun} jets by the medium created in central Pb-Pb collisions at $\sqrt{s_{NN}}=5.02$~TeV. 
For gluon jets, a monotonically decreasing trend with increasing $r_{g}$ is observed in cases with a larger $p^{\mathrm{jet}}_{T}$ cut. 
Then, as the $p^{\mathrm{jet}}_{T}$ cut value is reduced, the decrease rate becomes more gentle, and almost no modification can be seen for $p^{\mathrm{jet}}_{T}/E_{\mathrm{init}} > 0.2$. 
Thus, for gluon jets, the modification pattern is entirely brought by the effect from jet selection with $p^{\mathrm{jet}}_{T}$ cut: 
Jets with an originally large splitting radius tend to have their constituents reach the edge of the jet cone more easily, making them lose more energy and thus less likely to be triggered.

The selection bias effect with $p^{\mathrm{jet}}_{T}$ cut can be observed also for quark jets as large-$r_{g}$ suppression, with the same trend disappearing as the $p^{\mathrm{jet}}_{T}$ cut value is reduced. 
Additionally, in the case of quark jets, a prominent bump structure accompanied by slight suppression at small $r_{g}$ is observed in the mid-$r_{g}$ region. 
This feature is significantly different from that of gluon jets, providing evidence of direct modifications to the hard-branching structure within quark jets.

\begin{figure*}[htb]
\centering
\includegraphics[width=0.98\textwidth]{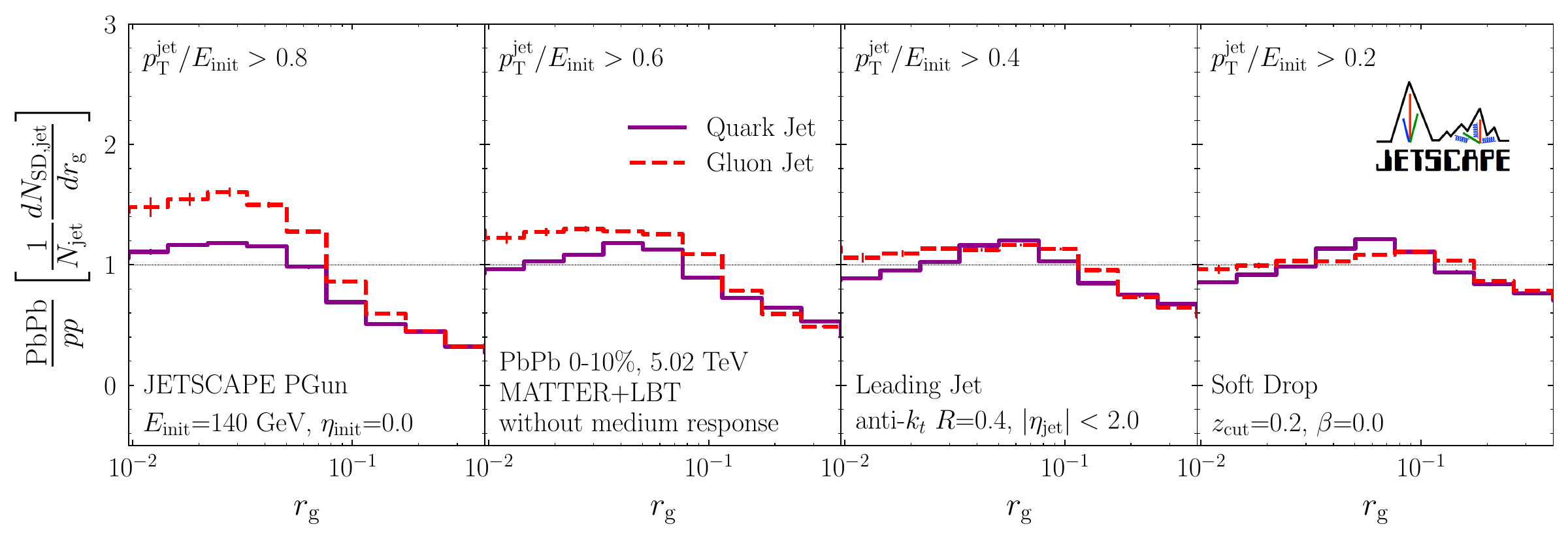}
\caption{Same as Fig.~\ref{fig:pgun_rg_in_medium} for \textsc{matter}$+$\textsc{lbt} simulations in which the medium effects from recoils and holes are artificially removed.}
\label{fig:pgun_rg_in_medium_no_recoil}
\end{figure*} 
In Fig.~\ref{fig:pgun_rg_in_medium_no_recoil}, we show the \textsc{matter}$+$\textsc{lbt} results with the medium response effects from recoils and holes artificially disabled. 
For gluon jets, small recoil contributions to hard branchings at narrow angles manifest as a minor suppression at very small $r_g$ relative to the full results. 
In quark jets, the pronounced bump seen with recoils and holes is largely diminished, demonstrating that the feature is primarily driven by medium response, although a slight broadening remains visible. 
Since the recoils and holes are discarded, explicitly violating energy-momentum conservation, the large-$r_{g}$ suppression does not disappear even when the $p^{\mathrm{jet}}_{T}$ cut value is reduced.

\begin{figure*}[htb]
\centering
\includegraphics[width=0.98\textwidth]{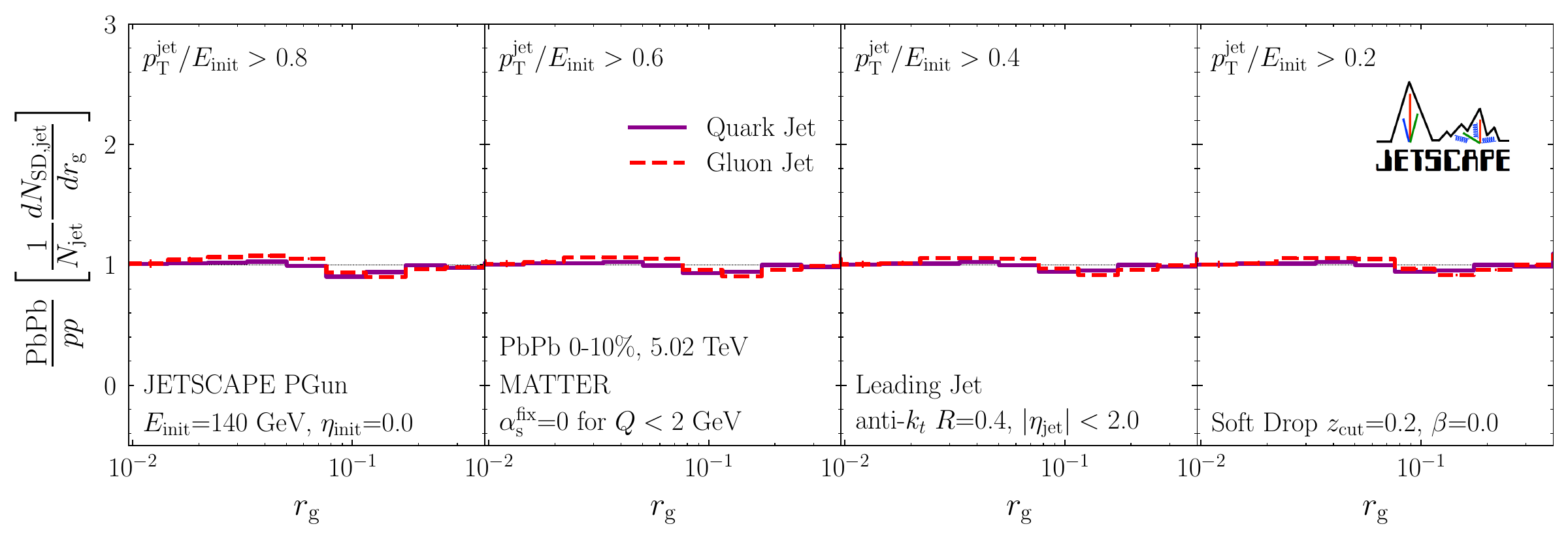} 
\caption{Same as Fig.~\ref{fig:pgun_rg_in_medium} for 
\textsc{matter} alone simulations, where the medium effect is turned off for jet partons with virtuality $Q<2$~GeV. }
\label{fig:pgun_rg_in_medium_lbt_off}
\end{figure*} 
Figure~\ref{fig:pgun_rg_in_medium_lbt_off} shows the \textsc{matter}-only results, in which the medium effect is turned off for $Q<2$~GeV. 
For both quark jets and gluon jets, $r_{g}$ distribution modification for any $p^{\mathrm{jet}}_{T}$ cut is nearly imperceptible. 
This indicates that the modifications observed in the full \textsc{matter}$+$\textsc{lbt} results are completely governed by the medium effect in the low-virtuality \textsc{lbt} phase.

In the case of quark jets, it is conceivable that soft-drop-reconstructed hard branchings modified by medium effects at low virtuality, primarily through the contribution of recoils, occur at larger angles than radiations predominantly driven by virtuality. 
This might be attributed to the relatively lower initial virtuality in quark jets, where the virtuality-driven splittings are less likely to have very large angles. 
Such large-angle medium-induced branchings can be identified as hard branchings by the soft-drop grooming, subsequently manifesting in the $r_{g}$ distribution as a shift from smaller $r_{g}$ to moderate $r_{g}$ ($\approx 0.1$), appearing as a bump structure in the full \textsc{matter}$+$\textsc{lbt} results. 
Conversely, for gluon jets, despite medium effects at low virtuality, splittings primarily driven by the initial virtuality are still predominantly identified as hard branchings by the soft-drop grooming. 
The medium effects at low virtuality do not alter the intrinsic hard structure of gluon jets while still resulting in significant jet energy loss.

\begin{figure*}[htb]
\centering
\includegraphics[width=0.98\textwidth]{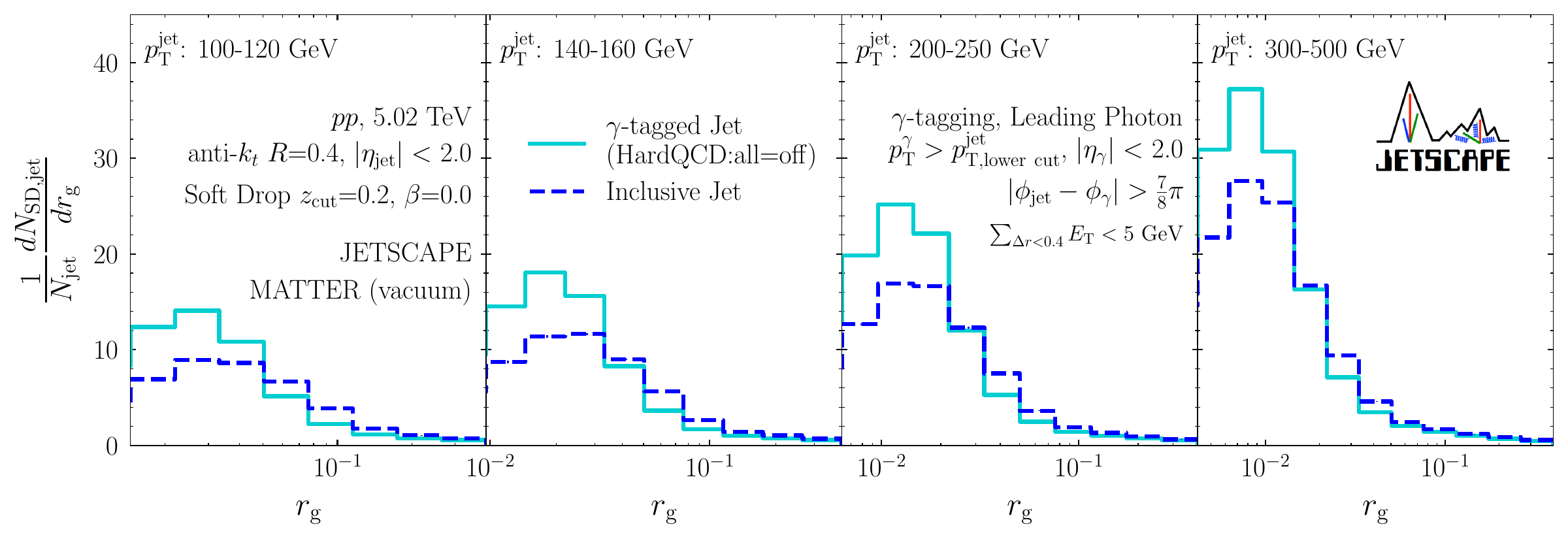}
\caption{Same as Fig.~\ref{fig:pgun_rg_vacuum} for jets from hard scatterings in $p$-$p$ collisions at $\sqrt{s}=5.02$~TeV, generated by \textsc{pythiagun} with ISR and MPI. 
The results are shown for $\gamma$-tagged jets (solid) from prompt photon-generating hard processes (\texttt{HardQCD:all=off}+\texttt{PromptPhoton:all=on}) and inclusive jets (dashed) from inclusive hard processes (\texttt{HardQCD:all=on}+\texttt{PromptPhoton:all=on}) generated at leading order by \textsc{pythia} 8 with different $p^{\mathrm{jet}}_{\mathrm{T}}$ triggers. 
For $\gamma$-tagged jets, the isolation requirement, the relative azimuth angle cut, and the additional cut of $p^{\mathrm{jet}}_{\mathrm{T}} < p^{\gamma}_{\mathrm{T}}$ are imposed. 
}
\label{fig:inclusive_gamma_tagged_rg_vacuum}
\end{figure*}
In Fig.~\ref{fig:inclusive_gamma_tagged_rg_vacuum}, $r_{g}$ distributions for $\gamma$-tagged jets and inclusive jets in $p$-$p$ collisions at $\sqrt{s}=5.02$~TeV are compared. 
Unlike the case with $z_{g}$, both $\gamma$-tagged jets and inclusive jets exhibit a distinct $p^{\mathrm{jet}}_{T}$ dependency: narrowing as $p^{\mathrm{jet}}_{T}$ increases. 
This can lead to an additional contribution to the selection bias with $p^{\mathrm{jet}}_{T}$ cut in the presence of jet energy loss. 
Even without any modification on the splitting angle, the energy loss causes jets from parent partons with higher initial $p_{T}$, which typically exhibit narrower splittings, to be triggered more likely. 
As a result, narrowing of the $r_{g}$ distribution is expected in the presence of the medium.

Furthermore, clear differences between $\gamma$-tagged jets and inclusive jets are also observed. 
Inclusive jets show a broader distribution since they have a larger fraction of gluon jets. 
As $p^{\mathrm{jet}}_{T}$ increases, this difference diminishes, owing to the decrease in the gluon jet fraction of inclusive jets.

\begin{figure*}[htb]
\centering
\includegraphics[width=0.98\textwidth]{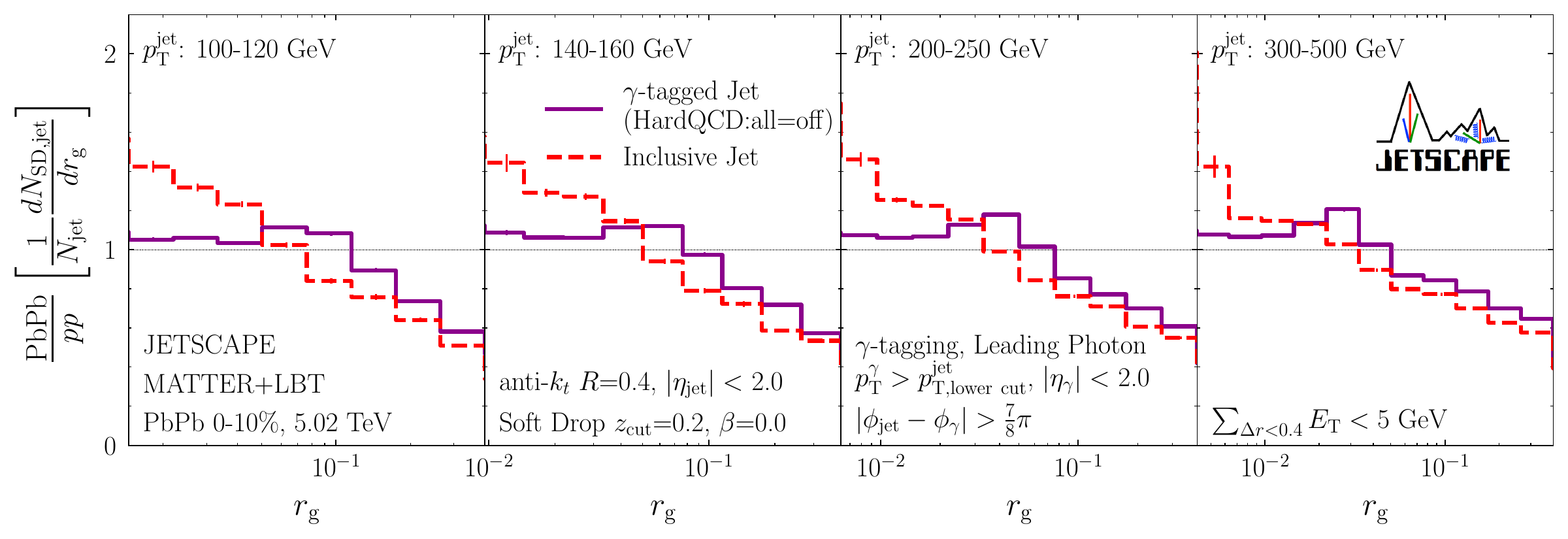}
\caption{Same as Fig.~\ref{fig:pgun_rg_in_medium} for jets from hard scatterings at $\sqrt{s_{NN}}=5.02$~TeV, generated by \textsc{pythiagun} with ISR and MPI. 
The results are shown for $\gamma$-tagged jets (solid) from prompt photon-generating hard processes (\texttt{HardQCD:all=off}+\texttt{PromptPhoton:all=on}) and inclusive jets (dashed) from inclusive hard processes (\texttt{HardQCD:all=on}+\texttt{PromptPhoton:all=on}) generated at leading order by \textsc{pythia} 8 with different $p^{\mathrm{jet}}_{\mathrm{T}}$ triggers. 
For $\gamma$-tagged jets, the isolation requirement, the relative azimuth angle cut, and the additional cut of $p^{\mathrm{jet}}_{\mathrm{T}} < p^{\gamma}_{\mathrm{T}}$ are imposed.
}
\label{fig:inclusive_gamma_tagged_rg_in_medium}
\end{figure*}
The $r_{g}$ distribution modifications for $\gamma$-tagged jets and inclusive jets in Pb-Pb collisions at $\sqrt{s_{NN}}=5.02$~TeV are compared in Fig.~\ref{fig:inclusive_gamma_tagged_rg_in_medium}. 
Here, it should be noted that two factors contribute to the energy loss selection bias effect of $p^{\mathrm{jet}}_{T}$ cut, causing the $r_{g}$ narrowing irrelevant to the actual structural modification of hard splittings. 
First, as confirmed by the results with a fixed $E_{\mathrm{init}}$, jets with originally wider splittings tend to lose more energy, resulting in less triggered jets as $r_{g}$ increases. 
In addition, as mentioned above for the $p$-$p$ case, jets with parent partons with larger initial $p_{T}$, which possess the narrower $r_{g}$ distribution, are triggered more likely in the presence of energy loss.

In the case of $\gamma$-tagged jets, at large $r_{g}$, the suppression due to the selection bias is evident for all the presented $p^{\mathrm{jet}}_{T}$ ranges. 
On the other hand, from small to mid $r_{g}~(\lessapprox 0.1)$, the modification pattern is significantly flattened. 
This flat structure can predominantly be attributed to the large fraction of quark jets in $\gamma$-tagged jets: balance between the $r_{g}$ broadening observed in quark jets and the selection bias effect. 
As $p^{\mathrm{jet}}_{T}$ increases, the broadening effect starts to slightly dominate, leading to subtle indications of a bump structure.

In the case of inclusive jets, across all presented $p^{\mathrm{jet}}_{T}$ ranges, the modification patterns are dominated by the selection bias effect, exhibiting a monotonically decreasing trend with increasing $r_{g}$. 
This monotonic modification pattern in inclusive jets is consistent with those presented in our previous study~\cite{JETSCAPE:2023hqn}, where our results show good agreement with experimental data from ALICE~\cite{ALargeIonColliderExperiment:2021mqf} and ATLAS~\cite{ATLAS:2022vii} simultaneously. 
Although the overall monotonic decreasing behavior remains, increasing the jet $p^{\mathrm{jet}}_{T}$ leads to a less steep slope at small $r_{g}$. 
This is attributed to the increase in the quark jet fraction, as seen in the narrowing of the $r_{g}$ distribution in $p$-$p$ collisions in Fig.~\ref{fig:inclusive_gamma_tagged_rg_vacuum}

\begin{figure*}[htb]
\centering
\includegraphics[width=0.98\textwidth]{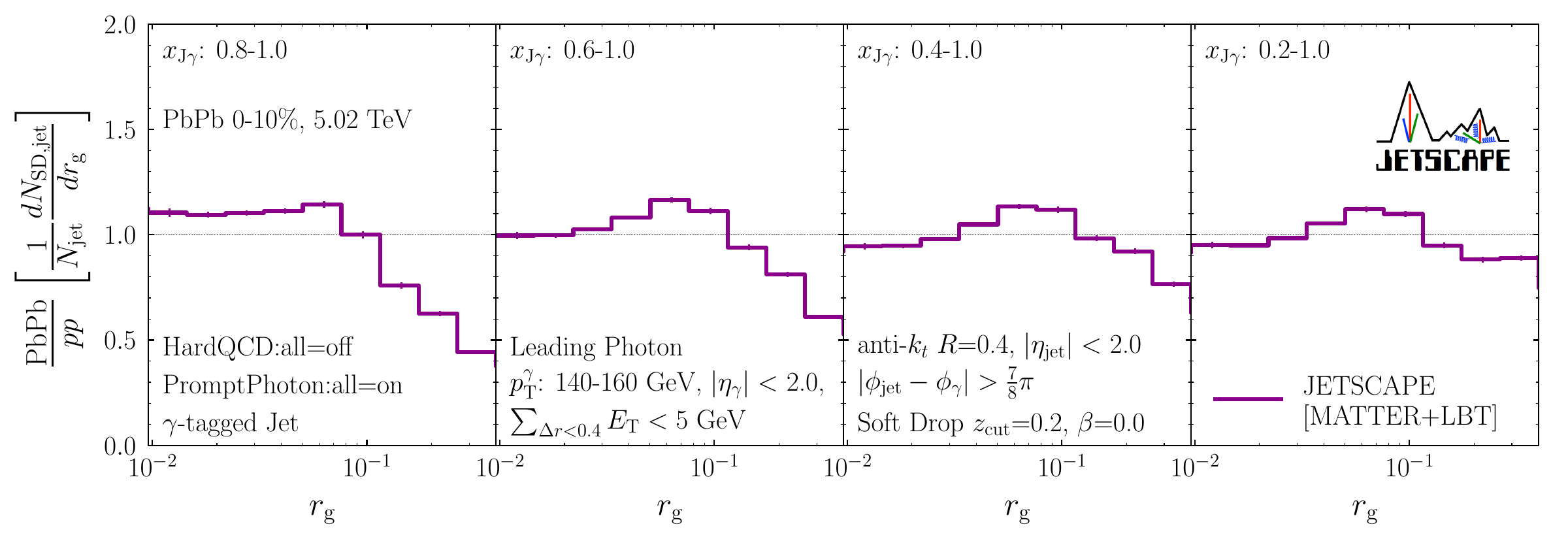}
\caption{Same as Fig.~\ref{fig:pgun_rg_in_medium} for $\gamma$-tagged jets from prompt photon-generating hard processes (\texttt{HardQCD:all=off}+\texttt{PromptPhoton:all=on}) generated at leading order by \textsc{pythia} 8 at $\sqrt{s_{NN}}=5.02$~TeV for different $x_{J\gamma}$ ranges. 
The photon of $140<p^{\gamma}_{\mathrm{T}} < 160$~GeV is triggered with the isolation requirement, and the relative azimuth angle cut.}
\label{fig:gamma_tagged_rg_xdep}
\end{figure*}
Figure~\ref{fig:gamma_tagged_rg_xdep} shows our prediction for the medium modification of the $r_{g}$ distribution for $\gamma$-tagged jets in central Pb-Pb collisions at $\sqrt{s_{NN}}=5.02$~TeV for different $x_{J\gamma}$ cuts. 
Unlike the $z_{g}$ distribution, the $r_{g}$ distribution exhibits a strong $x_{J\gamma}$ dependence. 
For the case with large $x_{J\gamma}$ cuts, the selection bias effect due to the $r_{g}$ dependence of energy loss appears prominently, resulting in a strong suppression at large $r_{g}$ values. 
Here, it should be noted that the effect of the strong $p^{\mathrm{jet}}_{T}$ dependence on the 
$r_{g}$ distribution of jets in vacuum is diminished, since, by setting an uppercut on $p^{\gamma}_{T}$, the feeddown from jets originating from parent partons with larger initial $p_{T}$ is suppressed.

The selection bias effect is counterbalanced by the strong broadening effects, attributable to the large fraction of quark jets, in the small-to-mid $r_{g}$ range. 
When the value of the lower $x_{J\gamma}$ cut is reduced to weaken the selection bias effect, the large $r_{g}$ suppression diminishes, and concurrently, even a bump structure due to broadening emerges at mid $r_{g}$ ($\lessapprox 0.1$). 
This manifestation of the broadening effect in the $r_{g}$ distribution represents a distinct characteristic of $\gamma$-tagged jets, in contrast with the behavior of inclusive jets, where only a monotonic pattern entirely dominated by selection bias can be seen. 
Particularly, by varying the $x_{J\gamma}$ cut, one can control the selection bias effect and study the broadening effect more quantitatively. 
Therefore, $\gamma$-tagged jets provide a modification pattern in the $r_{g}$ distribution that is more suitable for investigating the structural modifications on the soft-drop-reconstructed hard branchings.

\subsection{Relative transverse momentum of jet splittings}
\label{Subsection:ktG}
The relative transverse momentum of jet splittings, 
\begin{align}
k_{T,g} &= p_{T,2} \sin r_{g}, 
\label{eq:ktg}
\end{align}
has recently been measured in experiments to investigate the transverse structure of hard splittings in jets~\cite{Ehlers:2023jbf}. 
\begin{figure*}[htb]
\centering
\includegraphics[width=0.98\textwidth]{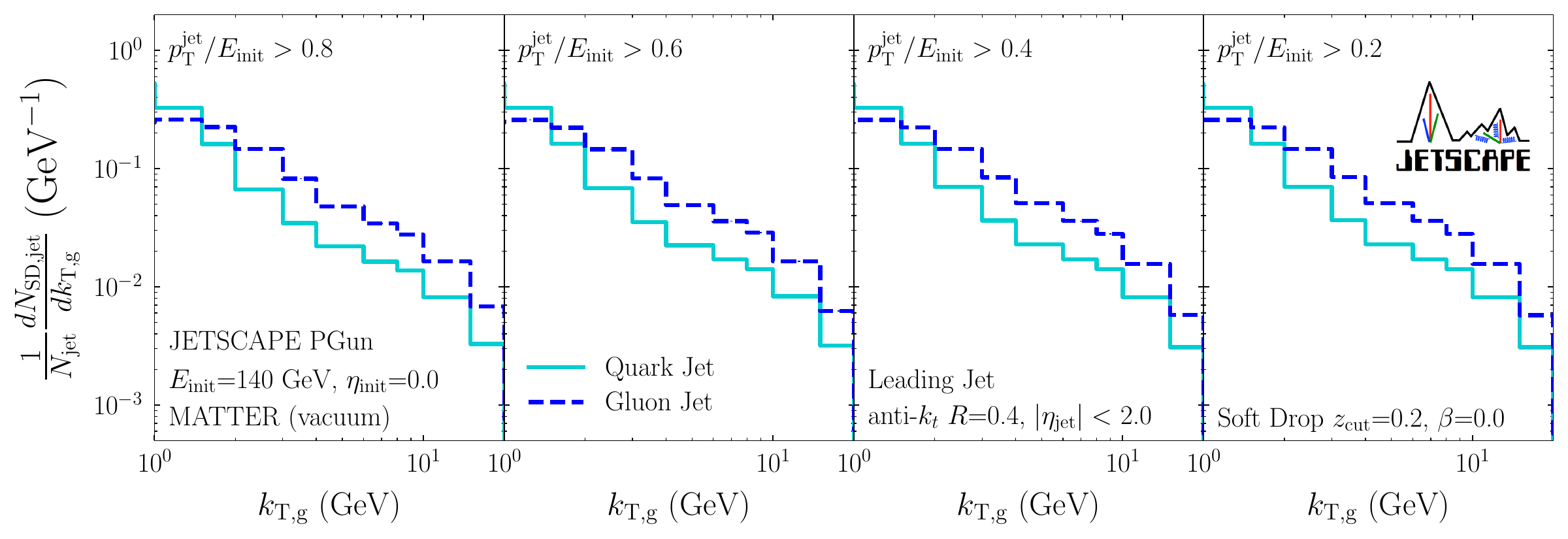}
\caption{Distributions of relative transverse momentum of jet splittings $k_{T,g}$ 
normalized by the number of all triggered jets for the leading jets in events generated with \textsc{pgun}. 
The jet shower evolution is performed by vacuum \textsc{matter} for the parent parton having $E_{\mathrm{init}} = 140$~GeV. 
Jets are reconstructed with $R=0.4$ at midrapidity $|\eta_{\mathrm{jet}}|<2.0$. 
The results are shown for quark jets (solid) and gluon jets (dashed) with different $p^{\mathrm{jet}}_{T}$ triggers, 
112, 84, 56, and 28~GeV. 
The soft-drop parameters are $z_{\mathrm{cut}}=0.2$ and $\beta = 0$. 
}
\label{fig:pgun_ktg_vacuum}
\end{figure*} 
We present the results of the $k_{T,g}$ distributions,
\begin{align}
\frac{1}{N_{\mathrm{jet}}}\frac{d N_{\mathrm{SD,jet}}}{d k_{T,g}},
\label{eq:ktg-dist-norm-njet}
\end{align}
for the vacuum \textsc{pgun} jets in Fig.~\ref{fig:pgun_ktg_vacuum}. 
Gluons are produced with greater virtuality than quarks, and thus, gluon jets exhibit a broader distribution with slightly larger values of $k_{T,g}$ than quark jets. 
Furthermore, no significant $p^{\mathrm{jet}}_{T}$ cut dependence can be seen in the vacuum case with fixed $E_{\mathrm{init}} = 140$~GeV for the jet cone size $R=0.4$.

\begin{figure*}[htb]
\centering
\includegraphics[width=0.98\textwidth]{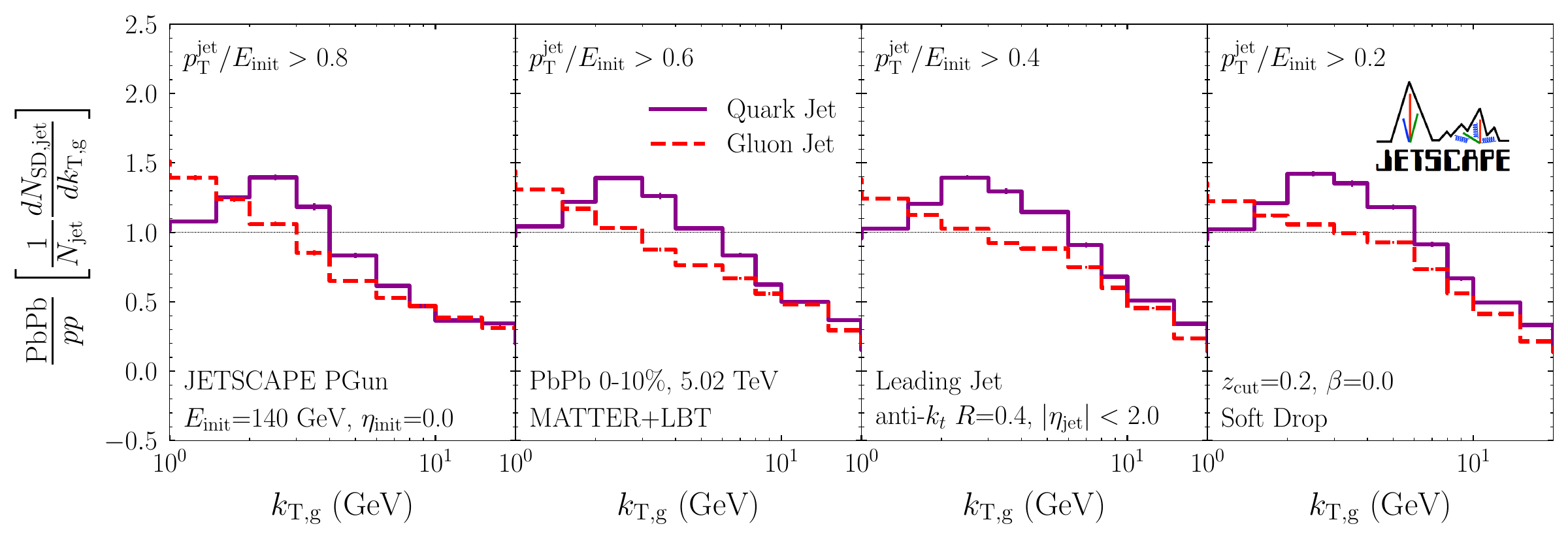} 
\caption{Ratios of $k_{T,g}$ distributions for the leading jets in events with the parent parton having a fixed initial energy $E_{\mathrm{init}} = 140$ GeV generated by \textsc{pgun}. 
The jet shower evolution in the QGP medium produced in 0\%--10\% Pb-Pb collisions at $\sqrt{s_{NN}}=5.02$~TeV is performed by \textsc{matter}$+$\textsc{lbt}. 
All the setups are the same as in Fig.~\ref{fig:pgun_zg_in_medium}. }
\label{fig:pgun_ktg_in_medium}
\end{figure*} 
The medium modifications in the $k_{T,g}$ distribution for the \textsc{pgun} jets are shown in Fig.~\ref{fig:pgun_ktg_in_medium}. 
In the results for both quark and gluon jets, suppression at large $k_{T,g}$ values is seen. 
As seen in Eq.~\eqref{eq:ktg}, $k_{T,g}$ increases with increasing $r_{g}$. 
Thus, one might naively think that this suppression is attributed to the same selection bias effect as in the $r_{g}$ distribution: larger energy loss in jets with larger $k_{T,g}$. 
However, this is not only the cause, as evinced by the persistence of the suppression at large $k_{T,g}$ even when the $p^{\mathrm{jet}}_{T}$ cut is lowered to reduce the effect of hard splitting angle ($r_{g}$) dependence on energy loss. 
This suppression at large $k_{T,g}$ is actually brought also by the loss of $p_{T,2}$: large-angle soft radiations in the later stage can be trimmed by either the jet cone or the soft-drop grooming.

The $p_{T,2}$ loss effect becomes apparent when focusing on the gluon jet with the smallest presented $p^{\mathrm{jet}}_{T}$ cut ($p^{\mathrm{jet}}_{T}/E_{\mathrm{init}} > 0.2$), taking into account the results of the $r_{g}$ modification shown in Fig.~\ref{fig:pgun_rg_in_medium}. 
In this instance, whether due to energy-loss effects or broadening, there is almost no modification observed in the $r_{g}$ distribution of gluon jets. 
Thus, from Eq.~\eqref{eq:ktg}, the modification in the $k_{T,g}$ must be attributed to the change in $p_{T,2}$, in particular, for the case with the small $p^{\mathrm{jet}}_{T}$ cut.

In the small to mid $k_{T,g}$ ($\lessapprox 2$ GeV) region, similar to what was observed in the modification of the $r_{g}$ distribution, a bump structure indicating broader hard branchings is observed for quark jets. 
For the gluon jet case, the modification pattern is dominated by the effect of $p_{T,2}$ loss, with the $r_{g}$ dependence of jet energy loss also exerting a slight influence in cases of large $p^{\mathrm{jet}}_{T}$ cuts.

\begin{figure*}[htb]
\centering
\includegraphics[width=0.98\textwidth]{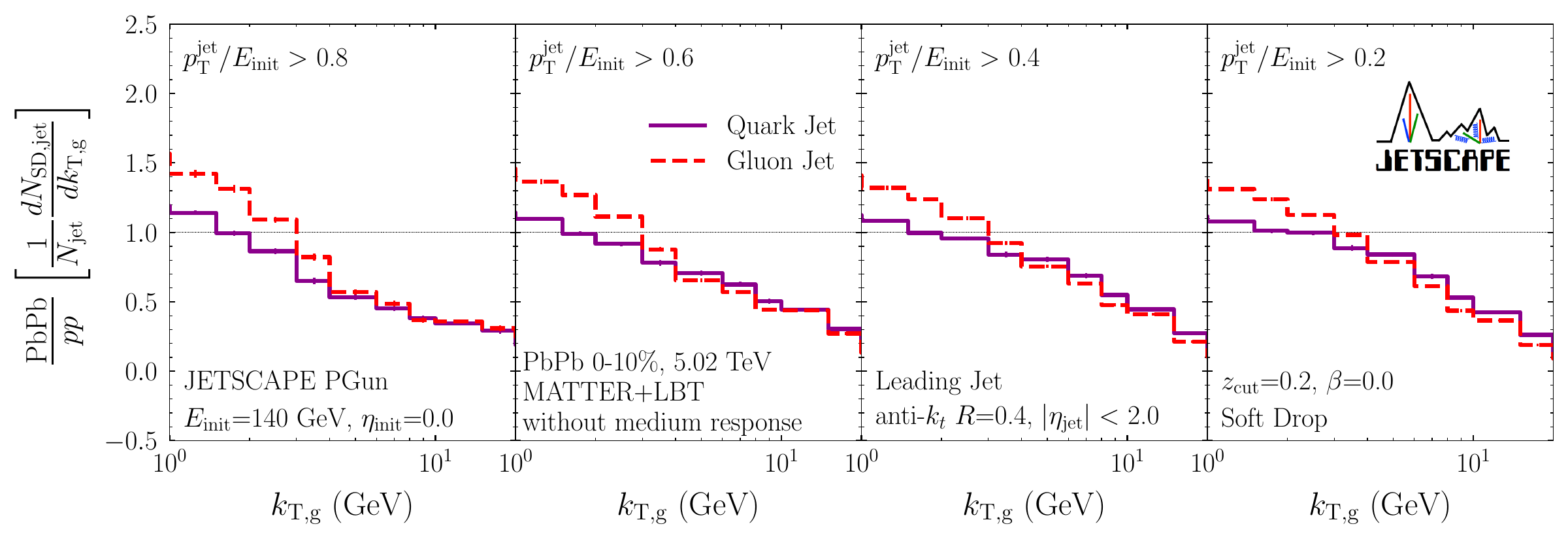}
\caption{Same as Fig.~\ref{fig:pgun_ktg_in_medium} for \textsc{matter}$+$\textsc{lbt} simulations in which the medium effects from recoils and holes are artificially removed. 
}
\label{fig:pgun_ktg_in_medium_no_recoil} 
\end{figure*} 
Figure~\ref{fig:pgun_ktg_in_medium_no_recoil} shows the \textsc{matter}$+$\textsc{lbt} results without the contributions from recoils and holes. 
For gluon jets, the results remain almost unchanged compared with those with the full medium-response contribution.
In contrast, for quark jets, the bump observed in the full medium-response results disappears, and their qualitative behavior becomes identical to that of gluon jets, exhibiting a monotonic decrease as a function of $k_{T}$. 
Furthermore, unlike in the case of $r_{g}$, no residual bump is visible, indicating that the bump in $k_{T,g}$ originates entirely from recoils, while the bump in $r_{g}$ observed even without recoils is caused by soft branchings with small $k_{T,g}$.

\begin{figure*}[htb]
\centering
\includegraphics[width=0.98\textwidth]{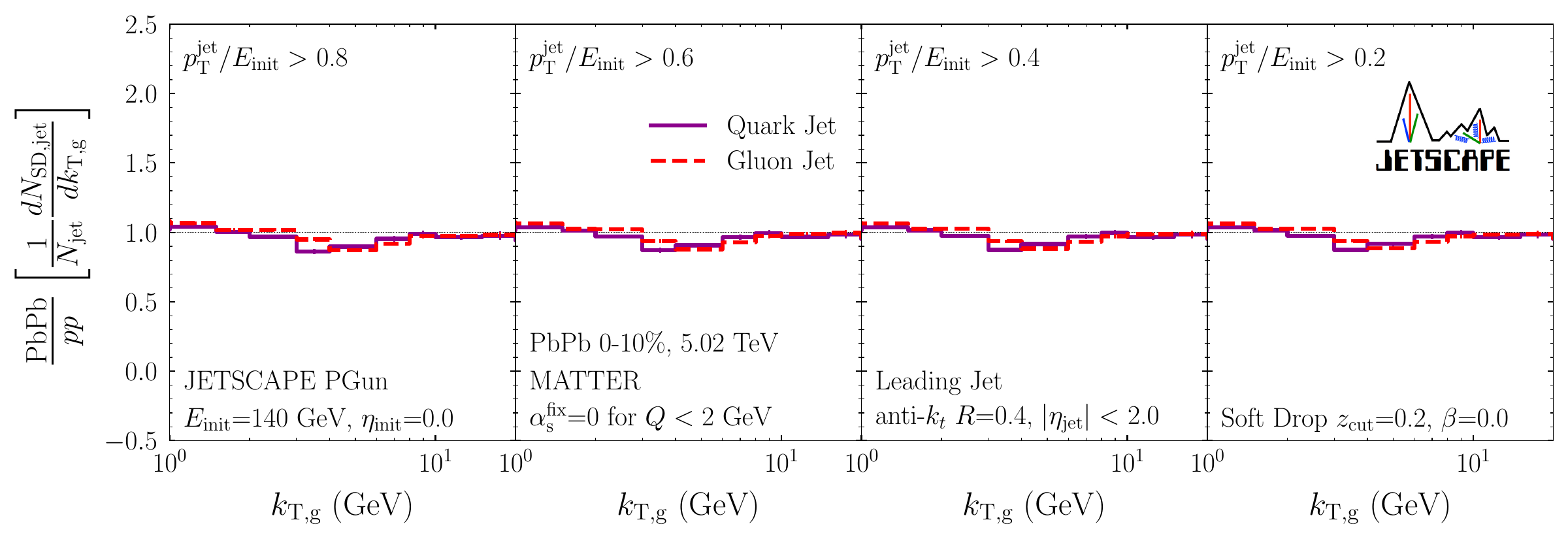} 
\caption{Same as Fig.~\ref{fig:pgun_ktg_in_medium} for \textsc{matter} alone simulations, where the medium effect is turned off for jet partons with virtuality $Q<2$~GeV. 
}
\label{fig:pgun_ktg_in_medium_lbt_off}
\end{figure*} 
The modification on the $k_{T,g}$ from the \textsc{matter} alone simulations is shown in Fig.~\ref{fig:pgun_ktg_in_medium_lbt_off}. 
By turning off the medium effect in the \textsc{lbt} phase, the loss of energy for the jet itself, as well as the $p_{T,2}$ loss, is significantly diminished, resulting in the disappearance of suppression at large $k_{T,g}$. 
Similar to what was observed in $z_{g}$ and $r_{g}$ distributions, the modification in the $k_{T,g}$ distribution is entirely governed by the medium effect in the low-virtuality \textsc{lbt} phase.

\begin{figure*}[htb]
\centering
\includegraphics[width=0.98\textwidth]{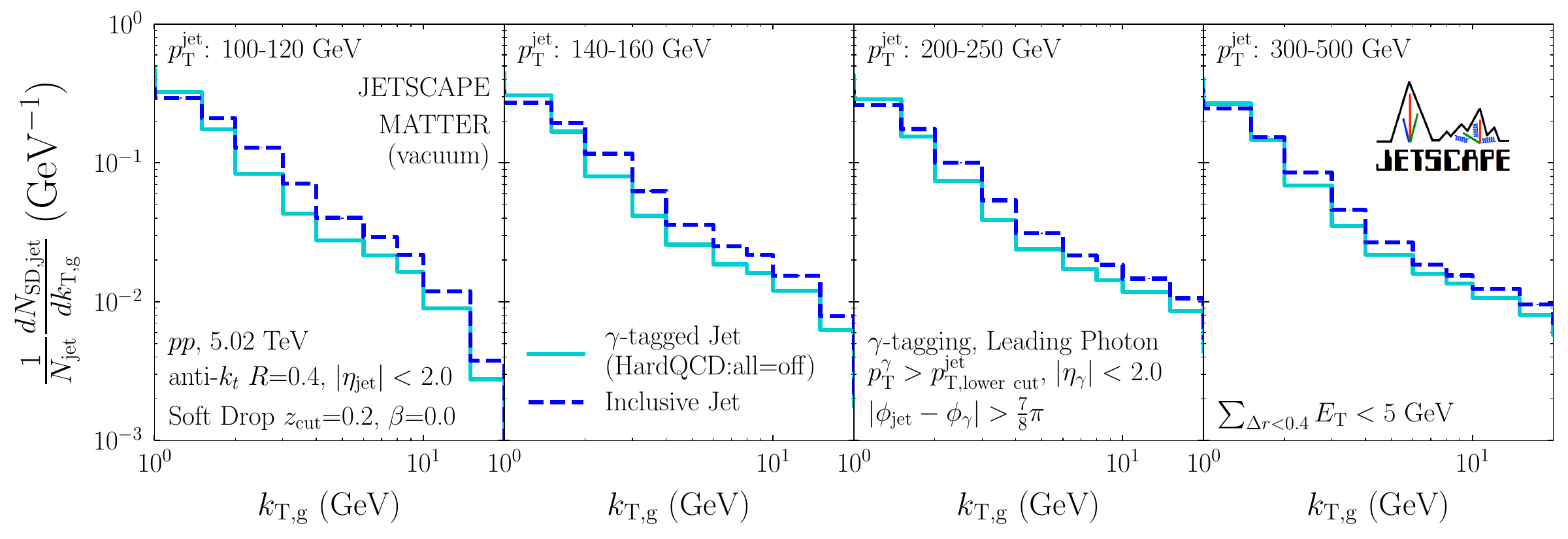}
\caption{Same as Fig.~\ref{fig:pgun_ktg_vacuum} for jets from hard scatterings in $p$-$p$ collisions at $\sqrt{s}=5.02$~TeV, generated by \textsc{pythiagun} with ISR and MPI. 
The results are shown for $\gamma$-tagged jets (solid) from prompt photon-generating hard processes (\texttt{HardQCD:all=off}+\texttt{PromptPhoton:all=on}) and inclusive jets (dashed) from inclusive hard processes (\texttt{HardQCD:all=on}+\texttt{PromptPhoton:all=on}) generated at leading order by \textsc{pythia} 8 with different $p^{\mathrm{jet}}_{\mathrm{T}}$ triggers. 
For $\gamma$-tagged jets, the isolation requirement, the relative azimuth angle cut, and the additional cut of $p^{\mathrm{jet}}_{\mathrm{T}} < p^{\gamma}_{\mathrm{T}}$ are imposed. 
}
\label{fig:inclusive_gamma_tagged_ktg_vacuum}
\end{figure*} 
In Fig.~\ref{fig:inclusive_gamma_tagged_ktg_vacuum}, the comparison between the $k_{T,g}$ distributions for $\gamma$-tagged jets and inclusive jets in $p$-$p$ collisions at $\sqrt{s}=5.02$~TeV is shown for the same $p^{\mathrm{jet}}_{T}$ cuts. 
Inclusive jets exhibit a broader transverse momentum in hard splittings compared with $\gamma$-tagged jets, attributed to a larger gluon jet fraction, but this difference diminishes as $p^{\mathrm{jet}}_{T}$ increases.

\begin{figure*}[htb]
\centering
\includegraphics[width=0.98\textwidth]{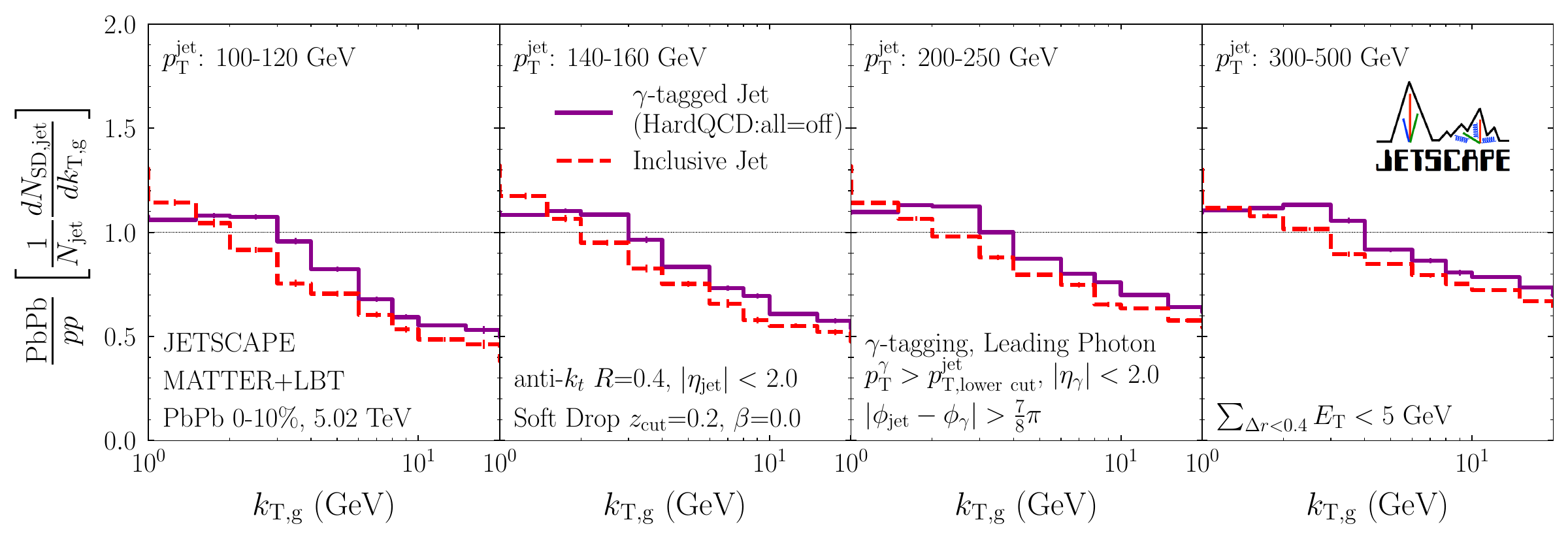}
\caption{Same as Fig.~\ref{fig:pgun_ktg_in_medium} for jets from hard scatterings at $\sqrt{s_{NN}}=5.02$~TeV, generated by \textsc{pythiagun} with ISR and MPI. 
The results are shown for $\gamma$-tagged jets (solid) from prompt photon-generating hard processes (\texttt{HardQCD:all=off}+\texttt{PromptPhoton:all=on}) and inclusive jets (dashed) from inclusive hard processes (\texttt{HardQCD:all=on}+\texttt{PromptPhoton:all=on}) generated at leading order by \textsc{pythia} 8 with different $p^{\mathrm{jet}}_{\mathrm{T}}$ triggers. 
For $\gamma$-tagged jets, the isolation requirement, the relative azimuth angle cut, and the additional cut of $p^{\mathrm{jet}}_{\mathrm{T}} < p^{\gamma}_{\mathrm{T}}$ are imposed.}
\label{fig:inclusive_gamma_tagged_ktg_in_medium}
\end{figure*}
Figure~\ref{fig:inclusive_gamma_tagged_ktg_in_medium} presents 
the medium modification of the $k_{T,g}$ distribution for $\gamma$-tagged jets and inclusive jets. 
For both $\gamma$-tagged jets and inclusive jets, the overall modification exhibits a narrowing pattern mostly dominated by the $p_{T,2}$ loss of the prongs. 
Although no pronounced bump structure is observed, $\gamma$-tagged jets exhibit a subtle indication of broadening in the hard branchings of quark jets due to recoils generated in the \textsc{lbt} phase, manifesting as a slight flattening on the small-$k_{T,g}$ side.

\begin{figure*}[htb]
\centering
\includegraphics[width=0.98\textwidth]{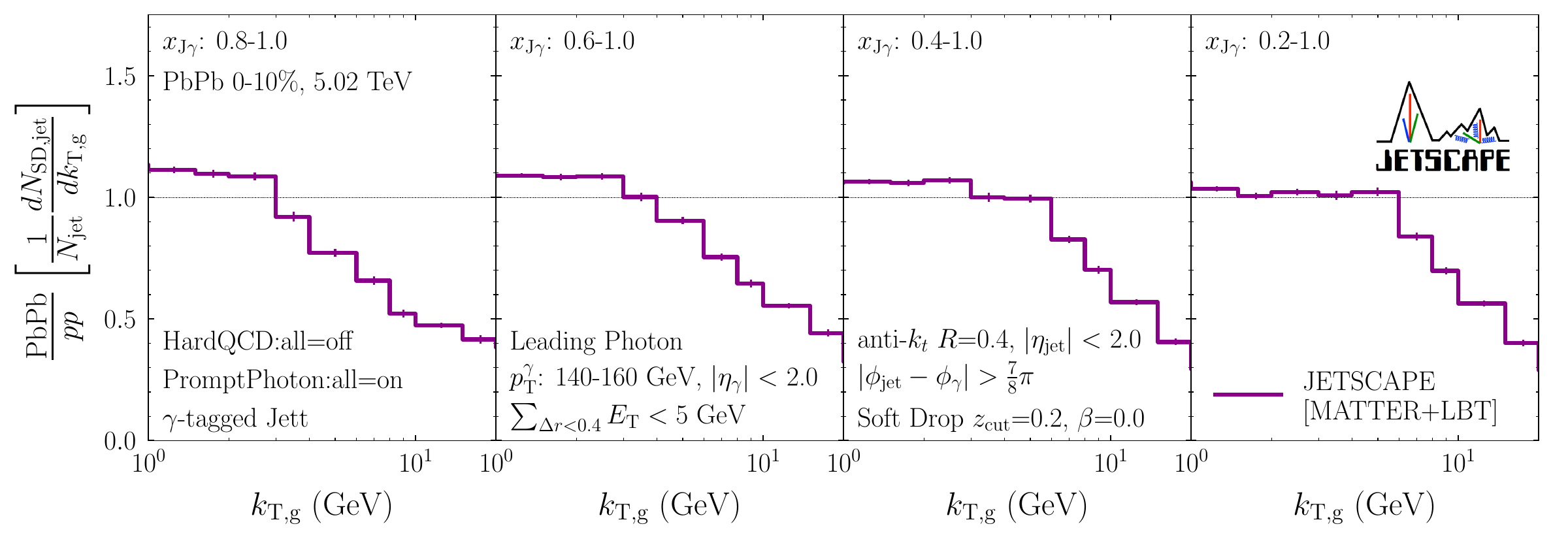}
\caption{Same as Fig.~\ref{fig:pgun_ktg_in_medium} for $\gamma$-tagged jets from prompt photon-generating hard processes (\texttt{HardQCD:all=off}+\texttt{PromptPhoton:all=on}) generated at leading order by \textsc{pythia} 8 at $\sqrt{s_{NN}}=5.02$~TeV for different $x_{J\gamma}$ ranges. 
The photon of $140<p^{\gamma}_{\mathrm{T}} < 160$~GeV is triggered with the isolation requirement, and the relative azimuth angle cut.} 
\label{fig:gamma_tagged_ktg_xdep}
\end{figure*}
Finally, the prediction for the $x_{J\gamma}$-dependent medium modification of the $k_{T,g}$ distribution for $\gamma$-tagged jets in central Pb-Pb collisions at $\sqrt{s_{NN}}=5.02$~TeV is shown in Fig.~\ref{fig:gamma_tagged_ktg_xdep}. 
Given that the suppression at large $k_{T,g}$ persists even with a reduced $x_{J\gamma}$ cut ($x_{J\gamma} > 0.2$), it is evident that jets with large $k_{T,g}$ cannot recover, even when jets with significant energy loss are included, due to the presence of $p_{T,2}$ loss. 
On the small-$k_{T,g}$ side, a very flat pattern appears as a result of the competition between broad branchings generated by recoils in quark jets during the \textsc{lbt} phase and other effects that narrow the distribution.

\subsection{Groomed jet mass}
\label{Subsection:mG}
Finally, we investigate the medium modification of the groomed jet mass distribution, 
\begin{align}
\frac{1}{N_{\mathrm{jet}}}\frac{d N_{\mathrm{SD,jet}}}{d m_{g}},
\label{eq:mg-dist-norm-njet}
\end{align}
where the groomed jet mass is defined as 
\begin{align}
m_{g} &= \sqrt{\left(p^0_{1}+p^0_{2}\right)^2-\left(\vec{p}_{1}+\vec{p}_{2}\right)^2}. 
\label{eq:mg}
\end{align}
\begin{figure*}[htb]
\centering
\includegraphics[width=0.98\textwidth]{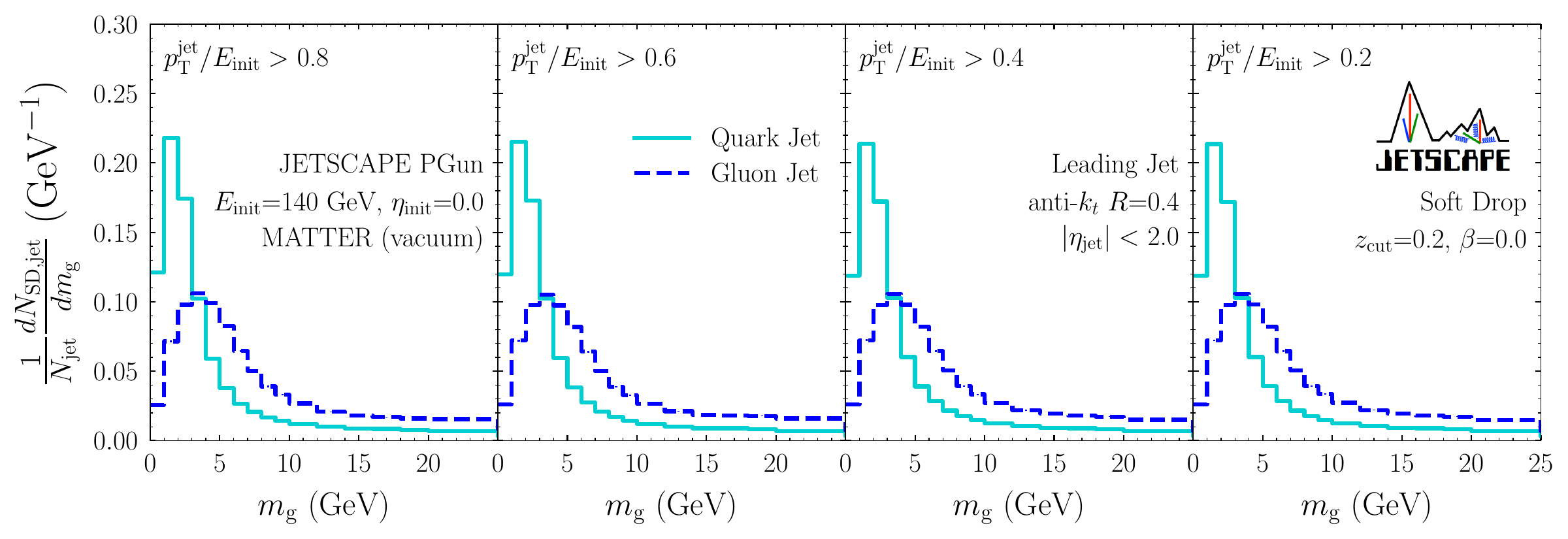}
\caption{Distributions of the relative transverse momentum of jet splittings $m_{g}$ 
normalized by the number of all triggered jets for the leading jets in events generated with \textsc{pgun}. 
The jet shower evolution is performed by vacuum \textsc{matter} for the parent parton having $E_{\mathrm{init}} = 140$~GeV. 
Jets are reconstructed with $R=0.4$ at midrapidity $|\eta_{\mathrm{jet}}|<2.0$. 
The results are shown for quark jets (solid) and gluon jets (dashed) with different $p^{\mathrm{jet}}_{T}$ triggers, 
112, 84, 56, and 28~GeV. 
The soft-drop parameters are $z_{\mathrm{cut}}=0.2$ and $\beta = 0$. 
}
\label{fig:pgun_mg_vacuum}
\end{figure*}
In Fig.~\ref{fig:pgun_mg_vacuum}, the $m_{g}$ distributions for the vacuum \textsc{pgun} jets are shown. 
Here, again, due to the larger virtuality at the initial hard scattering production, gluon jets have a broader distribution with larger $m_{g}$ than quark jets. 
For the vacuum case with fixed $E_{\mathrm{init}} = 140$~GeV, no significant $p^{\mathrm{jet}}_{T}$-cut dependence is seen for the jet cone size $R=0.4$.

\begin{figure*}[htb]
\centering
\includegraphics[width=0.98\textwidth]{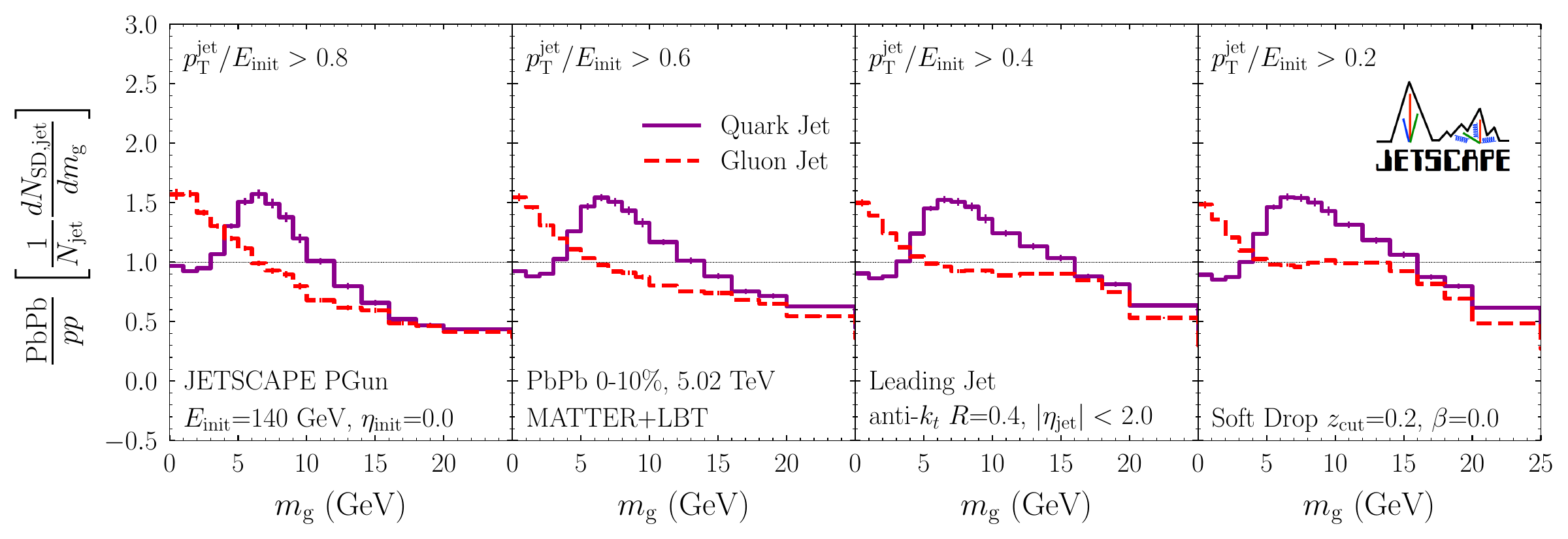} 
\caption{Ratios of $m_{g}$ distributions for the leading jets in events with the parent parton having a fixed initial energy $E_{\mathrm{init}} = 140$ GeV generated by \textsc{pgun}. 
The jet shower evolution in the QGP medium produced in 0\%--10\% Pb-Pb collisions at $\sqrt{s_{NN}}=5.02$~TeV is performed by \textsc{matter}$+$\textsc{lbt}. 
All the setups are the same as in Fig.~\ref{fig:pgun_zg_in_medium}.
}
\label{fig:pgun_mg_in_medium}
\end{figure*} 
The medium modification of $m_{g}$ distribution for jets from the \textsc{pgun} simulations with fixed $E_{\mathrm{init}}$ is shown in Fig.~\ref{fig:pgun_mg_in_medium}. 
In the case of quark jets, a clear bump structure is observed due to the shift of small $m_{g}$ to mid $m_{g}$ caused by medium effects. 
On the other hand, such behavior is not observed for gluon jets, which exhibit a monotonic trend. 
Besides, for both quark and gluon jets with large $p^{\mathrm{jet}}_{T}$ cuts, suppression at large $m_{g}$ is observed, attributed to the greater energy loss in jets with larger masses, which results in wider splittings. 
Then, with the small $p^{\mathrm{jet}}_{T}$ cuts, the suppression becomes slightly moderate at mid $m_{g}$ ($\approx 10$--$15$~GeV), yet still dominates the modification. 
This is because, analogous to the case for $k_{T,g}$, the jet energy loss---primarily due to large-angle energy emission outside of the jet cone---results in the loss of $m_{g}$~\cite{Majumder:2014gda}.

\begin{figure*}[htb]
\centering
\includegraphics[width=0.98\textwidth]{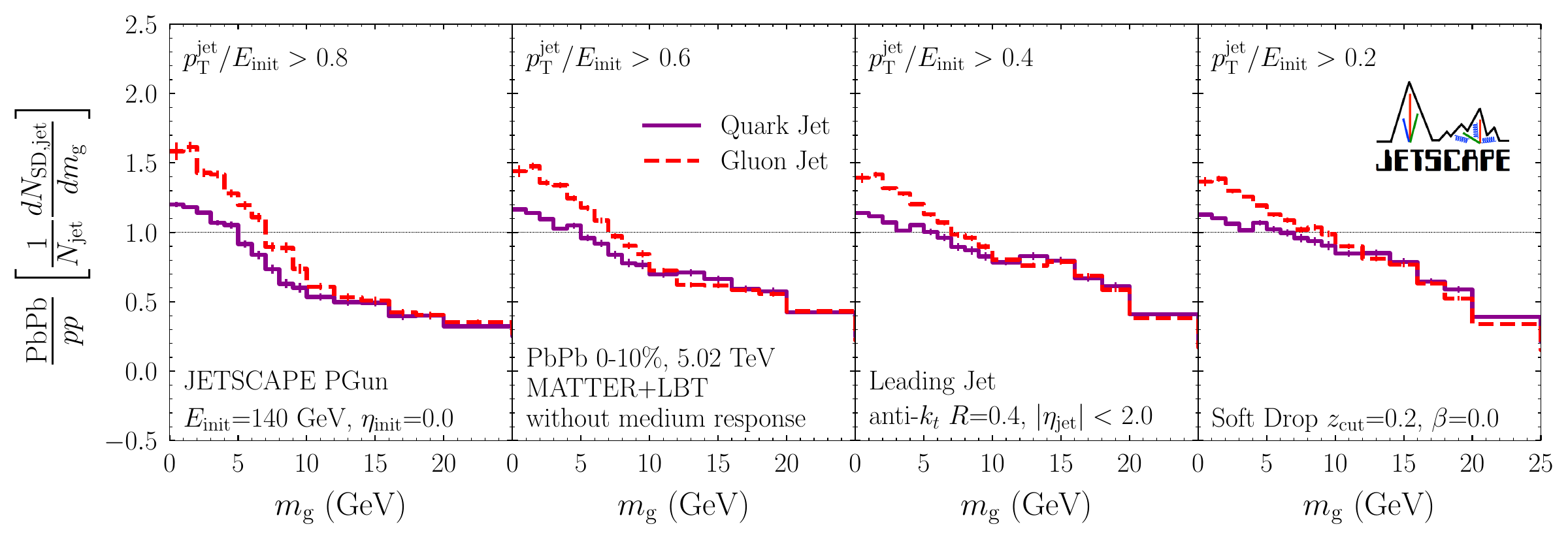}
\caption{Same as Fig.~\ref{fig:pgun_mg_in_medium} for \textsc{matter}$+$\textsc{lbt} simulations in which the medium effects from recoils and holes are artificially removed. }
\label{fig:pgun_mg_in_medium_no_recoil}
\end{figure*} 
Figure~\ref{fig:pgun_mg_in_medium_no_recoil} shows the results with the contributions from recoils and holes artificially removed. 
As in the case of $k_{T,g}$, the results for gluon jets remain almost unchanged compared with those with the full medium response. For quark jets, however, the bump observed with the full medium response disappears, and their qualitative behavior becomes identical to that of gluon jets, exhibiting a monotonic decrease as a function of $m_{g}$. 
The slight recoil-independent broadening seen in $r_{g}$ originates from soft branchings that do not produce a bump in $m_{g}$.

\begin{figure*}[htb]
\centering
\includegraphics[width=0.98\textwidth]{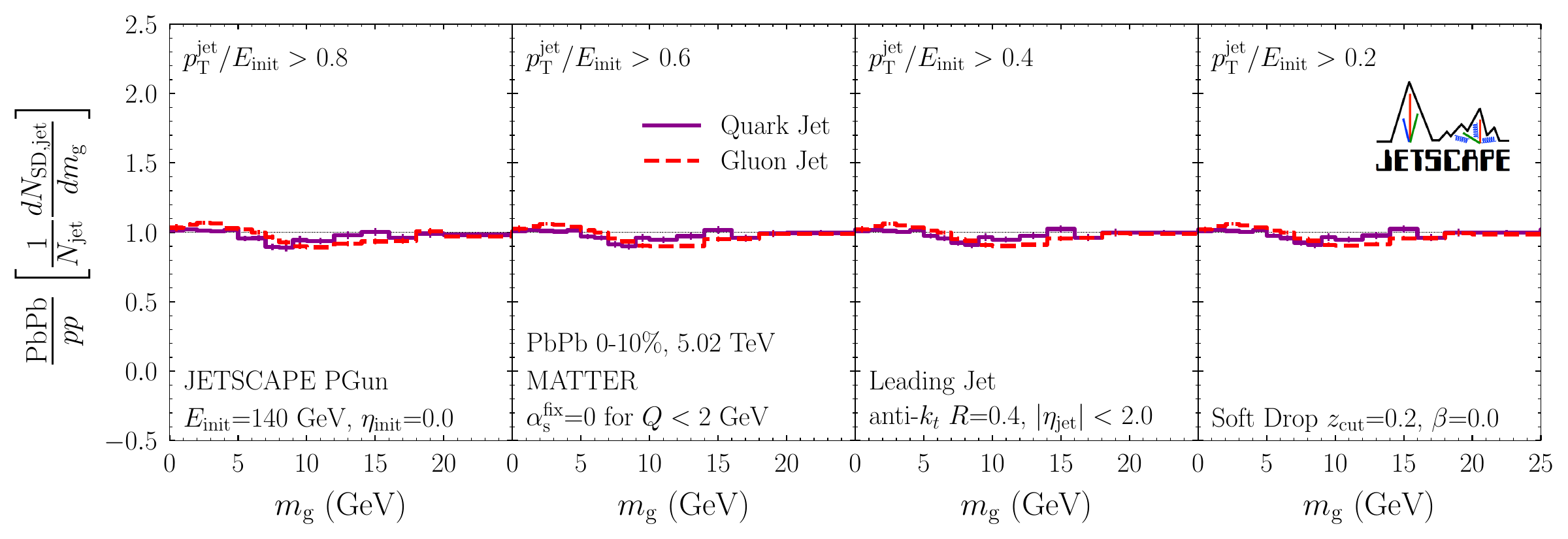} \caption{Same as Fig.~\ref{fig:pgun_mg_in_medium} for \textsc{matter} alone simulations, where the medium effect is turned off for jet partons with virtuality $Q<2$~GeV. 
}
\label{fig:pgun_mg_in_medium_lbt_off}
\end{figure*} 
In Fig.~\ref{fig:pgun_mg_in_medium_lbt_off}, the modification by the \textsc{matter} alone simulations is shown. 
Also, through $m_{g}$ distribution, it can be confirmed that no significant modification in the structure of jet hard splittings at high virtuality occurs for both gluon jets and quark jets. 
Thus, the modification pattern is entirely brought about by the evolution at low virtuality.

\begin{figure*}[htb]
\centering
\includegraphics[width=0.98\textwidth]{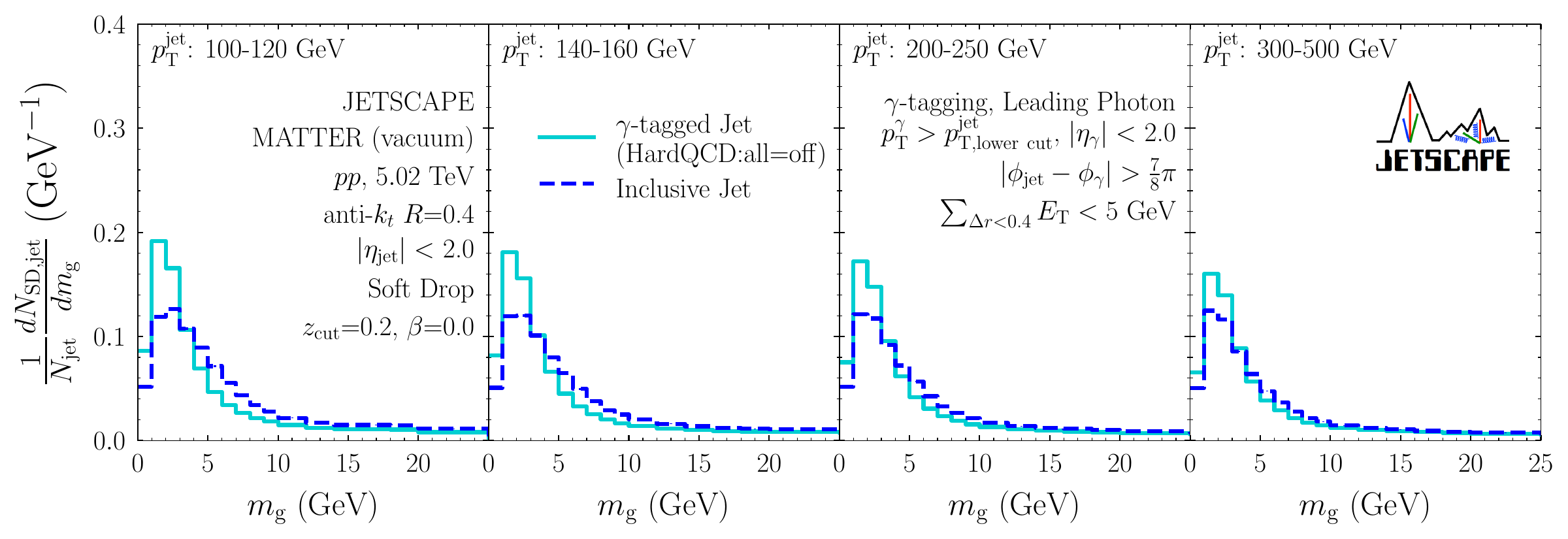}
\caption{Same as Fig.~\ref{fig:pgun_mg_vacuum} for jets from hard scatterings in $p$-$p$ collisions at $\sqrt{s}=5.02$~TeV, generated by \textsc{pythiagun} with ISR and MPI. 
The results are shown for $\gamma$-tagged jets (solid) from prompt photon-generating hard processes (\texttt{HardQCD:all=off}+\texttt{PromptPhoton:all=on}) and inclusive jets (dashed) from inclusive hard processes (\texttt{HardQCD:all=on}+\texttt{PromptPhoton:all=on}) generated at leading order by \textsc{pythia} 8 with different $p^{\mathrm{jet}}_{\mathrm{T}}$ triggers. 
For $\gamma$-tagged jets, the isolation requirement, the relative azimuth angle cut, and the additional cut of $p^{\mathrm{jet}}_{\mathrm{T}} < p^{\gamma}_{\mathrm{T}}$ are imposed. }
\label{fig:inclusive_gamma_tagged_mg_vacuum}
\end{figure*}
The $m_{g}$ distributions for $\gamma$-tagged jets and inclusive jets in $p$-$p$ collisions at $\sqrt{s}=5.02$~TeV are compared in Fig.~\ref{fig:inclusive_gamma_tagged_mg_vacuum}. 
The inclusive jet results exhibit a broader distribution in $m_{g}$ due to the larger gluon jet fraction, but the difference diminishes as $p^{\mathrm{jet}}_{T}$ increases, similarly to the $r_{g}$ and $k_{T,g}$ distributions. 
\begin{figure*}[htb]
\centering
\includegraphics[width=0.98\textwidth]{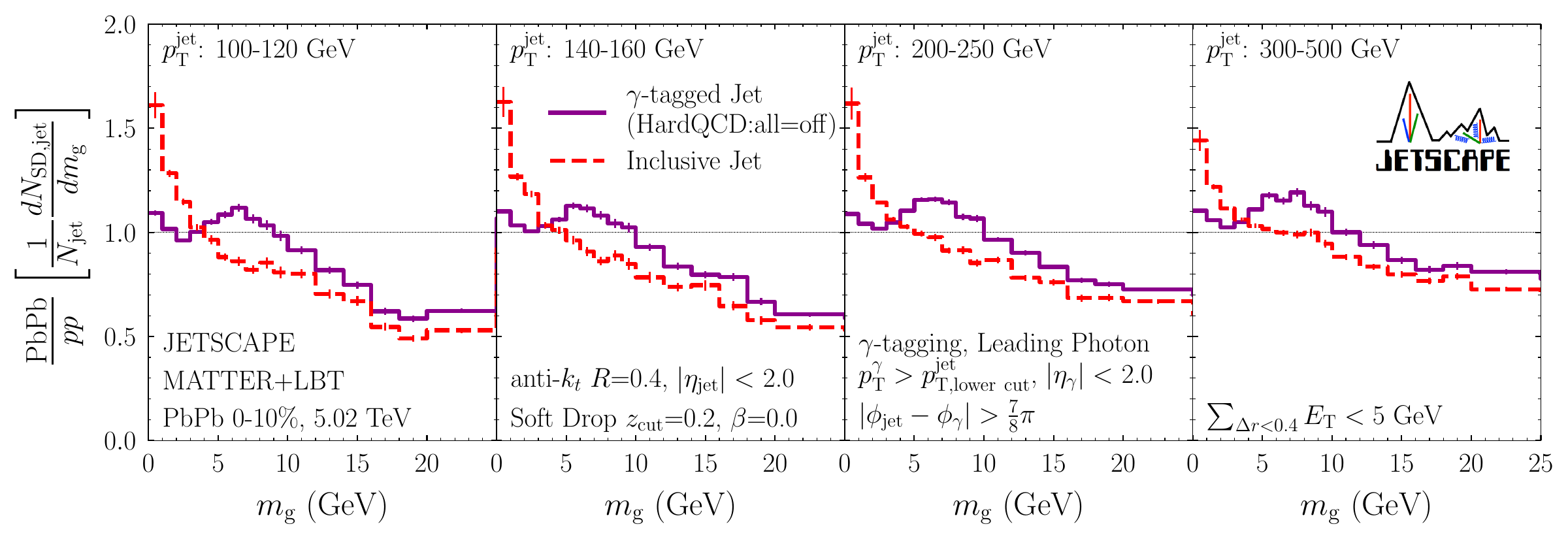}
\caption{Same as Fig.~\ref{fig:pgun_mg_in_medium} for jets from hard scatterings at $\sqrt{s_{NN}}=5.02$~TeV, generated by \textsc{pythiagun} with ISR and MPI. 
The results are shown for the inclusive jets (dashed) from inclusive hard processes (\texttt{HardQCD:all=on}+\texttt{PromptPhoton:all=on}) and $\gamma$-tagged jets (solid) from prompt photon-generating hard processes (\texttt{HardQCD:all=off}+\texttt{PromptPhoton:all=on}) generated at leading order by \textsc{pythia} 8 with different $p^{\mathrm{jet}}_{\mathrm{T}}$ triggers. 
For $\gamma$-tagged jets, the isolation requirement, the relative azimuth angle cut, and the additional cut of $p^{\mathrm{jet}}_{\mathrm{T}} < p^{\gamma}_{\mathrm{T}}$ are imposed.
}
\label{fig:inclusive_gamma_tagged_mg_in_medium}
\end{figure*}
The medium modification is shown in Fig.~\ref{fig:inclusive_gamma_tagged_mg_in_medium}. 
For inclusive jets, the modification pattern is dominated by the shift toward smaller $m_{g}$, attributed to larger energy loss with wider angle profiles. 
In the case of $\gamma$-tagged jets, a bump due to the mass gain for quark jets via recoils in the \textsc{lbt} phase appears on the small $m_{g}$ side, while a similar suppression to that of inclusive jets is observed at large $m_{g}$.

\begin{figure*}[htb]
\centering
\includegraphics[width=0.98\textwidth]{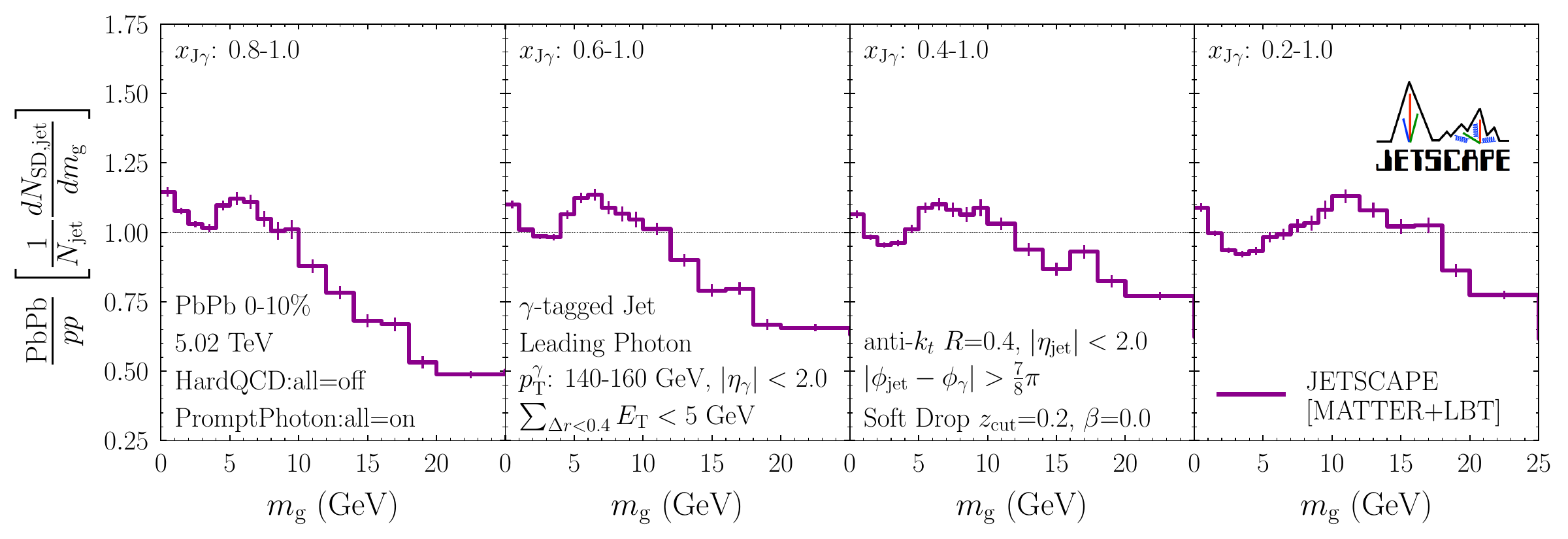}
\caption{Same as Fig.~\ref{fig:pgun_mg_in_medium} for $\gamma$-tagged jets from prompt photon-generating hard processes (\texttt{HardQCD:all=off}+\texttt{PromptPhoton:all=on}) generated at leading order by \textsc{pythia} 8 at $\sqrt{s_{NN}}=5.02$~TeV for different $x_{J\gamma}$ ranges. 
The photon of $140<p^{\gamma}_{\mathrm{T}} < 160$~GeV is triggered with the isolation requirement, and the relative azimuth angle cut.
} 
\label{fig:gamma_tagged_mg_xdep}
\end{figure*}
We present our prediction for the $x_{J\gamma}$-dependent medium modification of the $m_{g}$ distribution for $\gamma$-tagged jets in Fig.~\ref{fig:gamma_tagged_mg_xdep}. 
A bump structure associated with mass gain via recoils from the medium at small virtuality is observed in soft-drop-reconstructed hard branchings. 
At large $m_{g}$, jets with larger masses experience greater energy loss, leading to wider splittings and suppression at large $x_{J\gamma}$ cuts. 
Reducing the $x_{J\gamma}$ cut includes contributions from jets with significant energy loss, weakening the suppression, but it persists even at the smallest $x_{J\gamma}$ cut ($x_{J\gamma} > 0.2$). 
This is attributed to the loss of $m_{g}$ caused by energy emitted at large angles outside the jet cone or soft components groomed away by the soft-drop procedure.

\section{Comparison with experimental data}
\label{Section:VsExp}
In this section, we compare our analysis results, using the same event set of \textsc{jetscape} simulations as in the previous section, with existing experimental data from the LHC for benchmarking. 
Additionally, we note that similar comparisons with experimental results, using the same parameter settings and configurations, have also been detailed in our previous work~\cite{JETSCAPE:2023hqn}.

\subsection{Inclusive jet substructure}
\label{subsection:ExpIncl}
In this section, we present the results of groomed observables for inclusive jets. The initial hard processes for the simulations were generated using the \textsc{pythiagun} module with \texttt{HardQCD:all=on}.

\subsubsection{Relative transverse momentum of jet splittings}
\label{Subsubsection:ExpGroomedKt}
\begin{figure*}[htb]
\centering
\includegraphics[width=0.8\textwidth]{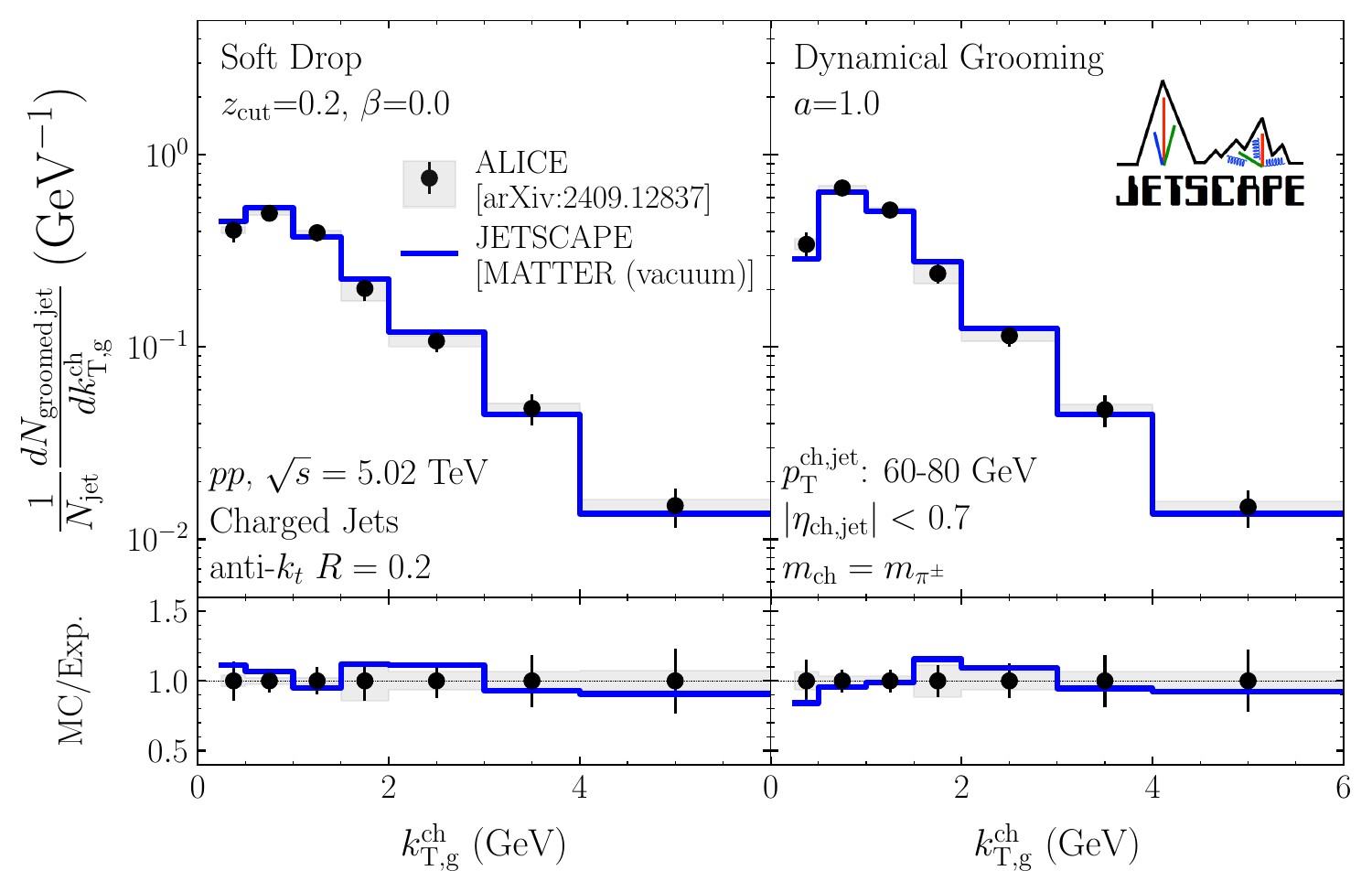}
\caption{Distributions of the relative transverse momentum of jet splittings, $k_{T,g}$ for charged jets with $60 < p^{\mathrm{ch,jet}}_T < 80$ GeV and $| \eta_{\mathrm{ch,jet}} | < 0.7$, using $R = 0.2$ in $p$-$p$ collisions at $\sqrt{s} = 5.02$ TeV, and the ratios for different grooming algorithms: soft drop with $z_{\mathrm{cut}}=0.2$ and $\beta=0$ (left), and dynamical grooming with $a=1.0$ (right). 
The solid lines and circles with statistical error bars show the results from vacuum \textsc{matter} of \textsc{jetscape} and the experimental data from the ALICE Collaboration~\cite{ALICE:2024fip}, respectively. 
The bands indicate the systematic uncertainties of the experimental data. 
}
\label{fig:alice_sd_ktg_pp}
\end{figure*}
\begin{figure*}[htb]
\centering
\includegraphics[width=0.8\textwidth]{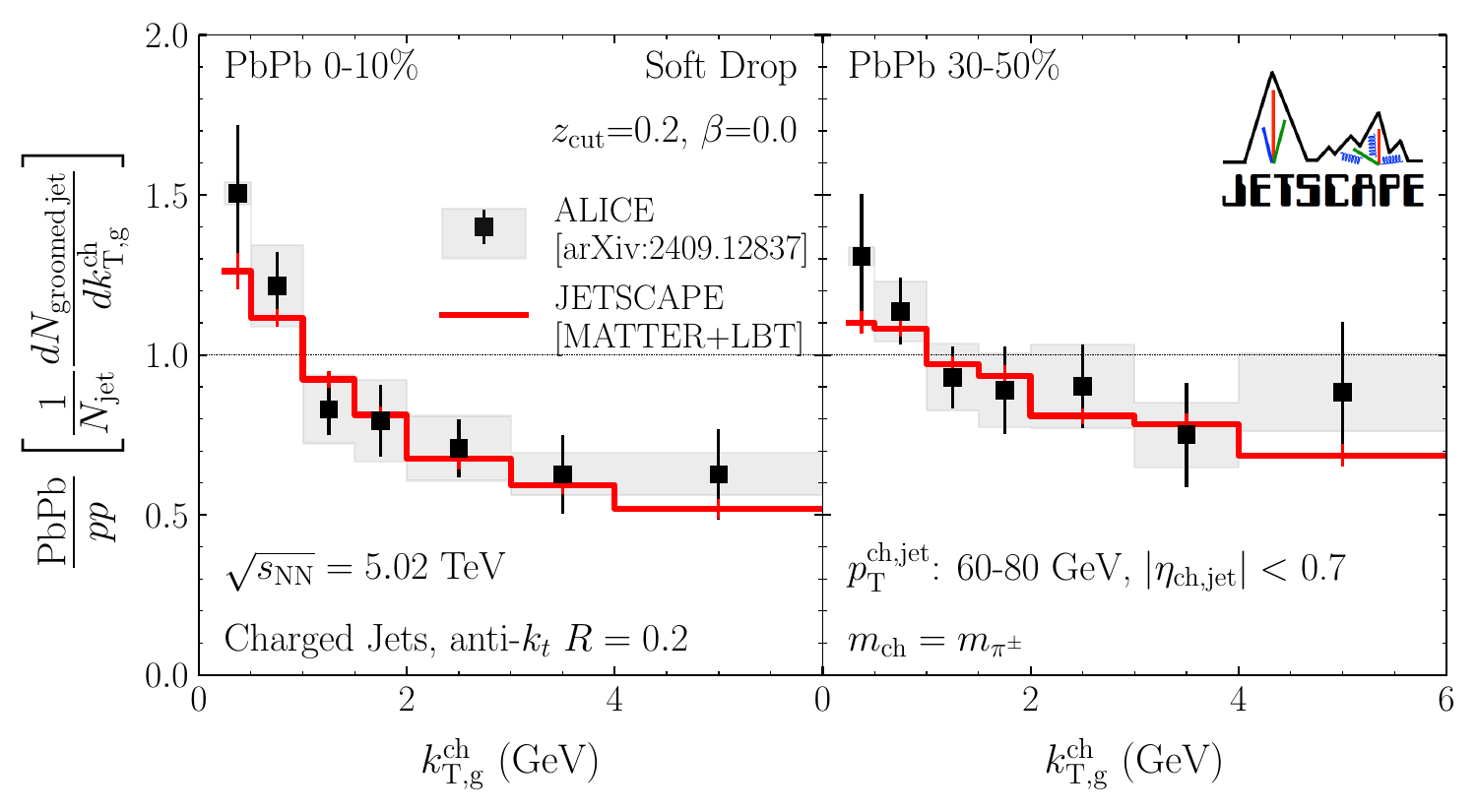}
\includegraphics[width=0.8\textwidth]{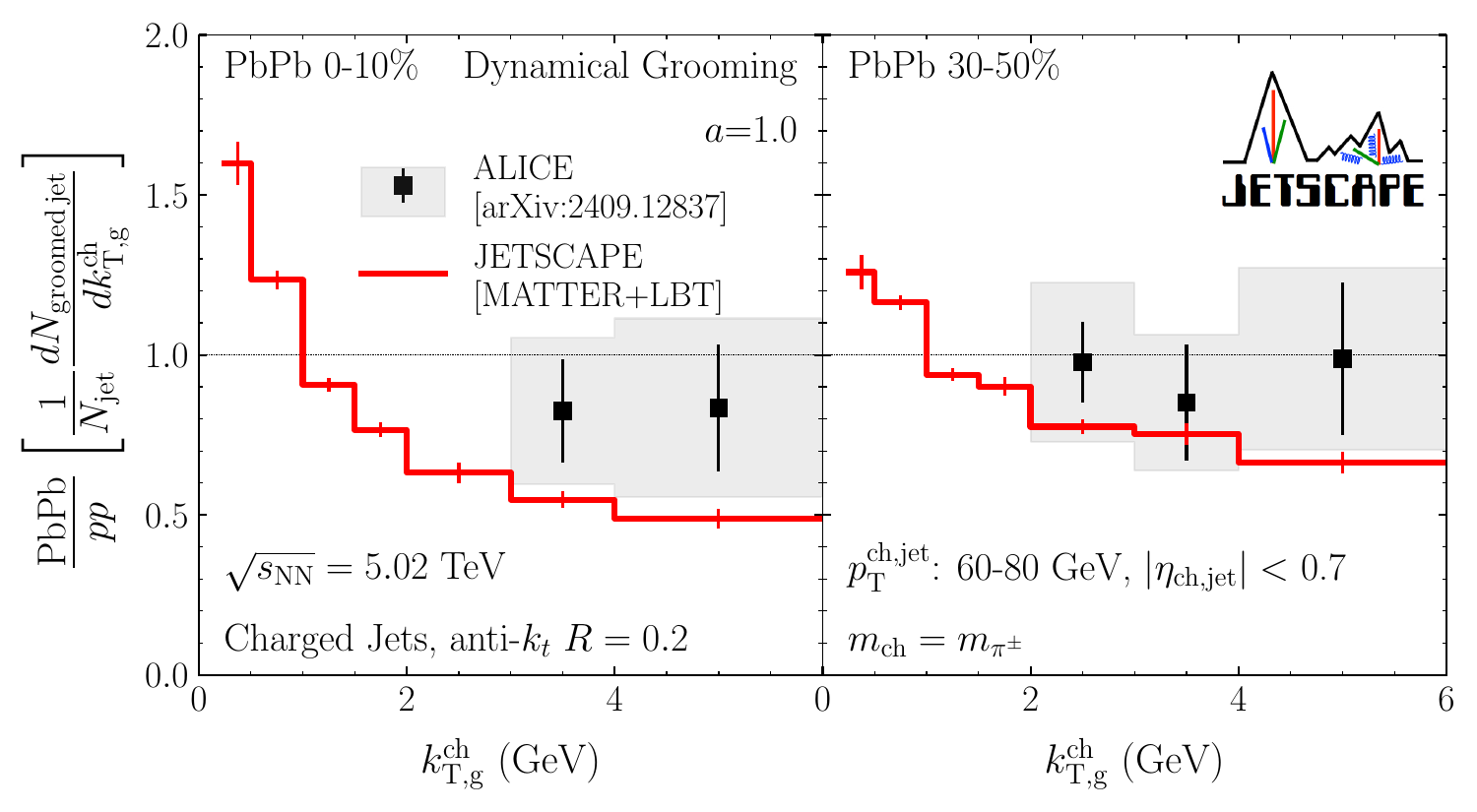}
\caption{Ratios of $k_{T,g}$ distributions between Pb-Pb and $p$-$p$ collisions at $\sqrt{s_{NN}}=5.02$~TeV for charged jets with $60 < p^{\mathrm{ch,jet}}_T < 80$ GeV and $| \eta_{\mathrm{ch,jet}} | < 0.7$, using $R = 0.2$ for different centralities, 
$0\%$--$10\%$ (left) and $30\%$--$50\%$ (right), and grooming algorithms, soft drop with $z_{\mathrm{cut}}=0.2$ and $\beta=0$ (upper) and dynamical grooming with $a=1.0$ (lower). 
The solid lines and squares with statistical error bars show the results from \textsc{matter}+\textsc{lbt} of \textsc{jetscape} and the experimental data from the ALICE Collaboration~\cite{ALICE:2024fip}, respectively. 
The bands indicate the systematic uncertainties of the experimental data. 
}
\label{fig:alice_sd_ktg_pbpb}
\end{figure*}
The distributions of $k_{T,g}$ obtained from soft-drop and dynamical grooming~\cite{Mehtar-Tani:2019rrk}, in $p$-$p$ collisions are shown in Fig.~\ref{fig:alice_sd_ktg_pp}, and their modifications in Pb-Pb collisions at $\sqrt{s_{NN}}=5.02$~TeV are presented in Fig.~\ref{fig:alice_sd_ktg_pbpb}. 
The results from our \textsc{jetscape} simulations are compared with the experimental results from ALICE~\cite{ALICE:2024fip}.

The soft-drop and dynamical grooming results from the JETSCAPE PP19 tune show consistency with experimental data for the $k_{T,g}$ distribution in $p$-$p$ collisions, remaining within the uncertainty range. 
For Pb-Pb collisions, the modification in the distributions aligns with the inclusive full jet results presented in the main text, demonstrating a monotonically decreasing behavior as a function of $k_{T,g}$, which is primarily governed by selection bias by the $p_{T}$ cut and the loss of $p_{T,2}$. 
They successfully capture the experimental trends, while dynamical grooming results exhibit slight over-suppression. 
Notably, the soft-drop results are in excellent agreement with experimental data within the uncertainties.

\subsubsection{Groomed jet mass}
\label{Subsubsection:BenchmarkGroomedMass}
\begin{figure*}[htb]
\centering
\includegraphics[width=0.8\textwidth]{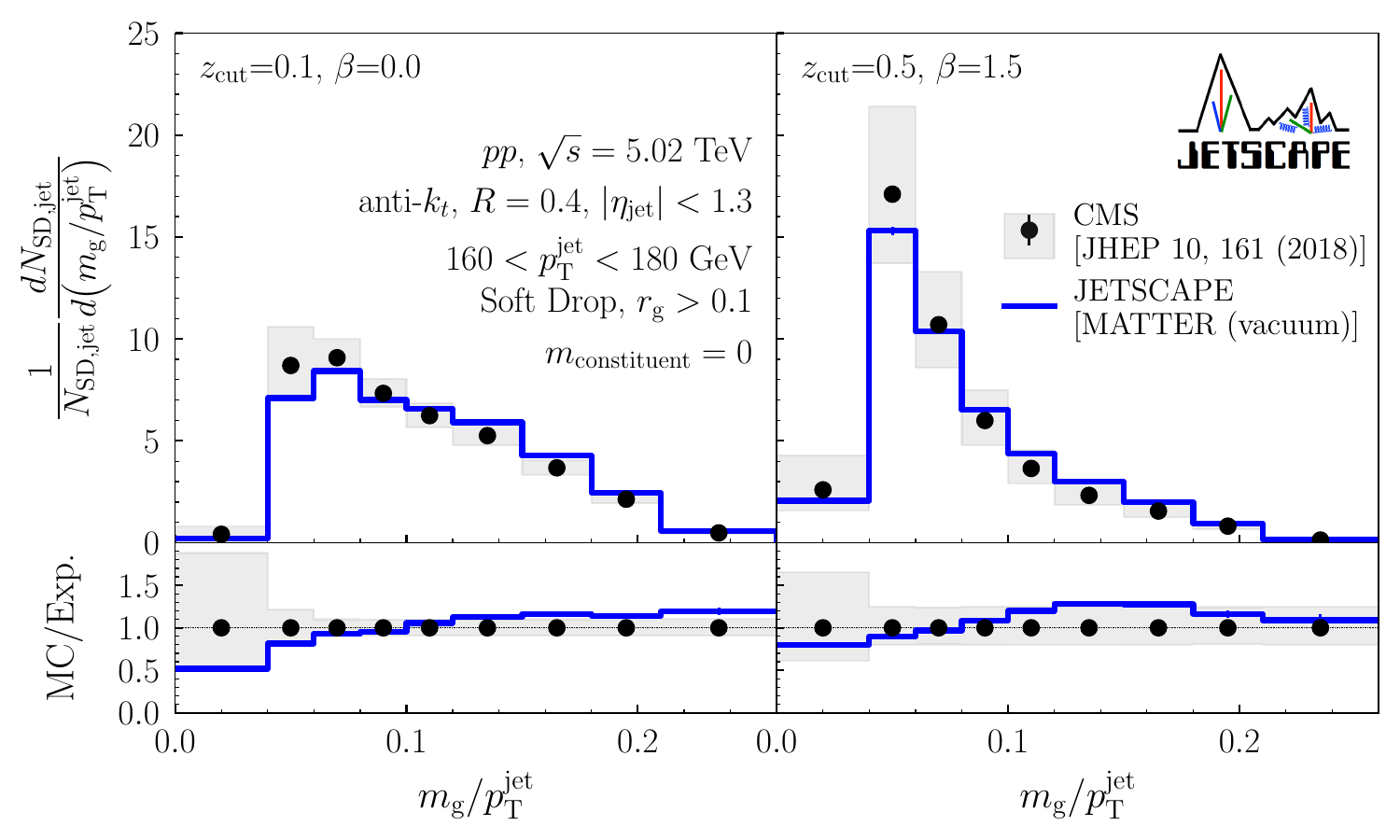}
\caption{Distributions of the ratio of the soft-drop groomed jet mass to the jet transverse momentum, $m_{g}/p^{\mathrm{jet}}_{T}$, for full jets with $160 < p^{\mathrm{jet}}_T < 180$ GeV and $| \eta_{\mathrm{jet}} | < 1.3$, using $R = 0.4$ in $p$-$p$ collisions at $\sqrt{s} = 5.02$ TeV for different grooming parameters: $z_{\mathrm{cut}} = 0.1$ and $\beta = 0$ (left), and $z_{\mathrm{cut}} = 0.5$ and $\beta = 1.5$ (right). 
The solid lines and circles with statistical error bars show the results from vacuum \textsc{matter} of \textsc{jetscape} and the unfolded experimental data from the CMS Collaboration~\cite{CMS:2018fof}, respectively. 
The bands indicate the systematic uncertainties of the experimental data. 
}
\label{fig:cms_mg_pp}
\end{figure*}
\begin{figure*}[htb]
\centering
\includegraphics[width=0.98\textwidth]{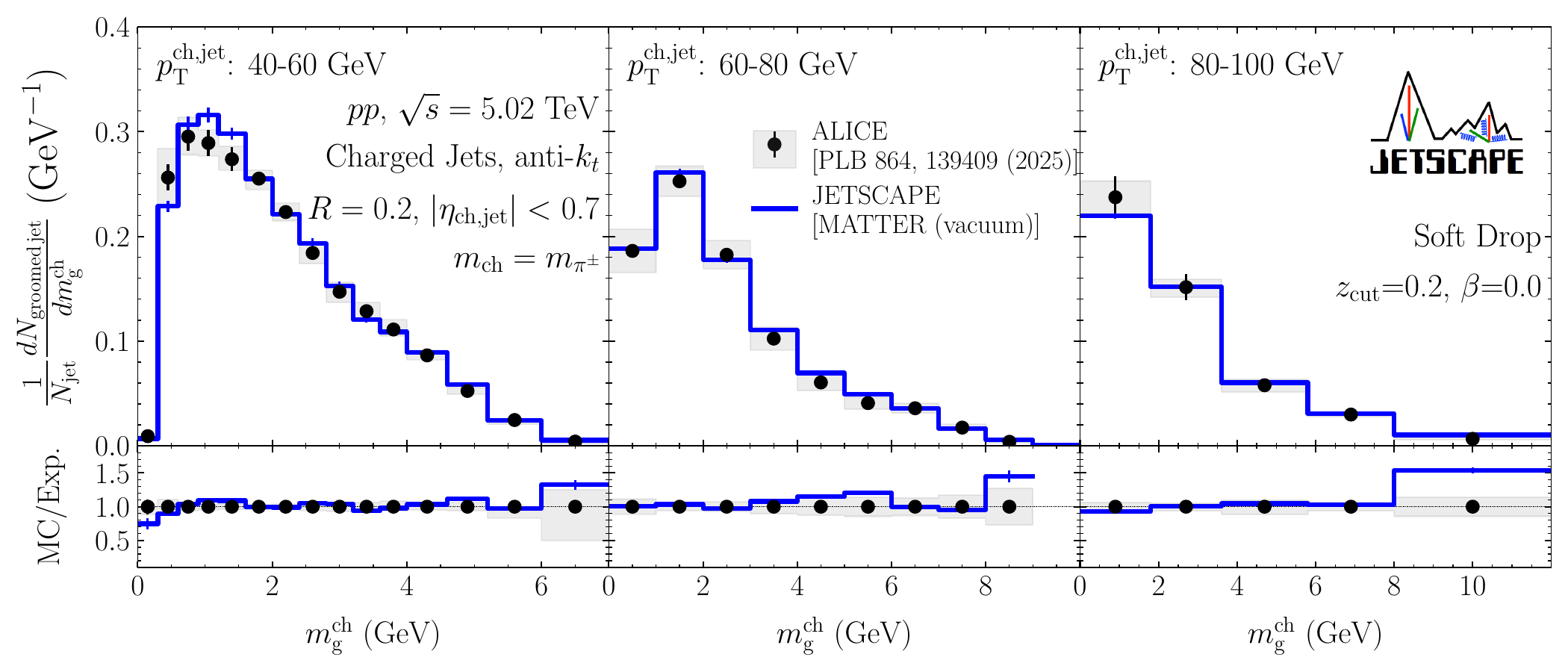}
\caption{Distributions of the soft-drop groomed mass, $m_{g}$ for charged jets with $R = 0.2$ and $| \eta_{\mathrm{ch,jet}} | < 0.7$ in $p$-$p$ collisions at $\sqrt{s} = 5.02$ TeV, and the ratios for different $p^{\mathrm{ch,jet}}_{T}$ range. 
The soft-drop parameters are $z_{\mathrm{cut}}=0.2$ and $\beta = 0$. 
The solid lines and circles with statistical error bars show the results from vacuum \textsc{matter} of \textsc{jetscape} and the experimental data from the ALICE Collaboration~\cite{ALICE:2024jtb:Manual}, respectively. 
The bands indicate the systematic uncertainties of the experimental data. 
}
\label{fig:alice_mg_pp}
\end{figure*}
\begin{figure*}[htb]
\centering
\includegraphics[width=0.98\textwidth]{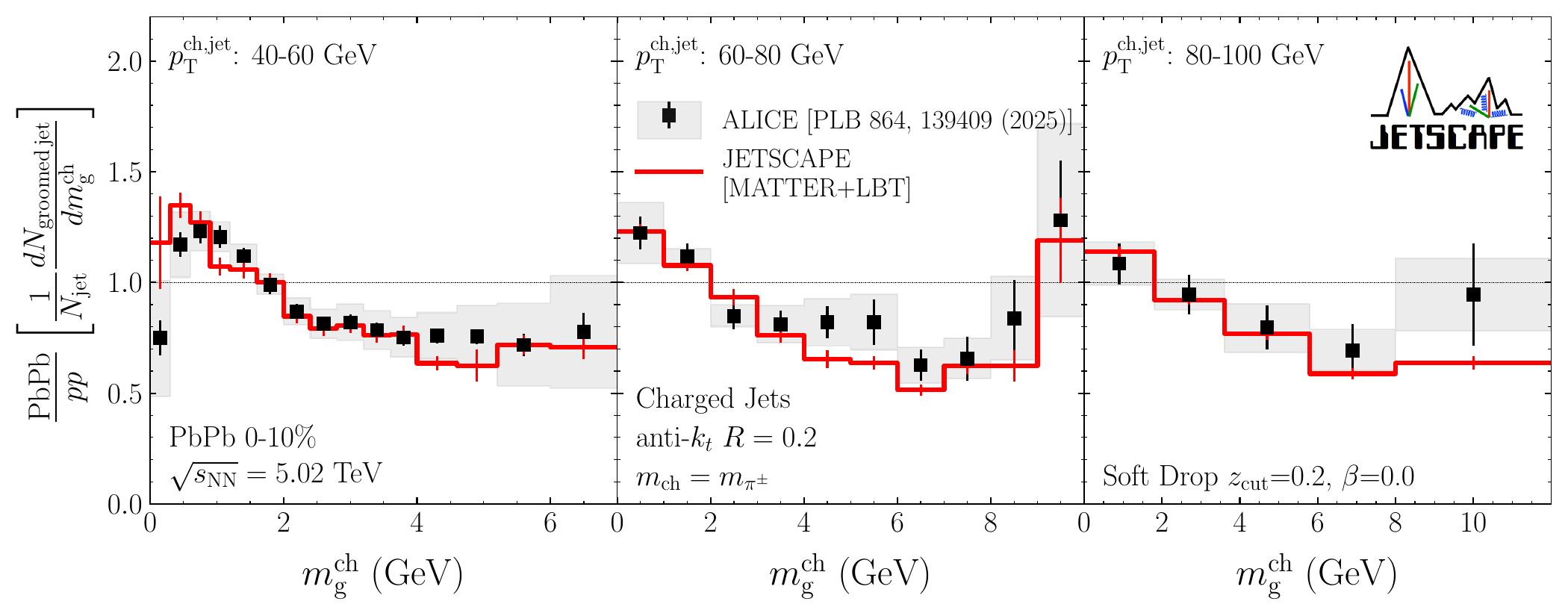}
\caption{Ratios of soft-drop $m_{g}$ distributions between $0\%$--$10\%$ Pb-Pb and $p$-$p$ collisions at $\sqrt{s_{NN}}=5.02$~TeV for charged jets with $R = 0.2$ and $| \eta_{\mathrm{ch,jet}} | < 0.7$ for different $p^{\mathrm{ch,jet}}_{T}$ range. 
The soft-drop parameters are $z_{\mathrm{cut}}=0.2$ and $\beta = 0$. 
The solid lines and squares with statistical error bars show the results from \textsc{matter}+\textsc{lbt} of \textsc{jetscape} and the experimental data from the ALICE Collaboration~\cite{ALICE:2024jtb:Manual}, respectively. 
The bands indicate the systematic uncertainties of the experimental data. 
}
\label{fig:alice_mg_pbpb}
\end{figure*}
Here, we present a comparison of our results for soft-drop-groomed jet mass $m_{g}$ with experimental results. 
The comparison of the unfolded $m_g / p_T$ distribution for full jets with $R = 0.4$ in $p$-$p$ collisions from CMS~\cite{CMS:2018fof} is shown in Fig.~\ref{fig:cms_mg_pp}, while the comparison of the $m_g$ distribution for charged jets with $R = 0.2$ in $p$-$p$ collisions from ALICE~\cite{ALICE:2024jtb:Manual} is shown in Fig.~\ref{fig:alice_mg_pp}. 
The results for $p$-$p$ collisions are in agreement with the experimental data within the uncertainty range, except for regions with significantly low jet counts, such as the tails of the distributions. 
It should be noted that, while our calculations are based on Monte Carlo simulations~\cite{Majumder:2014gda}, the masses of jet constituents were adjusted to align with experimental methodologies: massless for comparison with CMS data and the charged pion mass for comparison with ALICE data. 
This adjustment, while necessary to match experimental analysis techniques, has a non-negligible impact on the results, leading to noticeable changes in the distributions.

Figure~\ref{fig:alice_mg_pbpb} shows the modifications of the soft-drop-groomed jet mass distributions in $0\%$--$10\%$ Pb-Pb collisions, compared with the ALICE data~\cite{ALICE:2024jtb:Manual}. 
As with the inclusive full jet results presented in the main text, the behavior is dominated by the $p^{\mathrm{jet}}_{T}$ cut-induced selection bias and the loss of prong mass, leading to suppression at large $m_g$. 
The results capture the experimental trends and agree with the data within the uncertainty range across most $m_g$ regions. 
A slight enhancement observed in some $p^\mathrm{jet}_{T}$ ranges for the largest $m_g$ bin is primarily attributed to the very low jet counts in the tail of the distribution, particularly in $p$-$p$ collisions.

\subsection{Jet splitting radius of $\gamma$-tagged jet}
\label{subsection:ExpGammaTagged}
Here, we present the results for the jet splitting radius $r_{g}$ of $\gamma$-tagged jets from simulations conducted using the \textsc{pythiagun} module with \texttt{HardQCD:all=off} and \texttt{PromptPhoton:all=on}. 
\begin{figure*}[htb]
\centering
\includegraphics[width=0.8\textwidth]{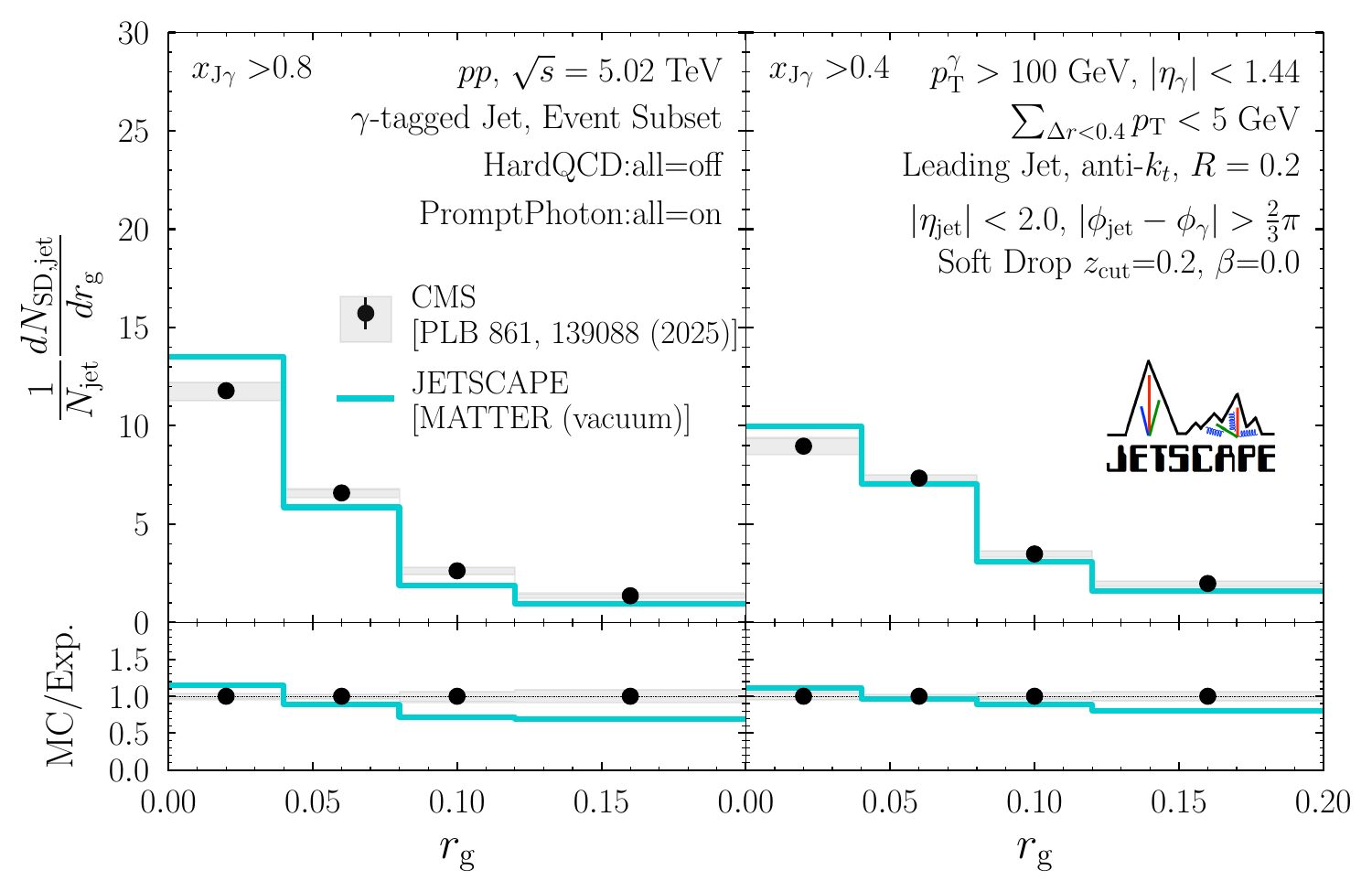}
\caption{Distributions of the soft-drop-groomed mass, $r_{g}$ for $\gamma$-tagged jets with $R = 0.2$ in $p$-$p$ collisions at $\sqrt{s} = 5.02$ TeV, and the ratios for $x_{J\gamma}>0.8$ (left) and $x_{J\gamma}>0.4$ (right). 
Leading full jets within $|\eta_{\mathrm{jet}}| < 2$ and $|\phi_{\mathrm{jet}} - \phi_{\gamma}| > 2\pi/3$ are paired with isolated photons satisfying $p^{\gamma}_T > 100~\mathrm{GeV}$ and $|\eta_{\gamma}| < 1.44$ as the trigger condition. 
The soft-drop parameters are $z_{\mathrm{cut}}=0.2$ and $\beta = 0$. 
The solid lines and circles with statistical error bars show the results from vacuum \textsc{matter} of \textsc{jetscape} and the experimental data from the CMS Collaboration~\cite{CMS:2024zjn:Manual}, respectively. 
The bands indicate the systematic uncertainties of the experimental data. 
}
\label{fig:cms_rg_pp}
\end{figure*}
\begin{figure*}[htb]
\centering
\includegraphics[width=0.8\textwidth]{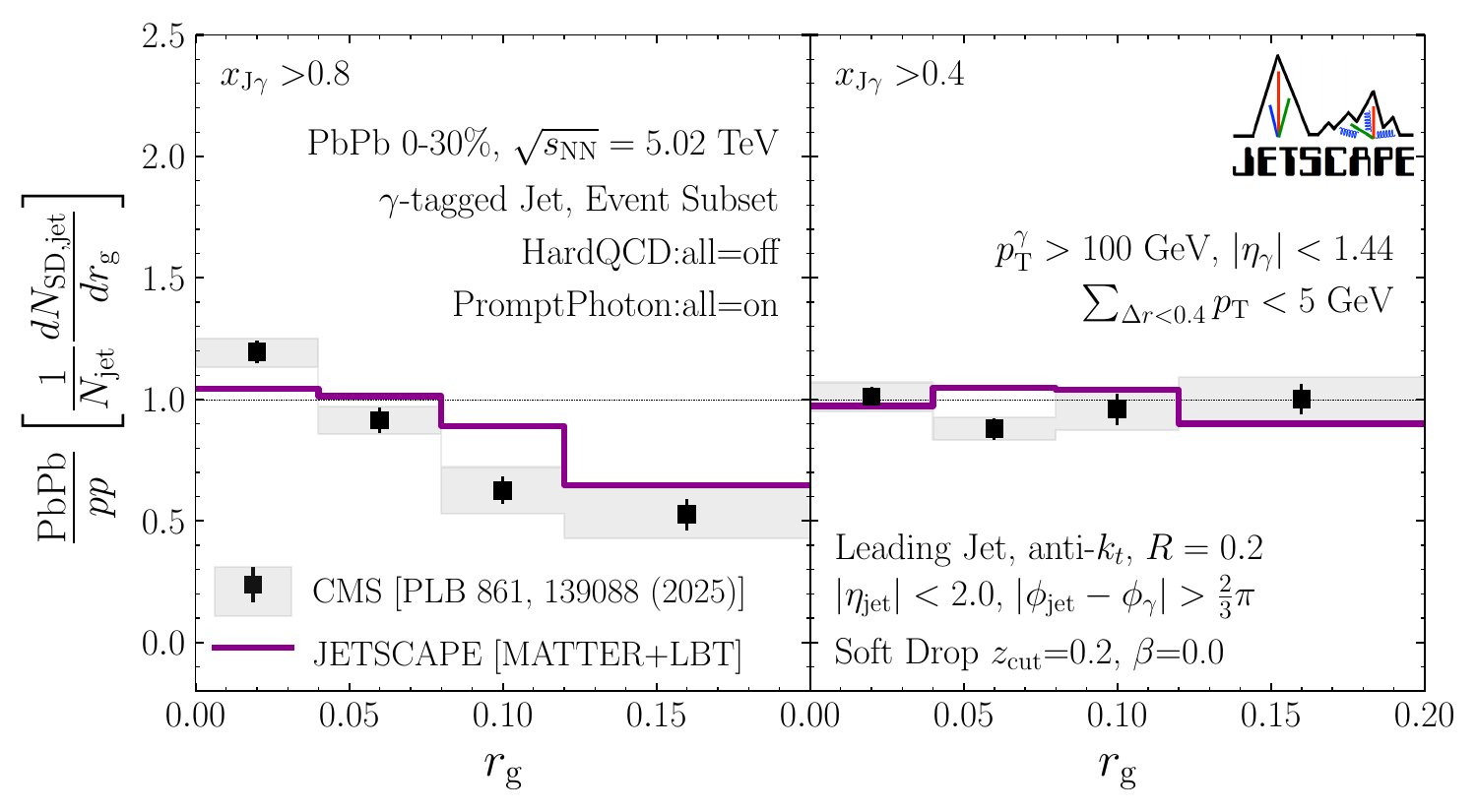} 
\caption{Ratios of soft-drop $r_{g}$ distributions between $0\%$--$30\%$ Pb-Pb and $p$-$p$ collisions at $\sqrt{s_{NN}}=5.02$~TeV. 
Full jets of highest $p^{\mathrm{jet}}_{T}$ within $|\eta_{\mathrm{jet}}| < 2$ and $|\phi_{\mathrm{jet}} - \phi_{\gamma}| > 2\pi/3$ are paired with isolated photons satisfying $p^{\gamma}_T > 100~\mathrm{GeV}$ and $|\eta_{\gamma}| < 1.44$ as the trigger condition. 
The soft-drop parameters are $z_{\mathrm{cut}}=0.2$ and $\beta = 0$. 
The solid lines and squares with statistical error bars show the results from \textsc{matter}+\textsc{lbt} of \textsc{jetscape} and the experimental data from the CMS Collaboration~\cite{CMS:2024zjn:Manual}, respectively. 
The bands indicate the systematic uncertainties of the experimental data. }
\label{fig:cms_rg_pbpb}
\end{figure*} 
Figures~\ref{fig:cms_rg_pp} and \ref{fig:cms_rg_pbpb} present comparisons with CMS data~\cite{CMS:2024zjn:Manual} for $r_{g}$ distributions in $p$-$p$ collisions and their modifications in $0\%$--$30\%$ Pb-Pb collisions, respectively. 
Results are shown for two lower cuts on $x_{J\gamma}$: $0.8$ and $0.4$. 
For $p$-$p$ collisions, the distributions obtained from the vacuum \textsc{matter} calculations are narrower than the experimental data for both $x_{J\gamma}$ cut values. 
This behavior is consistent with the trends observed in other Monte Carlo model calculations compared in Ref.~\cite{CMS:2024zjn:Manual}.

For the Pb-Pb results, at $x_{J\gamma} > 0.8$, suppression at large $r_g$ region, attributed to selection bias, is observed, similar to the inclusive jet results presented in the main text and in our previous study~\cite{JETSCAPE:2023hqn}, as well as the $\gamma$-tagged jet results with large $x_{J\gamma}$ cuts in the main text. 
While there is a sizable quantitative difference compared with the experimental data, the results qualitatively reproduce the observed trends.

On the other hand, for $x_{J\gamma} > 0.4$, as investigated in the $x_{J\gamma}$ dependence presented in the main text, the effect of selection bias is mitigated, leading to the disappearance of the suppression in the large $r_g$ region. 
This behavior closely matches the experimental observations. 
Furthermore, as discussed in the main text, the strong modification in splitting structures, such as the pronounced bump associated with the dominance of quark jets, are not clearly visible with the current binning resolution. 
Future measurements with higher precision are anticipated to reveal this feature more distinctly.

\section{Summary}
\label{Section:Summary}
In this work, we investigated medium modifications in jet substructure observables from soft-drop grooming, including the jet splitting momentum fraction ($z_{g}$), splitting radius ($r_{g}$), relative transverse momentum of splittings ($k_{T,g}$), and groomed jet mass ($m_{g}$), in $0\%$--$10\%$ Pb-Pb collisions at $\sqrt{s_{NN}} = 5.02$~TeV. 
These analyses utilized the \textsc{matter}$+$\textsc{lbt} multistage jet evolution model within the \textsc{jetscape} framework, using the JETSCAPEv3.5 AA22 tune~\cite{JETSCAPE:2022jer}, as employed in previous studies~\cite{JETSCAPE:2022hcb, JETSCAPE:2023hqn}. In particular, this effort should be compared with the closely related earlier companion paper Ref.~\cite{JETSCAPE:2024nkj}, which studied medium effects on $\gamma$-triggered jets. 
For the initial hard processes, events were generated with different configurations to isolate specific effects, primarily from the flavor of the parent partons, energy loss, and the selection bias introduced by the $p_{T}$ trigger.

The \textsc{pgun} simulations, in which a single hard parton with specified initial energy and flavor (light quark or gluon) evolves through shower development, reveal a clear flavor dependence in medium modifications to $r_{g}$, $k_{T,g}$, and $m_{g}$. 
Quark jets exhibit pronounced bump structures absent in gluon jets, with most of the quark-jet bump originating from prongs formed by recoil partons from medium scatterings that satisfy the soft-drop condition and are identified as hard branchings. 
When recoil effects are artificially disabled, quark jets show only a slight broadening in $r_{g}$ due to other medium effects, such as the transverse diffusion via scatterings and the medium-induced radiation, while the associated branchings remain soft, leading to no visible enhancement in $k_{T,g}$ or $m_{g}$ and resulting in a modification pattern qualitatively identical to gluon jets. 
The clear flavor dependence arises because quark jets have an intrinsically narrow structure and relatively fewer radiations, making it easier for medium effects to modify the hard prong structures reconstructed by the soft-drop algorithm, whereas the hard, vacuum-like branchings of gluon jets remain largely unaffected. 
The \textsc{pgun} results also highlight the selection bias, which dominates inclusive jet measurements and suppresses jets with broader structures. Lowering the $p^{\mathrm{jet}}_{T}$ cut mitigates this effect, fully eliminating the suppression in $r_{g}$ distributions and partially reducing it in $k_{T,g}$ and $m_{g}$ distributions. 
All these medium modifications were found to originate predominantly at low virtuality ($Q < 2$~GeV).

We also presented predictions for $\gamma$-tagged jets by performing simulations with the \textsc{matter}$+$\textsc{lbt} setup within the \textsc{jetscape} framework, incorporating a realistic initial hard process generated by \textsc{pythia}~8. These simulations revealed clear medium-induced modification in $r_{g}$, $k_{T,g}$, and $m_{g}$ of soft-drop-reconstructed hard branchings. 
Unlike inclusive jets, which exhibited smooth, monotonic suppression dominated by selection bias, $\gamma$-tagged jets displayed a flat structure at small $r_{g}$ and $k_{T,g}$ and a distinct bump in $m_{g}$, amplified by their quark-jet-dominated nature. 
Further analysis demonstrated that lowering the $x_{J\gamma}$ cut for $\gamma$-tagged jets effectively reduced selection bias, making the intrinsic modifications more apparent.

This study underscores the utility of $\gamma$-tagged jets, and similarly $Z$-tagged jets, as powerful probes for disentangling medium effects on jet substructure and advancing our understanding of jet-QGP interactions in high-energy heavy-ion collisions. 
Once measured with high precision in experiments, incorporating these observables into Bayesian analyses with theoretical models could provide stringent constraints on the parameters governing the fundamental mechanisms of jet-QGP medium interactions. 
Furthermore, if a bump-like structure indicating a broadening of the hard substructure is confirmed experimentally, our findings would support the interpretation that the most pronounced medium effects arise primarily from the relatively hard contributions of recoils, making these observables particularly valuable for detailed investigations of recoil dynamics~\cite{Milhano:2017nzm,DEramo:2018eoy,Pablos:2024muu}.

\acknowledgments
\label{Ack}
This work was supported in part by the National Science Foundation (NSF) within the framework of the JETSCAPE collaboration, under grant number OAC-2004571 (CSSI:X-SCAPE). It was also supported under  PHY-1516590 and PHY-1812431 (R.J.F., M.Ko., C.P. and A.S.); it was supported in part by the US Department of Energy, Office of Science, Office of Nuclear Physics under Grants No. \rm{DE-AC02-05CH11231} (X.-N.W. and W.Z.), No. \rm{DE-AC52-07NA27344} (A.A., R.A.S.), No. \rm{DE-SC0013460} (A.K., A.M., C.Sh., I.S., C.Si and R.D.), No. \rm{DE-SC0021969} (C.Sh. and W.Z.), No. \rm{DE-SC0024232} (C.Sh. and H.R.), No. \rm{DE-SC0012704} (B.S.), No. \rm{DE-FG02-92ER40713} (J.H.P. and M.Ke.), No. \rm{DE-FG02-05ER41367} (C.Si, D.S. and S.A.B.), No. \rm{DE-SC0024660} (R.K.E), and No. \rm{DE-SC0024347} (J.-F.P. and M.S.). The work was also supported in part by the National Science Foundation of China (NSFC) under Grant No. 11935007, No. 11861131009 and No. 11890714 (Y.H. and X.-N.W.), by the Natural Sciences and Engineering Research Council of Canada (C.G., S.J., and G.V.),  by the University of Regina President's Tri-Agency Grant Support Program (G.V.), by the Canada Research Chair program (G.V. and A.K.) reference number CRC-2022-00146, by the Office of the Vice President for Research (OVPR) at Wayne State University (Y.T.), by JSPS KAKENHI Grants No. 22K14041 and No. 25K07303 (Y.T.), and by the S\~{a}o Paulo Research Foundation (FAPESP) under Projects 2016/24029-6, 2017/05685-2 and 2018/24720-6 (M.L.). C.Sh., J.-F.P., and R.K.E. acknowledge a DOE Office of Science Early Career Award. I.~S. was funded as part of the European Research Council project ERC-2018-ADG-835105 YoctoLHC, and as a part of the Center of Excellence in Quark Matter of the Academy of Finland (project 346325). 
Calculations for this work used the Wayne State Grid, generously supported by the Office of the Vice President of Research (OVPR) at Wayne State University. 
The bulk medium simulations were performed using resources provided by the Open Science Grid (OSG) \cite{Pordes:2007zzb, Sfiligoi:2009cct}, which is supported by the National Science Foundation award \#2030508.
Data storage was provided in part by the OSIRIS project supported by the National Science Foundation under Grant No. OAC-1541335.

\section*{DATA AVAILABILITY}
All datasets for the current study were generated and analyzed using the open-source software framework \textsc{jetscapev3.5} with the AA22 tune, available at~\cite{JETSCAPECODE}. The results can be
reproduced using the software and methods described in the manuscript.

\bibliography{main,manual,misc}

\end{document}